\documentclass[prb,notitlepage,aps]{revtex4-2}
\usepackage{epsfig}
\usepackage{graphicx}
\usepackage{amsmath}
\usepackage{bm}
\usepackage{comment}
\usepackage{multirow}
\usepackage{color}
\usepackage{hyperref}
\hypersetup{
    colorlinks=true,
    linkcolor=blue,
    filecolor=magenta,      
    urlcolor=cyan,
    pdftitle={Superconductivity of anomalous pseudospin},
    pdfpagemode=FullScreen,
}

\begin{document}

\title{Superconductivity of anomalous pseudospin}
 
    \author{Han Gyeol Suh$^1$}
    \thanks{These authors contributed equally.}
    \author{Yue Yu$^{1,2}$}%
    \thanks{These authors contributed equally.}
    \author{Tatsuya Shishidou$^{1}$}%
    \author{Michael Weinert$^1$}
    \author{P. M. R. Brydon$^3$}
    \author{Daniel F. Agterberg$^{1}$}
    \thanks{Corresponding author \href{mailto:agterber@uwm.edu}{agterber@uwm.edu}}

\affiliation{$^1$Department of Physics, University of Wisconsin, Milwaukee, Wisconsin 53201, USA}
\affiliation{$^2$Department of Physics, Stanford University, 476 Lomita Mall, Stanford, CA 94305, USA}
\affiliation{$^3$Department of Physics and MacDiarmid Institute for
Advanced Materials and Nanotechnology, University of Otago, P.O. Box
56, Dunedin 9054, New Zealand}

\begin{abstract}

Spin-orbit coupling driven by broken inversion symmetry ($I$) is known to lead to unusual magnetic response of superconductors, including extremely large critical fields for spin-singlet superconductors. This unusual response is  also known to  appear in materials that have $I$, provided there is local $I$-breaking: fermions participating in superconductivity reside on crystal sites that lack $I$. Here we show that this unusual response exists even when the crystal sites preserve $I$. Indeed, we argue that the symmetry of Kramers degenerate fermionic pseudospin is more relevant than the local crystal site symmetry. We examine and classify non-symmorphic materials with momentum space spin-textures  that exhibit an anomalous pseudospin with different symmetry properties than usual spin-1/2.  We find that this anomalous pseudospin does not depend on the existence of local $I$ breaking crystal sites and it optimizes the unusual magnetic response traditionally associated with locally noncentrosymmetric superconductors, dramatically extending the range of relevant materials. We further show this anomalous pseudospin leads to fully gapped `nodal' superconductors and provides additional insight into the breakdown of Blount's theorem for pseudospin triplet superconductors. We apply our results to UPt$_3$, BiS$_2$-based superconductors, Fe-based superconductors, and paramagnetic UCoGe.
\vglue 0.5 cm

\end{abstract}
\maketitle
\section{Introduction} 
\noindent Momentum space spin-textures of electronic bands are known to underlie spintronic and superconducting properties of quantum materials \cite{Manchon:2015,Smidman:2017,Baltz:2018}. In the spintronics context, Rashba-like spin textures allow control of electronic spin through applied electric fields \cite{Manchon:2015, Baltz:2018}. In superconductors, these same spin textures lead to unusual and counter-intuitive magnetic response, such as the robustness of spin-singlet superconductivity to applied magnetic fields, pair density wave  states,  and singlet-triplet mixing \cite{Smidman:2017}. While such spin-textures are common when inversion symmetry ($I$) is broken, it has been realized that these can also occur when $I$ is present. This has lead to the notion of hidden spin-textures \cite{Zhang:2014} and locally non-centrosymmetric superconductivity \cite{Fischer:2023},  where $I$ related sectors each allow a Rashba-like spin-texture due to the local $I$ breaking. These spin-textures are of opposite sign on the two sectors, so that global inversion symmetry is restored. These hidden spin-textures allow the novel physics associated with spin-orbit coupling (SOC) to emerge even when $I$ is not broken. It further allows for new physics to emerge. One notable example is a field induced transition from an even-parity (pseudospin singlet) to odd-parity (pseudospin triplet) observed in CeRh$_2$As$_2$ \cite{Yoshida:2012,Khim:2021,Landaeta:2022,Cavanagh:2022}.  

In materials with inversion symmetry, we call the above mentioned strongly anisotropic Pauli limiting fields (and related anisotropic spin susceptibilities), fields far exceeding the usual Pauli limiting field, and field induced transitions between different superconducting states, unusual magnetic response. Key to observing this unusual magnetic response associated with the spin-textures in inversion symmetric materials, is that the $I$ related sectors are weakly coupled \cite{Cavanagh:2022,Yuan:2019,Pengke:2018,Fischer:2023}. Theoretical proposals for how to achieve this fall under two approaches: the first is to tailor weak coupling between the inversion related sectors, for example by separating two  inversion symmetry related layers so that the interlayer coupling is weak \cite{Yoshida:2012}; the second is to exploit symmetries that ensure that this inter-sector coupling vanishes. The symmetry based approach has been applied to points and lines in momentum space. Examples include two-dimensional (2D) transition metal dichalcogenides near the K-point \cite{Nakamura:2017} and non-symmorphic symmetries near the $X-M$ line in BaNiS$_2$ with space group 129 (P4/nmm) \cite{Yuan:2019}. Recently, we have generalized this to planes in momentum space through an analysis of the locally non-centrosymmetric superconductor CeRh$_2$As$_2$ \cite{Cavanagh:2022}.   In all these cases, the only energy splitting between the inversion-related sectors is due to SOC - a situation conceptually similar to materials with broken $I$, where the usual two-fold pseudopsin degeneracy is broken solely by SOC. Indeed, this suggests another route toward tailoring  unusual magnetic response of superconductors with inversion symmetry: Instead of emphasizing the local $I$ breaking, as has been done in the examples described above, it may be fruitful to identify electronic degeneracies that are broken solely by SOC. This is the approach we take here and we find it naturally leads to the desired unusual magnetic response. Furthermore, we find it does not require crystal site symmetries with local $I$ breaking, but rather is dictated by the symmetry of the Bloch fermion pseudospin. As pointed out by Anderson in 1984, \cite{Anderson:1984}, fermion pseudospin, derived from the two-fold Kramer's degeneracy originating from $TI$ symmetry (where $T$ is time-reversal symmetry), plays a fundamental role in superconductivity. Here we find that when the band degeneracy is lifted solely by SOC,  this pseudospin has different symmetry properties than usual spin-1/2 pseudospin.

Specifically, we identify electronic band degeneracies that are split {\it solely} by SOC in materials with both inversion, $I$ and time-reversal $T$, symmetries. This requires bands that are at least four-fold degenerate when SOC is ignored. Such band degeneracies are not generic and require symmetries beyond the usual two-fold pseudospin (or Kramers) degeneracy that arises from $TI$ symmetry.  As discussed in a variety of contexts \cite{Bradley:1972,Zhang:2018,Hirschmann:2021,Leonhardt:2021}, such degeneracies can arise in non-symmorphic crystal structures.  Here we focus on the largest momentum region in the 3D Brillouin zone that allows such degeneracies. This occurs on 2D momentum planes, which are often called nodal planes. More specifically, this is the largest region in momentum space for which the required four-fold electronic degeneracies can appear when SOC is ignored. Here we provide a complete list of space groups for which this occurs and provide symmetry based $kp$ theories for all time-reversal-invariant momenta (TRIM) on these nodal  planes. As discussed later, many relevant superconductors exhibit Fermi surfaces near these TRIM. We find that the SOC-split electronic states on these nodal planes generically exhibit a pseudospin that has a different symmetry than that of usual spin-1/2 fermions (this generalizes a result we found for space group P4/nmm in the context of locally non-centrosymmetric superconductor CeRh$_2$As$_2$ \cite{Cavanagh:2022}). Here we name this anomalous pseudospin and examine the consequences of this anomalous pseudospin on superconductivity. We find that this anomalous pseudospin plays a central role on the superconducting magnetic response and on the properties of spin-triplet superconductivity. Our results provide further insight on earlier nodal and topological classifications of superconductivity in non-symmorphic materials \cite{Norman:1995,Micklitz:2017,Micklitz:2017-2,Daido:2019,Yanase:2016,Sumita:2018,Kobayashi:2016,Kobayashi:2014,Micklitz:2009,Taufour:2022}.  Furthermore, all the non-symmorphic crystal structures we examine have Wyckoff positions with site symmetries that contain inversion symmetry. So, although unusual magnetic response is typically associated with locally noncentrosymmetric superconductors, our theory establishes that the local $I$  breaking is not an essential ingredient, and our classification may guide the experimental search for new materials where local $I$ breaking is not a feature. 

In this paper we begin by defining anomalous pseudospin on nodal momenta planes, we then characterize all possible symmetry based $kp$ theories near TRIM points on these nodal planes. Using these $kp$ theories, we analyze the magnetic response and nodal excitations of superconducting states formed from anomalous pseudospin. We apply this analysis to a series of materials that  exhibit Fermi surfaces that lie on or near these nodal planes.  More specifically we reveal how anomalous pseudospin: explains critical fields that far exceed the Pauli field in BiS$_2$-based materials \cite{Mizuguchi:2015} and the observed magnetic response 3D Fe-based superconductors \cite{Stewart:2011}; identifies which space groups and TRIM are ideal to find a  field induced even parity to odd parity transition akin to that observed in CeRh$_2$As$_2$ \cite{Khim:2021}; provides insight into the gap symmetry of UPt$_3$ \cite{Joynt:2002}; and shines new light on re-entrant  superconductivity in UCoGe \cite{Aoki:2019}.

\section{Anomalous pseudospin: symmetry origin}

Our aim is to exploit symmetry to find nodal plane band degeneracies that are lifted solely by SOC. As discussed below, once these band degeneracies are lifted, a two-fold pseudospin degeneracy will remain.  We find that generically, the  pseudospin that results from this procedure does not share the same symmetry properties as usual spin 1/2 and hence we name this anomalous pseudospin. 

Pseudospin describes the two-fold Kramers degeneracy that arises at each momentum point ${\bm k}$ when the product of  time-reversal $T$ and inversion $I$ symmetries, $TI$, is present. The product $TI$ is anti-unitary and for fermions satisfies $(TI)^2=-1$, ensuring at least a two-fold degeneracy. It is often the case that this pseudospin behaves as spin-1/2 under rotations \cite{Fu:2015}. However, when symmetries beyond $TI$ are present, it is possible that this is not the case. One example of this is the angular momentum $j_z=\pm 3/2$ electronic states that arise when cubic symmetry  or a three-fold rotation axis is present \cite{Brydon:2016,Smidman:2017,Wang:2019}. In the latter case, this gives rise to so-called type-II Ising superconductivity in 2D materials  \cite{Wang:2019,Falson:2020} where large in-plane critical fields appear when the Fermi surface is sufficiently close to momentum points with this three-fold rotation symmetry. A systematic analysis of the appearance of anomalous pseudospin for fermions near the $\Gamma$ point has been carried out \cite{Samokhin:2019-2,Samokhin:2020,Samokhin:2021}. In our case, the anomalous pseudospin appears on momentum planes in the Brillouin zone, allowing a larger phase space for the physical properties of anomalous pseudospin to manifest. 

To ensure the requisite band degeneracy on a nodal plane, consider the symmetry elements that keep a momentum point on the plane invariant (here taken to be normal to the $\bm{\hat{n}}$ axis). These are  $\{E,\tilde{M}_{\bm{\hat{n}}},TI,T\tilde{C}_{2,\bm{\hat{n}}}\}$, where $\tilde{M}_{\bm{\hat{n}}}$ is a translation mirror symmetry and $\tilde{C}_{2,\bm{\hat{n}}}$ is a translation two-fold rotation symmetry. Their point group rotation and translation component can be denoted using Seitz notation, for example $\tilde{M}_{\bm{\hat{n}}}=\{M_{\bm{\hat{n}}}|t_1,t_2,t_3\}$ where $M_{\bm{\hat{n}}}$ is a point group mirror symmetry along $\hat{n}$ and $(t_1,t_2,t_3)$ is a fractional translation vector (here the $t_3$ is the translation component parallel to $\hat{n}$). Since we are searching for a degeneracy that appears without SOC, we consider orbital or sublattice degrees of freedom for which $(TI)^2=1$. The only remaining symmetry that can enforce a two-fold degeneracy is $T\tilde{C}_{2,\bm{\hat{n}}}$, since this is anti-unitary, it must satisfy $(T\tilde{C}_{2,\bm{\hat{n}}})^2=-1$ to do so. Since $T$ commutes with rotations, this implies $\tilde{C}_{2,\bm{\hat{n}}}^2=-1$. When operating on orbital or sublattice degrees of freedom, $\tilde{C}_{2,\bm{\hat{n}}}^2$ is typically 1, suggesting it is not possible to have the required degeneracy. However, in non-symmorphic groups, $\tilde{C}_{2,\bm{\hat{n}}}$ can be a screw axis, for which it is possible to satisfy $\tilde{C}_{2,\bm{\hat{n}}}^2=-1$. In particular, using Seitz notation $\tilde{C}_{2,\bm{\hat{n}}}=\{C_{2\bm{\hat{n}}}|t_1,t_2,1/2\}$ (here $t_1$ and $t_2$ correspond to either a half in-plane translation vector or to no translation) we have $(\tilde{C}_{2,\bm{\hat{n}}})^2=\{E|0,0,1\}$. When operating on a state carrying momentum ${\bm k}$, $(\tilde{C}_{2,\bm{\hat{n}}})^2$ is represented by $e^{i{\bm k}\cdot \bm{\hat{n}}}$.  Hence if the nodal plane sits at momentum ${\bm k}\cdot \bm{\hat{n}}=\pi$, then $\tilde{C}_{2,\bm{\hat{n}}}^2=-1$ and a two-fold orbital or sublattice degeneracy is ensured. When spin-degeneracy is also included, these states are then four-fold degenerate when SOC is ignored. 

When SOC is included, it is possible to show that the $TI$ pseudospin partners have the same $M_{\bm{\hat{n}}}$ mirror eigenvalue (this result is generalization of that given in  Ref.~\cite{Cavanagh:2022} where $t_1=0$ and $t_2=0$ was used). That is, labeling the two Kramers degenerate states as $|+\rangle$ and $TI|+\rangle$, both  belong to the same eigenstate of $\tilde{M}_{\bm{\hat{n}}}$. As a consequence, all Pauli matrices $\tilde{\sigma}_i$ made from the two states $|+\rangle$ $TI|+\rangle$ must all be invariant under $\tilde{M}_{\bm{\hat{n}}}$. It is this feature that differs from usual spin-1/2. Of the three Pauli matrices $\sigma_i$, constructed from usual spin-1/2 states, two will be odd under $\tilde{M}_{\bm{\hat{n}}}$ and one will be even under $\tilde{M}_{\bm{\hat{n}}}$. It is this symmetry distinction between the anomalous pseudospin operators  ($\tilde{\sigma}_i$) and usual spin 1/2 operators ($\sigma_i$) that underlie the unusual superconducting properties discussed below. 

The above argument can also be applied to nodal lines generated by the symmetry elements $\{E,\tilde{C}_{2,\bm{\hat{n}}}, TI, T\tilde{M}_{\bm{\hat{n}}}\}$ with $(T\tilde{M}_{\bm{\hat{n}}})^2=-1$ when applied to orbital or sublattice degrees of freedom. In this case, repeating the same arguments above show that  SOC will also split the band degeneracy and lead to anomalous pseudospin. Here, due to the larger available momentum phase space, we restrict our analysis and classification to nodal planes and leave an analysis of nodal lines to a later work. For all space groups that host nodal planes, we develop symmetry-based $kp$ theories valid near all TRIM on these nodal planes. We emphasize these TRIM since Cooper pairs are formed by pairing states at momenta ${\bm k}$ and $-{\bm k}$ with the momentum origin given by a TRIM.  We then consider Fermi surfaces near these TRIM and discuss the resultant superconducting properties. Figure 1 illustrates our approach. Here, in green, we show the nodal planes and lines that exhibit anomalous pseudospin. Here we examine the properties of superconductivity for a Fermi surface near the $Z$ point, which is a TRIM on the nodal plane. The properties of superconductivity for a Fermi surface near the $\Gamma$ point, for which pseudospin is typically not anomalous, are described in earlier review articles  \cite{Sigrist:1991, Gorkov:2001}. We note that many superconducting materials, including the examples discussed in this paper, exhibit Fermi surfaces near nodal planes.

\begin{figure*}[tt]
	\centering
	\includegraphics[width=0.5\linewidth]{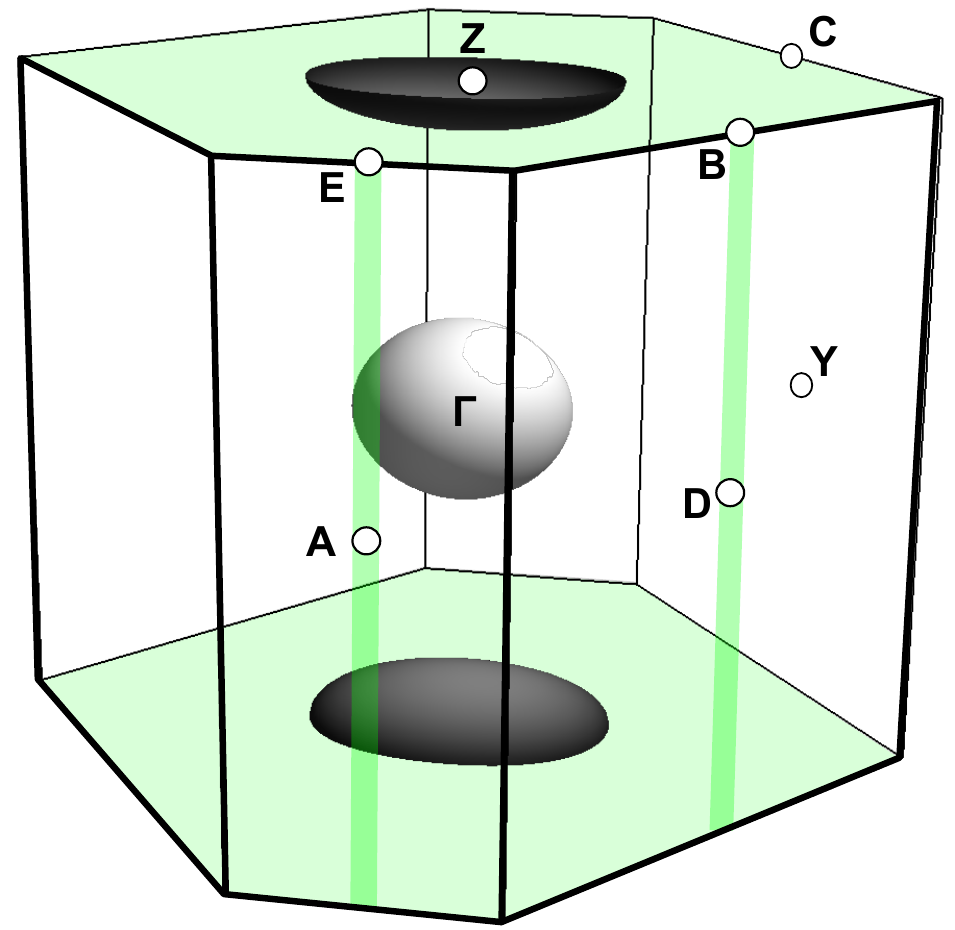}
	\caption{\label{fig1} Example from space group 14 where the green shading reveals the planes and lines in momentum space on which anomalous pseudospin exists. A Fermi surface located near the momentum plane $k_z=\pi$ (as depicted by the dark Fermi surface near the $
	Z$ point) will have its superconducting properties governed by pairing of anomalous pseudospin. However, Fermi surfaces far from these planes (such as that depicted near the $\Gamma$ point) will exhibit more usual superconducting properties.}
\end{figure*}


\section{Nodal plane space groups and single-particle $kp$ Hamiltonians}

Here we identify all space groups that allow anomalous pseudospin on nodal planes  and construct the corresponding symmetry-based $kp$-like Hamiltonians for all TRIM  on these planes. A key new result is that these $kp$ theories are of two types. Type 1 $kp$ theories have Hamiltonians of the same form generically examined in locally non-centrosymmetric superconductors and explicitly contain SOC terms that are odd in momentum $\bm{k}$. Type 2 $kp$ theories contain SOC terms that are {\it even} in momentum $\bm{k}$, and have not appeared in the context of locally non-centrosymmetric superconductors.

\subsection{Space groups with nodal planes}

To identify these nodal planes, all space groups containing inversion symmetry $I=\{I|0,0,0\}$ and the screw axis $\tilde{C}_{2,\bm{\hat{n}}}=\{C_{2\bm{\hat{n}}}|t_1,t_2,1/2\}$ (where $t_1=0,1/2$ and $t_2=0,1/2$) were identified. For these space groups, the nodal planes lie on the Brillouin zone boundary. Table 1 lists the resultant space groups, point groups, nodal planes, and types of $kp$ theories allowed for these space groups. As discussed in the previous section, the degeneracies of these nodal planes are generically lifted by SOC, yielding anomalous pseudospin.

\begin{table}
\begin{tabular}{|c|c|c|c|c|}
\hline
Crystal Type & Number & Name &Nodal planes & $kp$ theory classes\\\hline\hline
\multirow{2}{*}{Monoclinic $(C_{2h})$}&11& $P2_1/m$&$(u,1/2,w)$&
$C_{2h,1}^{\text{type}1}$\\ 
&14&$P2_1/c$&$(u,1/2,w)$&
$C_{2h,1}^{\text{type}1}$, $C_{2h,2}^{\text{type}2}$\\ 
\hline
\multirow{14}{*}{Orthorhombic $(D_{2h})$}&51&$Pmma$&$(1/2,v,w)$&
$D_{2h,3}^{\text{type}1}$\\ 
&52&$Pnna$&$(u,1/2,w)$&
$D_{2h,3}^{\text{type}1}$, $D_{2h,4}^{\text{type}2}$, 8-fold\\ 
&53&$Pmna$&$(u,v,1/2)$&
$D_{2h,3}^{\text{type}1}$, $D_{2h,4}^{\text{type}2}$\\ 
&54&$Pcca$&$(1/2,v,w)$&
$D_{2h,3}^{\text{type}1}$, 8-fold\\ 
&55&$Pbam$&$(1/2,v,w),(u,1/2,w)$&
$D_{2h,2}^{\text{type}2}$, $D_{2h,3}^{\text{type}1}$\\ 
&56&$Pccn$&$(1/2,v,w),(u,1/2,w)$&
$D_{2h,1}^{\text{type}1}$, $D_{2h,2}^{\text{type}2}$, $D_{2h,3}^{\text{type}1}$, 8-fold\\ 
&57&$Pbcm$&$(u,v,1/2),(u,1/2,w)$&
$D_{2h,3}^{\text{type}1}$, 8-fold\\ 
&58&$Pnnm$&$(1/2,v,w),(u,1/2,w)$&
$D_{2h,1}^{\text{type}1}$, $D_{2h,2}^{\text{type}2}$, $D_{2h,3}^{\text{type}1}$, $D_{2h,4}^{\text{type}2}$\\ 
&59&$Pmmn$&$(1/2,v,w),(u,1/2,w)$&
$D_{2h,1}^{\text{type}1}$, $D_{2h,3}^{\text{type}1}$\\ 
&60&$Pbcn$&$(1/2,v,w),(u,v,1/2)$&
$D_{2h,3}^{\text{type}1}$, $D_{2h,4}^{\text{type}2}$, 8-fold\\ 
&61&$Pbca$&$(1/2,v,w),(u,v,1/2),(u,1/2,w)$&
$D_{2h,3}^{\text{type}1}$, 8-fold\\ 
&62&$Pnma$&$(1/2,v,w),(u,v,1/2),(u,1/2,w)$&
$D_{2h,1}^{\text{type}1}$, $D_{2h,3}^{\text{type}1}$, 8-fold\\ 
&63&$Cmcm$&$(u,v,1/2)$&
$C_{2h,1}^{\text{type}1}$, $D_{2h,3}^{\text{type}1}$\\ 
&64&$Cmce$&$(u,v,1/2)$&
$C_{2h,2}^{\text{type}2}$, $D_{2h,3}^{\text{type}1}$\\ 
\hline
\multirow{8}{*}{Tetragonal $(D_{4h})$}&127&$P4/mbm$&$(u,1/2,w)$&
$D_{2h,3}^{\text{type}1}$, $D_{4h,2}^{\text{type}2}$, $D_{4h,4}^{\text{type}2}$\\ 
&128&$P4/mnc$&$(u,1/2,w)$&
$D_{2h,3}^{\text{type}1}$, $D_{2h,4}^{\text{type}2}$, $D_{4h,2}^{\text{type}2}$, $D_{4h,4}^{\text{type}2}$, $D_{4h,5}^{\text{type}1}$, 8-fold\\ 
&129&$P4/nmm$&$(u,1/2,w)$&
$D_{2h,3}^{\text{type}1}$, $D_{4h,1}^{\text{type}1}$, $D_{4h,3}^{\text{type}1}$\\ 
&130&$P4/ncc$&$(u,1/2,w)$&
$D_{2h,3}^{\text{type}1}$, $D_{4h,1}^{\text{type}1}$, $D_{4h,3}^{\text{type}1}$, 8-fold\\ 
&135&$P4_2/mbc$&$(u,1/2,w)$&
$D_{2h,3}^{\text{type}1}$, $D_{4h,2}^{\text{type}2}$, $D_{4h,4}^{\text{type}2}$, 8-fold\\ 
&136&$P4_2/mnm$&$(u,1/2,w)$&
$D_{2h,3}^{\text{type}1}$, $D_{2h,4}^{\text{type}2}$, $D_{4h,1}^{\text{type}1}$, $D_{4h,2}^{\text{type}2}$, $D_{4h,3}^{\text{type}1}$, $D_{4h,4}^{\text{type}2}$\\ 
&137&$P4_2/nmc$&$(u,1/2,w)$&
$D_{2h,3}^{\text{type}1}$, $D_{4h,1}^{\text{type}1}$, $D_{4h,3}^{\text{type}1}$, $D_{4h,5}^{\text{type}1}$, 8-fold\\ 
&138&$P4_2/ncm$&$(u,1/2,w)$&
$D_{2h,3}^{\text{type}1}$, $D_{4h,1}^{\text{type}1}$, $D_{4h,2}^{\text{type}2}$, $D_{4h,3}^{\text{type}1}$, $D_{4h,4}^{\text{type}2}$, 8-fold\\ 
\hline
\multirow{1}{*}{Hexagonal $(C_{6h})$}&176&$P6_3/m$&$(u,v,1/2)$&
$C_{2h,1}^{\text{type}1}$, $C_{6h}^{\text{type}1}$, 8-fold\\ \hline
\multirow{2}{*}{Hexagonal $(D_{6h})$}&193&$P6_3/mcm$&$(u,v,1/2)$&
$D_{2h,3}^{\text{type}1}$, $D_{6h}^{\text{type}1}$, 8-fold\\ 
&194&$P6_3/mmc$&$(u,v,1/2)$&
$D_{2h,3}^{\text{type}1}$, $D_{6h}^{\text{type}1}$, 8-fold\\ 
\hline
Cubic $(T_{h})$&205&$Pa\overline{3}$&$(u,1/2,w)$&
$D_{2h,3}^{\text{type}1}$, 8-fold\\ 
\hline
\end{tabular}
\caption{Space groups with nodal planes} \label{tab:sgmomenta}
\end{table}

\subsection{Symmetry based $kp$ theories near TRIM}

Understanding the consequences of anomalous pseudospin on superconductivity requires a theory for the normal state. Cooper pairs rely on the degeneracy between states of momenta ${\bm k}$ and $-{\bm k}$ and this degeneracy is ensured by both $T$ and $I$ symmetries. For this reason, we develop symmetry-based $kp$ theories expanded around TRIM. To derive these $kp$-like Hamiltonians, we have used the real representations for the TRIM given in the Bilbao Crystallographic server \cite{Aroyo:2006,Aroyo2:2006,Stokes:2013}. For these TRIM, we initially consider space group irreducible representations that do not include spin, which, for simplicity,  we name orbital representations. These representations are either 2-fold or 4-fold degenerate (when spin is added, these become 4-fold and 8-fold degenerate respectively). The full $kp$-like Hamiltonians are only listed for the 2-fold degenerate  representations. We present a partial classification of the 4-fold degenerate orbital representations near the end of this paper.

In constructing the $kp$ theories for the 2-fold orbital degenerate TRIM points, we choose \(\tau_i\) to be  Pauli matrices that encode the orbital degrees of freedom, and \(\sigma_i\) to be  spin Pauli matrices. We take $T=\tau_0 (i\sigma_y)K$ where $K$ is the complex conjugation operator, hence the $\tau_2$ operator is odd under time-reversal. For a given doubly degenerate space group representation on a TRIM, constructing its direct product leads to  four irreducible point group representations. These four representations each  correspond to an orbital operator \(\tau_i\), and this partially dictates the momentum dependencies of symmetry allowed terms in the {\it kp} Hamiltonian. We present our results for the $kp$ Hamiltonians in Table 2. The first row of each box gives the type of the $kp$ theory class and the point group representations of the orbital operators that are given by Pauli matrices $\tau_i$. In this decomposition, the square brackets correspond to the antisymmetric $\tau_2$ operator and remaining terms correspond to $\tau_0$, $\tau_1$, and $\tau_3$. The second row of a box gives the $kp$ Hamiltonian, and the last part of a box lists the space groups and TRIM points representations that belong to the $kp$ Hamiltonian class. We have tabulated the $kp$ Hamiltonians for 122 TRIM points and  we find that only 13 different $kp$ theories appear. These are of two types, which we call type 1 and type 2. Type 1 $kp$ theories have degenerate even and odd parity orbital basis functions. These Hamiltonians have a structure similar to those examined in the context of locally non-centrosymmetric superconductors \cite{Fischer:2023}. However, we note that local $I$ breaking on crystal sites is not required to generate Type 1 $kp$ theories. These Hamiltonians apply to all Wyckoff position site symmetries and the non-symmorphic groups we consider all include site symmetries that include $I$. For site symmetries that include $I$, the degenerate even and odd parity basis functions for type 1  Hamiltonians originate from the combination of non-symmorphic symmetries and  Bloch momenta at the zone boundary. Type 2 $kp$ theories have two degenerate orbital basis functions with the same parity symmetry. These Hamiltonian have a structure unlike that seen in locally noncentrosymmetric superconductors, yet as we show below, they exhibit a similar magnetic response.

The generic form of these {\it kp} theories are
\begin{equation}
H(\bm{k})=\varepsilon_{0,\bm{k}}+t_{1,\bm{k}}\tau_1+t_{\alpha,\bm{k}}\tau_\alpha+\tau_\beta (\bm{\lambda}_{\bm{k}}\cdot \bm{\sigma}) =  \varepsilon_{0,\bm{k}}+H_{\delta}(\bm{k})~,
\end{equation}
\begin{equation}\label{eq:typeindex}
    (I,\tau_\alpha,\tau_\beta) =
    \begin{cases}
(\tau_1 , \tau_2 , \tau_3) &\text{for type 1}~,\\
(\tau_0 , \tau_3 , \tau_2) &\text{for type 2}~,
\end{cases}
\end{equation}
where $H_{\delta}(\bm{k})=H(\bm{k})-\varepsilon_{0,\bm{k}}$ and \(\alpha\) and \(\beta\) are type indices will be used the remaining context. For parity mixed, type 1, $kp$ theories, the degeneracy at TRIM points is not broken by SOC. This is because the non-symmorphic symmetries combined with topological arguments imply  these TRIM must have an odd number of Dirac lines passing through them  \cite{Zhao:2016}. These Dirac lines lie in the nodal plane. Elsewhere in the nodal plane, SOC lifts the 4-fold degeneracy. We will discuss some consequences of these Dirac lines later. 
The non trivial inversion symmetry for type 1, $I=\tau_1$, implies the parity of the momentum functions that $\varepsilon_{0,\bm{k}}=\varepsilon_{0,-\bm{k}}$, $t_{1,\bm{k}}=t_{1,-\bm{k}}$, $t_{2,\bm{k}}=-t_{2,-\bm{k}}$, and $\bm{\lambda}_{\bm{k}}=-\bm{\lambda}_{-\bm{k}}$. This form of Hamiltonian has often been used to understand locally non-centrosymmetric superconductors \cite{Smidman:2017} and hidden spin polarization in inversion symmetric materials \cite{Pengke:2018}. In these contexts, the orbital degrees of freedom reside on different sectors that are related by inversion symmetry and there is typically no symmetry requirement that ensures the SOC dominates. The $\tau_3$ matrix is odd under inversion symmetry, allowing the odd-parity SOC  $\bm{\lambda}_{\bm{k}}$ to appear. Many superconductors of interest have Fermi surfaces near type 1 TRIM points, examples include: Fe-based superconductors,  which often have electron pockets near the $M$ point in space group 129 (classes $D_{4h,1}^{\text{type}1}$ or $D_{4h,3}^{\text{type}1}$) \cite{Stewart:2011}, in this context the high $T_c$ superconductor monolayer FeSe is of interest, since it only has Fermi surfaces near the $M$ point \cite{Huang:2017}; CeRh$_2$As$_2$ which exhibits a field induced transition from an even parity to an odd-parity superconducting state \cite{Khim:2021,Landaeta:2022} and has Fermi surfaces near the $M$ point in space group 129 (classes $D_{4h,1}^{\text{type}1}$ or $D_{4h,3}^{\text{type}1}$); BiS$_2$-based superconductors \cite{Mizuguchi:2015} which has superconductivity that survives to very high fields and which has electron pockets near the $X$ point in space group 129 (class $D_{2h,3}^{\text{type}1}$); the odd-parity heavy fermion superconductor UPt$_3$ \cite{Joynt:2002} which has a pancake-like Fermi surface at $k_z=\pi/c$ in space group 193 (class $D_{6h}^{\text{type}1}$); and the ferromagnetic superconductor UCoGe \cite{Aoki:2019} with space group 62 and a Fermi surface near the $T$ point (class $D_{2h,1}^{\text{type}1}$).


\begin{table}
\centering
\renewcommand{\arraystretch}{1.3}
\small
\begin{minipage}[h]{0.49\linewidth}
\begin{tabular}{|c|c|}
\hline
\textbf{~Class~} & \textbf{Symmetry} \\ \hline
\multicolumn{2}{|c|}{\textbf{Hamiltonian}} \\ \hline
\multicolumn{2}{|c|}{\textbf{Space Group Momenta}} \\ \hline
\hline
$C_{2h,1}^{\text{type}1}$ & \(A_g + B_g + [A_u]+ B_u\) \\ \hline 
\multicolumn{2}{|c|}{\(H=\epsilon_0+ (t_{1x}k_x+t_{1z}k_z)k_y\tau_1+ t_2k_y\tau_2\)}\\ 
\multicolumn{2}{|c|}{\(~+\tau_3[\lambda_xk_y\sigma_x+(\lambda_{yx}k_x+\lambda_{yz}k_z)\sigma_y+\lambda_zk_y\sigma_z]\)}\\ \hline
 \multicolumn{2}{|c|}{\(11(C_1, D_1, E_1, Z_1)\),~\(14(C_1, Z_1)\),}\\
 \multicolumn{2}{|c|}{\(63(R_1(yz))\), \(176(L_1(yz))\)}\\\hline \hline 

$C_{2h,2}^{\text{type}2}$&\(A_g + 2B_g + [A_g]\) \\ \hline 
\multicolumn{2}{|c|}{\(H=\epsilon_0+(t_{1x}k_x+t_{1z}k_z)k_y \tau_1+(t_{3x}k_x+t_{3z}k_z)k_y\tau_3\)}\\ 
\multicolumn{2}{|c|}{\(+\tau_2[(\lambda_{xx}k_x +\lambda_{xz}k_z)k_y\sigma_x+\lambda_y\sigma_y+(\lambda_{zx}k_x+\lambda_{zz}k_z)k_y\sigma_z]\)}\\ \hline
 \multicolumn{2}{|c|}{\(14(D_1^\pm D_2^\pm, E_1^\pm E_2^\pm)\), \(64(R_1^\pm R_2^\pm(yz))\)}\\ \hline \hline 

 $D_{2h,1}^{\text{type}1}$ & \(A_g + B_{1g} + [A_u] + B_{1u} \) \\ \hline 
\multicolumn{2}{|c|}{\(H=\epsilon_0+ t_1k_x k_y\tau_1+ t_2k_x k_y k_z\tau_2\)}\\ 
\multicolumn{2}{|c|}{\(~+\tau_3[ \lambda_xk_y\sigma_x+ \lambda_y k_x \sigma_y+ \lambda_z k_x k_y k_z\sigma_z]\)}\\ \hline
 \multicolumn{2}{|c|}{\(56(S_{1,2})\), \(58(R_{1,2})\)}\\ 
 \multicolumn{2}{|c|}{\(59(S_{1,2}, R_{1,2})\), \(62(T_{1,2}(xz))\)}\\\hline \hline

$D_{2h,2}^{\text{type}2}$&\(A_g + 2B_{1g} + [A_g]\) \\ \hline 
\multicolumn{2}{|c|}{\(H=\epsilon_0+t_1k_x k_y\tau_1+t_3k_x k_y\tau_3\)}\\ 
\multicolumn{2}{|c|}{\(~+\tau_2[\lambda_x k_y k_z \sigma_x+ \lambda_yk_x k_z\sigma_y+\lambda_zk_x k_y\sigma_z]\)}\\ \hline
 \multicolumn{2}{|c|}{\(55(S_{1}^{\pm}S_{2}^{\pm}, S_{3}^{\pm}S_{4}^{\pm}, R_{1}^{\pm}R_{2}^{\pm}, R_{3}^{\pm}R_{4}^{\pm})\)}\\ 
 \multicolumn{2}{|c|}{\(56(R_{1}^{\pm}R_{2}^{\pm}, R_{3}^{\pm}R_{4}^{\pm})\), \(58(S_{1}^{\pm}S_{2}^{\pm}, S_{3}^{\pm}S_{4}^{\pm})\)}\\ \hline \hline 

$D_{2h,3}^{\text{type}1}$ & \(A_g + B_{2g} + [B_{3u}] + B_{1u}\) \\ \hline 
\multicolumn{2}{|c|}{\(H=\epsilon_0+ t_1k_x k_z\tau_1+ t_2k_x\tau_2\)}\\ 
\multicolumn{2}{|c|}{\(~+\tau_3[ \lambda_x k_y\sigma_x+ \lambda_yk_x \sigma_y+ \lambda_z k_x k_y k_z\sigma_z]\)}\\ \hline
 \multicolumn{2}{|c|}{\(51(X_{1,2}, S_{1,2}, U_{1,2}, R_{1,2})\), \(52(R_{1,2}(xy), Y_{1,2}(xyz))\)}\\
 \multicolumn{2}{|c|}{\(53(Z_{1,2}(zyx), T_{1,2}(zyx))\), \(54(X_{1,2}, S_{1,2})\)}\\ 
 \multicolumn{2}{|c|}{\(55(U_{1,2}(yz), X_{1,2}(yz), Y_{1,2}(xyz), T_{1,2}(xyz))\)}\\ 
 \multicolumn{2}{|c|}{\(56(X_{1,2}, Y_{1,2}(xy))\)}\\ 
 \multicolumn{2}{|c|}{\(57(S_{1,2}(xyz), Y_{1,2}(xyz), Z_{1,2}(zyx), U_{1,2}(zyx))\)}\\ 
 \multicolumn{2}{|c|}{\(58(X_{1,2}(yz), Y_{1,2}(xyz))\)}\\ 
 \multicolumn{2}{|c|}{\(59(X_{1,2}, U_{1,2}, T_{1,2}(xy), Y_{1,2}(xy))\), \(60(X_{1,2}, Z_{1,2}(zyx))\)}\\ 
 \multicolumn{2}{|c|}{\(61(X_{1,2}, Y_{1,2}(xyz), Z_{1,2}(zyx))\)}\\ 
 \multicolumn{2}{|c|}{\(62(X_{1,2}, Z_{1,2}(xz), Y_{1,2}(xyz))\)}\\ 
 \multicolumn{2}{|c|}{\(63(T_{1,2}(zyx), Z_{1,2}(zyx))\), \(64(T_{1,2}(zyx), Z_{1,2}(zyx))\)}\\ 
 \multicolumn{2}{|c|}{\(127(X_{1,2}(xyz), R_{1,2}(xyz))\), \(128(X_{1,2}(xyz))\)}\\ 
 \multicolumn{2}{|c|}{\(129(X_{1,2}(xy), R_{1,2}(xy))\), \(130(X_{1,2}(xy))\)}\\ 
 \multicolumn{2}{|c|}{\(135(X_{1,2}(xyz), R_{1,2}(xyz))\), \(136(X_{1,2}(xyz))\)}\\ 
 \multicolumn{2}{|c|}{\(137(R_{1,2}(xy), X_{1,2}(xy))\), \(138(X_{1,2}(xy))\)}\\
 \multicolumn{2}{|c|}{\(193(L_{1,2})\), \(194(L_{1,2} (xy))\)}\\
 \multicolumn{2}{|c|}{\(205(X_{1,2}(xyz))\)}\\ \hline
\end{tabular}
\end{minipage}
\begin{minipage}[h]{0.49\linewidth}
\centering
\begin{tabular}{|c|c|}
\hline
$D_{2h,4}^{\text{type}2}$ & \(A_g + B_{1g} + B_{3g} + [B_{2g}]\) \\ \hline 
\multicolumn{2}{|c|}{\(H=\epsilon_0+t_1k_x k_y\tau_1+t_3 k_y k_z\tau_3\)}\\ 
\multicolumn{2}{|c|}{\(~+\tau_2[\lambda_x k_x k_y\sigma_x+\lambda_y\sigma_y+ \lambda_z k_y k_z \sigma_z]\)}\\ \hline
 \multicolumn{2}{|c|}{\(52(T_{1}^{\pm})\), \(53(U_{1}^{\pm}(yz), R_{1}^{\pm}(yz))\)}\\ 
 \multicolumn{2}{|c|}{\(58(T_{1}^{\pm}, U_{1}^{\pm}(xy))\), \(60(S_{1}^{\pm}(xy))\)}\\ 
 \multicolumn{2}{|c|}{\(128(R_{1}^{\pm})\), \(136(R_{1}^{\pm})\)}\\ \hline \hline 

$D_{4h,1}^{\text{type}1}$ & \(A_{1g} + B_{2g} + [A_{1u}] + B_{2u}\) \\ \hline 
\multicolumn{2}{|c|}{\(H=\epsilon_0+ t_1k_x k_y\tau_1+ t_2k_x k_y k_z ( k_x^2- k_y^2)\tau_2\)}\\ 
\multicolumn{2}{|c|}{\(~+\tau_3[ \lambda_x(k_x\sigma_y+k_y \sigma_x)+ \lambda_3k_x k_y k_z \sigma_z]\)}\\ \hline
 \multicolumn{2}{|c|}{\(129(M_{1,2}, A_{1,2})\), \(130(M_{1,2})\)}\\ 
 \multicolumn{2}{|c|}{\(136(A_{3,4})\), \(137(M_{1,2})\), \(138(M_{1,2})\)}\\\hline \hline

$D_{4h,2}^{\text{type}2}$& \(A_{1g} + 2B_{2g} + [A_{1g}]\) \\ \hline 
\multicolumn{2}{|c|}{\(H=\epsilon_0+t_1k_x k_y\tau_1+ t_3k_x k_y\tau_3\)}\\ 
\multicolumn{2}{|c|}{\(~+\tau_2[\lambda_x(k_y k_z\sigma_x+ k_x k_z \sigma_y)+ \lambda_zk_x k_y( k_x^2- k_y^2)\sigma_z]\)}\\ \hline
 \multicolumn{2}{|c|}{\(127(M_{1}^{\pm}M_{4}^{\pm}, M_{2}^{\pm}M_{3}^{\pm}, A_{1}^{\pm}A_{4}^{\pm}, A_{2}^{\pm}A_{3}^{\pm})\)}\\ 
 \multicolumn{2}{|c|}{\(128(M_{1}^{\pm}M_{4}^{\pm}, M_{2}^{\pm}M_{3}^{\pm})\), \(135(M_{1}^{\pm}M_{4}^{\pm}, M_{2}^{\pm}M_{3}^{\pm})\)}\\ 
 \multicolumn{2}{|c|}{\(136(M_{1}^{\pm}M_{4}^{\pm}, M_{2}^{\pm}M_{3}^{\pm})\), \(138(A_{1}^{\pm}A_{4}^{\pm}, A_{2}^{\pm}A_{3}^{\pm})\)}\\ \hline \hline 
 
$D_{4h,3}^{\text{type}1}$ & \(A_{1g} + B_{2g} + [B_{1u}] + A_{2u}\) \\ \hline 
\multicolumn{2}{|c|}{\(H=\epsilon_0+t_1k_x k_y\tau_1+ t_2k_x k_y k_z\tau_2\)}\\ 
\multicolumn{2}{|c|}{\(~+\tau_3[ \lambda_x(k_x\sigma_y-k_y \sigma_x)+ \lambda_z k_x k_y k_z( k_x^2- k_y^2)\sigma_z]\)}\\ \hline
 \multicolumn{2}{|c|}{\(129(M_{3,4}, A_{3,4})\), \(130(M_{3,4})\)}\\ 
 \multicolumn{2}{|c|}{\(136(A_{1,2})\), \(137(M_{3,4})\), \(138(M_{3,4})\)}\\ \hline \hline 

$D_{4h,4}^{\text{type}2}$&\(A_{1g} + A_{2g} + B_{2g} + [B_{1g}]\) \\ \hline 
\multicolumn{2}{|c|}{\(H=\epsilon_0+t_1k_x k_y( k_x^2- k_y^2)\tau_1+t_3k_x k_y\tau_3\)}\\ 
\multicolumn{2}{|c|}{\(~ +\tau_2[\lambda_x(k_y k_z\sigma_x+ k_x k_z \sigma_y)+\lambda_zk_x k_y\sigma_z]\)}\\ \hline
 \multicolumn{2}{|c|}{\(127(M_{5}^{\pm}, A_{5}^{\pm})\), \(128(M_{5}^{\pm})\)}\\ 
 \multicolumn{2}{|c|}{\(135(M_{5}^{\pm})\), \(136(M_{5}^{\pm})\), \(138(A_{5}^{\pm})\)}\\ \hline \hline
 
$D_{4h,5}^{\text{type}1}$ & \(A_{1g} + A_{2g} + [B_{1u}] + B_{2u}\) \\ \hline 
\multicolumn{2}{|c|}{\(H=\epsilon_0+ t_1k_x k_y(k_x^2 - k_y^2)\tau_1+ t_2k_x k_y k_z\tau_2\)}\\ 
\multicolumn{2}{|c|}{\(~+\tau_3[ \lambda_x(k_x\sigma_y+k_y \sigma_x)+ \lambda_zk_x k_y k_z\sigma_z]\)}\\ \hline
 \multicolumn{2}{|c|}{\(128(A_{1,2})\), \(137(A_{1,2})\)}\\ \hline \hline
 
$C_{6h}^{\text{type}1}$& \(A_{g} + B_{g} + [A_{u}] + B_{u}\) \\ \hline 
\multicolumn{2}{|c|}{\(H=\epsilon_0+(t_{1x}k_x(k_x^2-3k_y^2)+t_{1y}k_y (3k_x^2-k_y^2))k_z\tau_1\)}\\ 
\multicolumn{2}{|c|}{\(~+ t_2k_z\tau_2+\tau_3[ \lambda_xk_z(2k_xk_y\sigma_x+(k_x^2-k_y^2) \sigma_y)\)}\\ 
\multicolumn{2}{|c|}{\(~+ (\lambda_{zx}k_x(k_x^2- 3k_y^2)+\lambda_{zy}k_y(3k_x^2- k_y^2))\sigma_z]\)}\\ \hline
 \multicolumn{2}{|c|}{\(176(A_1)\)}\\ \hline \hline 
 
$D_{6h}^{\text{type}1}$ & \(A_{1g} + B_{2g} + [A_{2u}] + B_{1u} \) \\ \hline 
\multicolumn{2}{|c|}{\(H=\epsilon_0+t_1k_x k_z(k_x^2-3k_y^2)\tau_1+ t_2k_z\tau_2\)}\\ 
\multicolumn{2}{|c|}{\(+\tau_3[ \lambda_xk_z(2k_xk_y\sigma_x+(k_x^2-k_y^2) \sigma_y)+ \lambda_zk_y( 3k_x^2- k_y^2)\sigma_z]\)}\\ \hline
 \multicolumn{2}{|c|}{\(193(A_{1,2})\), \(194(A_{1,2}(xy))\)}\\ \hline
\end{tabular}
\end{minipage}
\caption{Classification of $kp$ theories. Subscript numbering of momenta represents different real representations on the same momentum point, and a permutation of the axes is denoted by the cyclic notation. For example, \(128(X_{1,2}(xyz))\) represents that there are two representations $X_1$ and $X_2$ on $X=(0,1/2,0)$ space group 128, and their local theory is obtained by \(D^{\text{type}1}_{2h,3}\) Hamiltonian under \(x\rightarrow y \rightarrow z \rightarrow x\) relabelling. The representation convention is following Bilbao Crystallographic server\footnote{\url{https://www.cryst.ehu.es/} Representations and Applications $\rightarrow$ Point and Space Groups $\rightarrow$ - Representations $\rightarrow$ SG Physically irreducible representations given in a real basis}\cite{Aroyo:2006,Aroyo2:2006,Stokes:2013} except for the $L$ point in 193 and 194.
} \label{tab:kp}
\end{table}

For type 2 {\it kp} theories, the 4-fold degeneracy is sometimes already split into 2  at the TRIM point when SOC is added, unlike what occurs for type 1 {\it kp} theories. This happens in classes $C_{2h,2}^{\text{type}2}$ and $D_{2h,1}^{\text{type}2}$.  For the other type 2 classes, this degeneracy at the TRIM point is not split. In these cases, an even number of Dirac lines pass through the TRIM point. These Dirac lines lie in the nodal plane. Since $I=\tau_0$ for type 2,
all terms in the Hamiltonian are even parity, that is, unchanged under $\bm{k}\rightarrow -\bm{k}$. One example where type 2 $kp$ theories apply is in strain induced superconductivity in RuO$_2$\cite{Ruf:2021,Uchida:2020}. Without strain, RuO$_2$ is thought to be a non-superconducting altermagnet \cite{Smejkal:2022}. When strain is applied,
bands near the $X$-$M$-$R$-$A$  Brillouin zone face are most strongly affected \cite{Ruf:2021}. RuO$_2$ has space group 136 with the $R$ and $M$ points belonging to classes $D_{2h,4}^{\text{type}2}$, $D_{4h,2}^{\text{type}2}$, or $D_{4h,4}^{\text{type}2}$.  Later we discuss the ferromagnetic superconductor UCoGe with  space group 62 \cite{Aoki:2019}. In this example, we highlight the role of 8-fold degenerate points which exhibit some properties similar to that found for type 2 TRIM points.

Type 1 and type 2 {\it kp} Hamiltonians share some common features that play an important role in understanding the properties of the superconducting states.  The first is that the non-symmorphic symmetry dictates that these  Hamiltonians are best described as two-band systems with eigenenergies given by 
\begin{equation}\label{eq:energydispersiondef}
E_{\pm}(\bm{k})=\varepsilon_{0,\bm{k}}\pm \sqrt{t_{1,\bm{k}}^2+t_{\alpha,\bm{k}}^2+|\bm{\lambda}_{\bm{k}}|^2}=\varepsilon_{0,\bm{k}}\pm\varepsilon_{\delta,\bm{k}}~,
\end{equation}
where $\alpha$ is the type index in Eq.~\ref{eq:typeindex}. 
The second feature is that both simplify dramatically on the nodal plane, where only the coefficient functions $\varepsilon_{0,\bm{k}}$ and ${\bm{\lambda}}_{\bm{k}}\cdot \bm{\hat{n}}$ are non-vanishing (that is $t_{1,\bm{k}}=t_{2,\bm{k}}=t_{3,\bm{k}}=|{\bm{\lambda}}_{\bm{k}}\times \bm{\hat{n}}|=0$). This property is a direct consequence of the anomalous pseudopspin. The symmetry arguments discussed in the previous section enforce this condition. In particular, for momenta on the nodal plane, the mirror operator through the nodal plane, $U_{M}$, takes the from $U_{M}=-i\tau_\beta(\bm{\sigma}\cdot{\bm{\hat{n}}})$. The requirement that these Hamiltonians obey time-reversal and inversion symmetries and  commute with $U_{M}$ leads to this simple form of the {\it kp} theories in the nodal plane. The final important property of these {\it kp} Hamiltonians is that the SOC terms are often the leading order terms in the {\it kp} expansions, that is, they appear with the lowest powers of $k_i$. This is the case for classes $C_{2h,2}^{type2}$, $D_{2h,1}^{type1}$, $D_{2h,4}^{type2}$, $D_{4h,2}^{type1}$, $D_{4h,3}^{type1}$, and $D_{4h,5}^{type1}$.  This feature ensures that there exists a limit in which the SOC is the dominant single-particle interaction on the Fermi surface and hence the unusual magnetic superconducting response we later discuss must exist.  

\section{Superconducting states}

In the previous section, complete symmetry-dictated {\it kp} theories were found for anomalous pseudospin. These theories are complete in the sense that they include all operators of the form $\tau_i\sigma_j$ allowed by symmetry. For superconductivity, the orbital degree of freedom enlarges the corresponding space of possible gap functions compared to the usual even-parity (pseudospin-singlet) $\tilde{\Delta}(\bm k)=\psi_{\bm{k}}(i\sigma_y)$ and odd-parity (pseudospin-triplet) $\tilde{\Delta}(\bm k) =\bm{d}_{\bm{k}}\cdot \bm{\sigma} (i\sigma_y)$ states that appear in single-band theories \cite{Sigrist:1991,Gorkov:2001}. Nevertheless, it is possible to understand some general properties of the allowed pairing states.

To deduce the symmetry properties of possible pairing channels in this larger space of electronic states, it is useful to define gap function differently than usual \cite{Blount:1985,Samokhin:2019}. In particular, we take
\begin{equation}
\mathcal{H}=\sum_{i,j,{\bm k}}H_{ij}(\bm k)c^{\dagger}_{{\bm k},i}c_{{\bm k},j}+\frac{1}{2}\sum_{i,j,{\bm k}}[{\Delta}_{ij}(\bm k)c^{\dagger}_{{\bm k},i}\tilde{c}^{\dagger}_{{\bm k},j}+h.c.]. \label{HMF}
\end{equation}
where $i,j$ are combined spin and orbital indices, $h.c.$ means Hermitian conjugate, $c_{\bm{k}}(c^{\dagger}_{\bm{k}})$ is the Fermionic spin-half particle creation(annihilation) operator, and $\tilde{c}_{\bm{k}}(\tilde{c}^{\dagger}_{\bm{k}})$ is the time reversed partner of $c_{\bm{k}}(c^{\dagger}_{\bm{k}})$. In the usual formulation $\tilde{c}^\dagger_{k,j}$ is replaced $c_{-k,j}^{\dagger}$ which leads to a different gap function $\tilde{\Delta}_{ij}$ and to difficulties in interpreting the symmetry transformation properties of this gap function \cite{Blount:1985,Samokhin:2019}. For a single-band, these new gap functions become $\Delta(\bm k)=\psi_{\bm{k}}\sigma_0$ for even-parity and $\Delta(\bm k)=\bm{d}_{\bm{k}}\cdot \bm{\sigma}$ for odd-parity. 
The key use of Eq.~\ref{HMF}  is that the $\Delta_{ij}(\bm k)$ transform under rotations in the same way as the $H_{ij}(\bm k)$, allowing the symmetry properties of the gap functions to be deduced. The disadvantage of this approach is that the antisymmetry of the gap functions that follow from the Pauli exclusion principle is not as readily apparent compared to the usual formulation \cite{Blount:1985,Samokhin:2019}.

Enforcing the Pauli exclusion principle leads to eight types of gap functions that generalize the pseudospin-singlet and pseudospin-triplet of single-band gap functions. Six of these are simple generalizations of the single-band gap functions: $\tau_i \psi_{\bm{k}}$ and $\tau_i (\bm{d}_{\bm{k}}\cdot \bm{\sigma})$ for $i=0$, 1, and 3 where $\psi_{-\bm{k}}=\psi_{\bm{k}}$ and $\bm{d}_{-\bm{k}}=-\bm{d}_{\bm{k}}$. Two are new gap functions: $\tau_2 (\bm{\psi}_{\bm{k}}\cdot\bm{\sigma})$ and $\tau_2 d_{\bm{k}}$ with $\bm{\psi}_{-\bm{k}}=\bm{\psi}_{\bm{k}}$ and $d_{-\bm{k}}=-d_{\bm{k}}$. It is possible to determine whether these gap functions are either even or odd-parity and this depends upon whether the {\it kp} Hamiltonian is type 1 or type 2. These gap functions and their parity symmetry are listed in Table \ref{tab:pairings}. Without further consideration of additional symmetries, the gap function will in general be a linear combination of all the even (or odd) parity gap functions.

To gain an understanding of the relative importance of these pairing states it is useful to project these gaps onto the band basis. Such a projection is meaningful if the energy separation between the two bands is much larger than the gap magnitude. For many of the {\it kp} Hamiltonians, due to the presence of Dirac lines, there will exist regions in momentum space for which this condition is not satisfied.  However, these regions represent a small portion of the Fermi surface when the SOC energies are much larger than the gap energies, so that an examination of the projected gap is still qualitatively useful in this limit.  Provided the superconducting state does not break time-reversal symmetry, the projected gap magnitude on band $a$  can be found through \cite{Cavanagh2:2022}
\begin{equation}
\tilde{\Delta}_{\pm}^2 = \frac{\text{Tr}[
    |\lbrace H_{\delta},\Delta \rbrace|^2 
  P_{\pm}]}{\text{Tr}[|H_\delta|^2]}.\label{eq:h_Fit}
\end{equation}
where $P_{\pm}(\bm{k})=
\frac{1}{2}(1 \pm
H_{\delta}(\bm{k})/\varepsilon_{\delta,\bm{k}})$ which is a projection operator onto $\pm$ band by the energy dispersion Eq.~\ref{eq:energydispersiondef}. This projected gap magnitude is related to superconducting fitness \cite{Ramires:2016,Ramires:2018}: if it vanishes, the corresponding gap function is called unfit and will have a $T_c=0$ in the weak coupling limit. Table \ref{tab:pairings} gives the projected gap functions for the pairing states discussed above. The projection generally reduces the size of the gap, with the exception of the usual even-parity $\tau_0\psi_{\bm{k}}$ state (interestingly, the odd-parity $\tau_0 (\bm{d}_{\bm{k}}\cdot \bm{\sigma})$ state has a gap that is generically reduced). This reduction strongly suppresses the $T_c$ of the pairings state, where it enters exponentially in the weak-coupling limit. We later examine the different  {\it kp} classes to identify fit gap functions since the $T_c$ of these states will be the largest, given a fixed attractive interaction strength.  

On the nodal plane, the projected gap functions, shown in Table \ref{tab:pairings},  simplify considerably since only $\varepsilon_{0}$ and ${\bm{\lambda}}_{\bm{k}}\cdot \bm{\hat{n}}$ are non-zero. For both type 1 and type 2 Hamiltonians, this leads to two gap functions that are fully fit, that is, not reduced by the projection. For type 1 Hamiltonians, these fully fit states are $\tau_0\psi_{\bm{k}}$ and $\tau_3 \psi_{\bm{k}}$. The state $\tau_0\psi_{\bm{k}}$ is even-parity and the state $\tau_3 \psi_{\bm{k}}$ is odd-parity and, as discussed later, these two states play an important role in the appearance of a field-induced transition from even to odd parity superconductivity as observed in CeRh$_2$As$_2$. For gap functions described by vectors, for example $\bm{d}_{\bm k}$, the projected gaps on the nodal plane are of the form  $|\bm{d}_{\bm{k}}\cdot \bm{\hat{n}}|^2$ or   $|\bm{d}_{\bm{k}}\times \bm{\hat{n}}|^2$. This is qualitatively different than the usual odd-parity single-band gap, where the gap magnitude is $|\bm{d}_{\bm{k}}|^2$. The latter requires that all three components of $\bm{d}_{\bm{k}}$ must vanish to have nodes. For the projected gaps on the nodal planes, this requirement is less stringent: only one or two components of $\bm{d}_{\bm{k}}$ need to vanish to have nodes. This is closely related to the violation of Blount's theorem on the nodal planes. 

\subsection{Gap projection and the violation of Blount's theorem}

Blount's theorem states that time-reversal symmetric odd-parity superconductors cannot have line nodes when SOC is present \cite{Blount:1985}. Key to Blount's theorem is the assumption that pseudsopsin shares the same symmetry properties as usual spin \cite{Blount:1985}.
The violation of Blount's theorem in non-symmorphic space groups has been demonstrated  through an examination of Cooper pair representations formed from antisymmetric direct products of the relevant fermions states.\cite{Norman:1995,Micklitz:2009,Micklitz:2017,Micklitz:2017-2,Yanase:2016,Yanase:2017,Kobayashi:2014}.  Here we use an alternate approach that exploits the completeness of the $kp$ Hamiltonian space and the inclusion of  all gap functions in this space that are allowed by symmetry to directly compute the general form of the superconducting excitation spectrum. This approach closely links the anomalous pseudopsin to the violation of Blount's theorem.   

\begin{table}
\renewcommand{\arraystretch}{1.3}
\begin{tabular}{|c||c|c|c||c|c|c|}
\hline
&    \multicolumn{3}{c||}{Type 1} & \multicolumn{3}{c|}{Type 2} \\
    \hline
 Gap function &    Inversion & Gap projection & Gap on nodal plane & Inversion & Gap projection & Gap on nodal plane\\\hline
 
$\tau_0\psi$ & $+$ & $|\psi|^2$ & $|\psi|^2$ & $+$ & $|\psi|^2$ & $|\psi|^2$\\\hline 

$\tau_0(\bm{d}\cdot\bm{\sigma})$ & $-$ & $\dfrac{(t_1^2+t_2^2)|\bm{d}|^2+|\bm{d}\cdot\bm{\lambda}|^2}{t_1^2+t_2^2+|\bm{\lambda}|^2}$ & $|\bm{d}\cdot \bm{\hat{n}}|^2$ & $-$ & $\dfrac{(t_1^2+t_2^2)|\bm{d}|^2+|\bm{d}\cdot\bm{\lambda}|^2}{t_1^2+t_2^2+|\bm{\lambda}|^2}$ & $|\bm{d}\cdot \bm{\hat{n}}|^2$ \\\hline

$\tau_3\psi$ & $-$ & $\dfrac{|\bm{\lambda}|^2|\psi|^2}{t_1^2+t_2^2+|\bm{\lambda}|^2}$ & $|\psi|^2$ & $+$ & $\dfrac{t_3^2|\psi|^2}{t_1^2+t_3^2+|\bm{\lambda}|^2}$ & $0$\\ \hline

$\tau_3 (\bm{d}\cdot\bm{\sigma})$ & $+$ & $\dfrac{|\bm{d}\cdot\bm{\lambda}|^2}{t_1^2+t_2^2+|\bm{\lambda}|^2}$ & $|\bm{d}\cdot \bm{\hat{n}}|^2$ & $-$ & $\dfrac{t_3^2|
\bm{d}|^2+|\bm{d}\times\bm{\lambda}|^2}{t_1^2+t_3^2+|\bm{\lambda}|^2}$ & $|\bm{d}\times \bm{\hat{n}}|^2$ \\\hline

$\tau_1\psi$ & $+$ & $\dfrac{t_1^2|\psi|^2}{t_1^2+t_2^2+|\bm{\lambda}|^2}$ & $0$ & $+$ &$\dfrac{t_1^2|\psi|^2}{t_1^2+t_3^2+|\bm{\lambda}|^2}$ & $0$\\\hline

$\tau_1 (\bm{d}\cdot\bm{\sigma})$ & $-$ & $\dfrac{t_1^2|\bm{d}|^2+|\bm{d}\times\bm{\lambda}|^2}{t_1^2+t_2^2+|\bm{\lambda}|^2}$ & $|\bm{d}\times \bm{\hat{n}}|^2$ & $-$ & $\dfrac{t_1^2|\bm{d}|^2+|\bm{d}\times\bm{\lambda}|^2}{t_1^2+t_2^2+|\bm{\lambda}|^2}$ & $|\bm{d}\times \bm{\hat{n}}|^2$\\\hline

$\tau_2 d$ & $+$ & $\dfrac{t_2^2|d|^2}{t_1^2+t_2^2+|\bm{\lambda}|^2}$ & $0$ & $-$ & $\dfrac{|\bm{\lambda}|^2|d|^2}{t_1^2+t_3^2+|\bm{\lambda}|^2}$ & $|d|^2$\\\hline

$\tau_2 (\bm{\psi}\cdot\bm{\sigma})$ & $-$ & $\dfrac{t_2^2|\bm{\psi}|^2+|\bm{\psi}\times\bm{\lambda}|^2}{t_1^2+t_2^2+|\bm{\lambda}|^2}$ & $|\bm{\psi}\times \bm{\hat{n}}|^2$ & $+$ & $\dfrac{|\bm{\psi}\cdot\bm{\lambda}|^2}{t_1^2+t_3^2+|\bm{\lambda}|^2}$ & $|\bm{\psi}\cdot \bm{\hat{n}}|^2$\\\hline
\end{tabular}
\caption{Classification of allowed pairing states for the {\it kp} theories. For both type I and II TRIMs we give the symmetry under inversion, the gap projection onto the Fermi surface, and the gap on the nodal plane. The momentum subscript indices \(\bm{k}\) of the coefficient functions are omitted here.}
  \label{tab:pairings}
  \end{table}  


The existence of anomalous pseudospin requires the presence of the translation mirror symmetry $\tilde{M}_{\bm{\hat{n}}}$. Consequently, the gap function can be classified as even or odd under this symmetry. Momenta on the nodal plane are invariant under $\tilde{M}_{\bm{\hat{n}}}$. Hence, for these momenta, $U_{M}^{\dagger}\Delta(\bm{k})U_{M}=\pm \Delta(\bm{k})$ where the $+$ ($-$) holds for a mirror-even (mirror-odd) gap function. For our basis choice $U_{M}=-i\tau_\beta(\bm{\sigma}\cdot{\bm{\hat{n}}})$. Importantly, for both types the {\it kp} theories on the nodal plane are given by $H(\bm{k})=\varepsilon_{0,\bm{k}}+iU_{M}(\bm{\lambda}_{\bm{k}}\cdot \bm{\hat{n}})$. 
This defines the two bands $E_{\pm}(\bm{k})=\varepsilon_{0,\bm{k}} \pm |\bm{\lambda}_{\bm{k}}\cdot\bm{\hat{n}}|$. Written in the band basis, we can divide the pairing potential into intraband and interband components. On the nodal plane the intraband gap functions are explicitly given by
\begin{equation}
     P_{\pm}\Delta P_{\pm} = \frac{1}{4}(-U_{M} \pm i~\text{sgn}(\bm{\lambda}_{\bm k}\cdot\bm{\hat{n}}))\{U_{M},\Delta\}\,,
\end{equation}
while the interband components are
\begin{equation}
P_{\pm}\Delta P_{\mp} = \frac{1}{4}(-U_{M} \pm i~\text{sgn}(\bm{\lambda}_{\bm k}\cdot\bm{\hat{n}}))[U_{M},\Delta]
\end{equation}
We observe that since a mirror-even gap function satisfies $[U_{M},\Delta]=0$, the interband gap components must vanish on the nodal plane, i.e. the pairing only involves particles from the same band. The general form of the BdG energy dispersion relation is then
\begin{equation}
    \pm' \sqrt{(\varepsilon_{0,{\bm k}} \pm |\bm{\lambda}_{\bm k}\cdot\bm{\hat{n}}|)^2 + |\Delta_{\pm\pm}|^2}~,
\end{equation}
where intraband gap magnitude \(|\Delta_{\pm\pm}|^2=\frac{1}{4}\text{Tr}[|P_{\pm}\Delta P_{\pm}|^2]\) and \(\pm'\) is the particle-hole symmetry index which is independent of band index \(\pm\). Since there is no requirement that $|\Delta_{\pm\pm}|^2=0$, line nodes are therefore not expected on the nodal plane, but rather we should generically find two-gap behavior with different size gaps on the two bands.
In contrast, for the mirror-odd gap functions we have $\{U_{M},\Delta\}=0$, so there is no intraband pairing on the nodal plane. The general form of the eigenenergies for this interband pairing state is then
\begin{equation}
    \pm^\prime \left(\pm |\bm{\lambda}_{\bm k}\cdot\bm{\hat{n}}| + \sqrt{\epsilon_{0,\bm{k}}^2 + |\Delta_{\pm\mp}|^2}\right)\,,
\end{equation}
where intraband gap magnitude \(|\Delta_{\pm\mp}|^2=\frac{1}{4}\text{Tr}[|P_{\pm}\Delta P_{\mp}|^2]\). The gap has line nodes provided $|\bm{\lambda}_{\bm{k}}\cdot\bm{\hat{n}}|^2>|\Delta_{\pm\mp}|^2$. This result depends only on the mirror-odd symmetry of the gap, and not on the parity symmetry. Since gaps that are odd under both mirror and parity symmetry are allowed, this result shows that odd-parity gaps can have line nodes, thus demonstrating a violation of Blount's theorem. 

The origin of these nodes due to purely interband pairing implies that the nodes are shifted off the Fermi surface \cite{PhysRevB.94.174518}. If the spin-orbit coupling is too weak, i.e. $|\bm{\lambda}_{\bm{k}}\cdot\bm{\hat{n}}|^2<|\Delta_{\pm\mp}|^2$, the nodes can annihilate with each other and are absent. This possibility has been discussed in the context of even parity superconductivity in monolayer FeSe \cite{Agterberg:2017} and odd-parity superconductivity in UPt$_3$ \cite{Yanase:2017}. 
The analysis above is valid even when Dirac lines pass through the TRIM points, as is the case in most of the derived {\it kp} theories. On the Dirac lines, the condition $|\bm{\lambda}_{\bm{k}}\cdot\bm{\hat{n}}|^2<|\Delta_{\pm\mp}|^2$ must occur and the spectrum is therefore gapped. In Appendix A we present exact expressions for the energy eigenstates on the nodal plane for all possible combinations of mirror and parity gap symmetries.

 \subsection{Unconventional pairing states from electron-phonon interactions}

 To highlight how the pairing of anomalous pseudospin can differ from the single-band superconductivity, it is instructive to consider an attractive $U$ Hubbard model. Such a model is often used to capture the physics of electron-phonon driven $s$-wave superconductivity in single-band models.  Here we show that this coupling also allows unconventional pairings states. In particular, odd-parity states in type 1 {\it kp} Hamiltonians. Such a state has recently likely  been observed in CeRh$_2$As$_2$. 

 Here we consider a local Hubbard-$U$ attraction on each site of the lattice and do not consider any longer range Coulomb interactions.  These sites are defined by their Wyckoff positions. Importantly, for the non-symmorphic groups we have considered here, each Wyckoff position has a multiplicity greater than one. Here we limit our discussion to Wyckoff positions with multiplicity two, which implies that there are two inequivalent atoms per unit cell. An attractive $U$ on these sites stabilizes a local spin-singlet Cooper pair. Since there are two sites per unit cell this implies that there are two stable superconducting degrees of freedom per unit cell. These two superconducting states can be constructed by setting the phase of Cooper pair wavefunction on each site to be the same or opposite. Since only local interactions are included, both these two states will have the same pairing interaction. 
 The in-phase state is a usual $s$-wave $\tau_0 \psi_{\bm{k}}$ state. Identifying the other, out of phase, superconducting state requires an understanding of the relationship between the basis states for the $kp$ Hamiltonians and orbitals located at the Wyckoff positions. In general, this will depend on the specific orbitals included in the theory. However, the condition that the resultant pairing states must be spin-singlet and local in space (hence momentum independent) allows only two possibilities for this additional pairing state: it is either a $\tau_1\psi_{\bm{k}}$ or a $\tau_3\psi_{\bm{k}}$ pairing state.  Of these states, for two reasons, the $\tau_3\psi_{\bm{k}}$ state for type 1 Hamiltonains is of particular interest. The first reason is that this state is odd-parity and therefore offers a route towards topological superconductivity \cite{Fu:2010,Qin:2022}. The second reason is that of the four possible states ($\tau_1\psi_{\bm{k}}$ or  $\tau_3\psi_{\bm{k}}$  for type 1 or type 2 Hamiltonians), this is the only state that is fully fit on the nodal plane (as can be seen in Table III, the other three states have zero gap projection on the nodal plane). This implies that for type 1 Hamiltonians, the odd-parity $\tau_3\psi_{\bm{k}}$ and the $s$-wave $\tau_0\psi_{\bm{k}}$ states can have comparable $T_c$ since they both have the same pairing interaction. In practice, the $\tau_3\psi_{\bm{k}}$ state will have a lower $T_c$ than the $\tau_0\psi_{\bm{k}}$ state since it will not be fully fit away from the nodal plane. Table III reveals that this projection is given by the ratio ${|\bm{\lambda}_{\bm{k}}}|^2/(t_{1,\bm{k}}^2+t_{2,\bm{k}}^2+|\bm{\lambda}_{\bm{k}}|^2)$. For classes  $D_{2h,1}^{\text{type}1},D_{4h,1}^{\text{type}1}, D_{4h,3}^{\text{type}1}$, and $D_{4h,5}^{\text{type}1}$, this ratio is nearly one since the SOC terms are the largest in the {\it kp} Hamiltonian. This suggests that these classes offer a promising route toward stabilizing odd-parity superconductivity. We stress that because ${|\bm{\lambda}_{\bm{k}}|}^2/(t_{1,\bm{k}}^2+t_{2,\bm{k}}^2+|\bm{\lambda}_{\bm{k}}|^2)$ is slightly less than one, the $T_c$ of the odd-parity $\tau_3\psi_{\bm{k}}$ will be comparable but less than that of the usual $s$-wave state. However, as we discuss later, the $\tau_3\psi_{\bm{k}}$ state can be stabilized over the usual $s$-wave $\tau_0\psi_{\bm{k}}$ state in an applied field. The identification of classes $D_{2h,1}^{\text{type}1},D_{4h,1}^{\text{type}1}, D_{4h,3}^{\text{type}1}$, and $D_{4h,5}^{\text{type}1}$ that maximize the $T_c$ of odd-parity pairing from electron-phonon interactions allows the earlier theory for a field induced even to odd parity transition CeRh$_2$As$_2$ \cite{Cavanagh:2022} (with space group 129) to be generalized to many other space groups.

While the above odd-parity state is only relevant for type 1 Hamiltonians, for type 2 Hamiltonians,  the usual $s$-wave interaction can develop a novel structure. In particular, for the classes $C_{2h,2}^{\text{type}2}$ and $D_{2h,4}^{\text{type}2}$, Table II shows that the state $\tau_2\sigma_y$ is maximally fit and has $s$-wave symmetry. Consequently, this state will admix with the usual $s$-wave $\tau_0\psi $ state. The theory describing this admixture formally resembles that of a Hund pairing mechanism proposed to explain the appearance of nodes in the likely  $s$-wave  superconductor KFe$_2$As$_2$ \cite{Vafek:2017}. The results of this analysis and a follow up analysis \cite{Cheung:2019} allow some of the properties of this state to be understood. An important conclusion of these works is that an $s$-wave superconducting state can emerge even when pairing for the usual $s$-wave state is repulsive (that is for the Hubbard $U>0$). This holds if two conditions are met: the effective interaction for the $\tau_2\sigma_y$ state is attractive (to first approximation, this effective interaction does not depend upon $U$ \cite{Vafek:2017,Cheung:2019}) and the two bands that emerge in the {\it kp} theory both cross the chemical potential. This $s$-wave  pairing state naturally leads to nodes.

\section{Role of Magnetic Fields}

The role of anomalous pseudopsin is perhaps most unusual  in response to magnetic fields. In many superconductors, there has been a push to drive up the magnetic field at which these are operational. Ising superconductors are one class of materials for which this has been successful, the in-plane critical field far surpasses the Pauli field, opening the door to applications \cite{Wang:2021}. Another relevant example is the field induced transition from an even parity to an odd-parity state observed in CeRh$_2$As$_2$ \cite{Khim:2021,Landaeta:2022}. 

Recently, a powerful method to examine the response of superconductors to time-reversal symmetry-breaking fields has been developed by the projection onto the band-basis\cite{Cavanagh2:2022}. The form of the {\it kp} theories we have developed allows for the direct application of this projection method. The response of superconductivity to time-reversal symmetry-
breaking is described by a time-reversal symmetry-breaking interaction $H_{\bm{h}}(\bm{k})$. A common form of TRSB Hamiltonian, and the one we emphasize here, is the Zeeman field interaction term, which is represented by
\begin{equation}\label{eq:Zeeman}
H_{\bm{h}}(\bm{k})=\tau_0(\bm{h}\cdot \bm{\sigma})~,
\end{equation}
where $\bm{h}$ is a magnetic field parameter in the system. We note that our qualitative results apply to a broader range of TRSB Hamiltonians. In particular, this is true if the TRSB field shares the same symmetry properties as a Zeeman field (for example if  $H_{\bm{h}}(\bm{k})$ describes the coupling between orbital angular momentum and an applied field). 

The theory introduces two parameters that quantify the response of superconductivity to time-reversal symmetry-breaking. The first  parameter is an effective $g$-factor given by
\begin{equation}
\tilde{g}_{\pm,\bm k,\bm h}^2 = \frac{2\text{Tr}[ 
    |\lbrace
        H_{\delta},H_{\bm{h}}\rbrace|^2 P_{\pm}]}{\text{Tr}[|H_\delta|^2]\text{Tr}[|H_{\bm{h}}|^2]}~.\label{eq:h_g}
\end{equation} 
The second parameter is the field-fitness, given by 
\begin{equation}
	\tilde{F}_{\pm,{\bm k},\bm{h}}=\frac{\text{Tr}[
            |\lbrace\lbrace H_{\delta},\tilde{\Delta}\rbrace,\lbrace
            H_{\delta},H_{\bm{h}}\rbrace\rbrace|^2
            P_{\pm}
          ]}{2\text{Tr}[ |\lbrace
            H_{\delta},H_{\bm{h}}\rbrace|^2 P_{\pm}
          ]\text{Tr}[ |\lbrace
            H_{\delta},\tilde{\Delta}\rbrace|^2P_\pm]}~. \label{eq:FieldFitnessFunction_1}
	\end{equation}
This field-fitness function ranges in value from zero to one. When the field-fitness is zero, the superconducting state is not suppressed by the time-reversal symmetry breaking perturbation. With these two parameters, the response of superconductivity to applied fields and the temperature dependence of magnetic susceptibility in the superconducting state can be determined. With the choice of the time-reversal symmetry-breaking field as the Zeeman field, Eq.~\ref{eq:Zeeman}, one finds 
\begin{equation}
\tilde{g}_{\pm,\bm k,\bm h}^2=\frac{t_{1,{\bm k}}^2+t_{\alpha,{\bm k}}^2+(\bm{\lambda}_{\bm k}\cdot\bm{\hat{h}})^2}{t_{1,{\bm k}}^2+t_{\alpha,{\bm k}}^2+\bm{\lambda}_{\bm k}^2} \label{Eq:g}
\end{equation}
where $\alpha$ is a type index that is 2 for type 1 and 3 for type 2. This agrees with results in \cite{Skurativska:2021} derived for Hamiltonians that resemble type 1 Hamiltonians. We note that the band index \(\pm\) and the magnitude of field \(\bm{h}\) in the field-fitness and the $g$-factor do not change the outcome, thus they will be omitted in the subsequent sections and they will be denoted by \(\tilde{F}_{\bm k,\bm{\hat h}}^2\) and \(\tilde{g}_{\bm k,\bm{\hat h}}^2\).

\subsection{Even parity superconductors}

It can be shown that the field-fitness parameter in Eq.~\ref{eq:FieldFitnessFunction_1} is 1 for all even parity states. Consequently, the magnetic response is governed solely by the generalized $g$-factor given in Eq.~\ref{Eq:g}. For momenta on the nodal plane, where $t_{1,{\bm k}}=t_{2,{\bm k}}=t_{3,{\bm k}}=\bm{\lambda}_{\bm k}\times \bm{\hat{n}}=0$, 
the $g$-factor vanishes for magnetic fields orthogonal to $\bm{\hat{n}}$.
This is a direct consequence of the anomalous pseudospin, since the symmetries of the Pauli matrices formed from anomalous pseudospin do not allow any coupling to a Zeeman field perpendicular to $\bm{\hat{n}}$.  An immediate consequence is that superconductivity survives to much stronger fields than expected for these field orientations. However, momenta that do not sit on the nodal plane  also contribute to the superconducting state and their contribution needs to be included as well. To quantify this, we  solve for the Pauli limiting field within weak coupling theory at zero temperature. For an isotropic $s$-wave superconductor, we find
\begin{equation}
\ln{\frac{h_{P,\bm{\hat{h}}}}{h_{0}}}=-\langle \ln{|\tilde{g}_{{\bm k},\bm{\hat{h}}}|}\rangle_{\bm{k}}
\label{Eq:Pauli}
\end{equation}
for field along direction $\hat{{\bm h}}$, where $h_{0}$ is the usual Pauli limiting field (found when the SOC is ignored), and $\langle \cdot \rangle_{\bm k}$ means an average over the Fermi surface weighted by the density of states.
Below, we apply this formula to BiS$_2$-based superconductors. We note that the spin susceptibility in the superconducting state can also be expressed using $\tilde{g}_{{\bm k},\bm{\hat{h}}}$ as well \cite{Cavanagh2:2022}, and this shows that a non-zero spin susceptibility is predicted at zero temperature whenever the critical field surpasses $h_0$.

\subsubsection{Enhanced in plane field Pauli  for BiS$_2$-based superconductors}

\begin{figure*}[tt]
	\centering
	\includegraphics[width=0.7\linewidth]{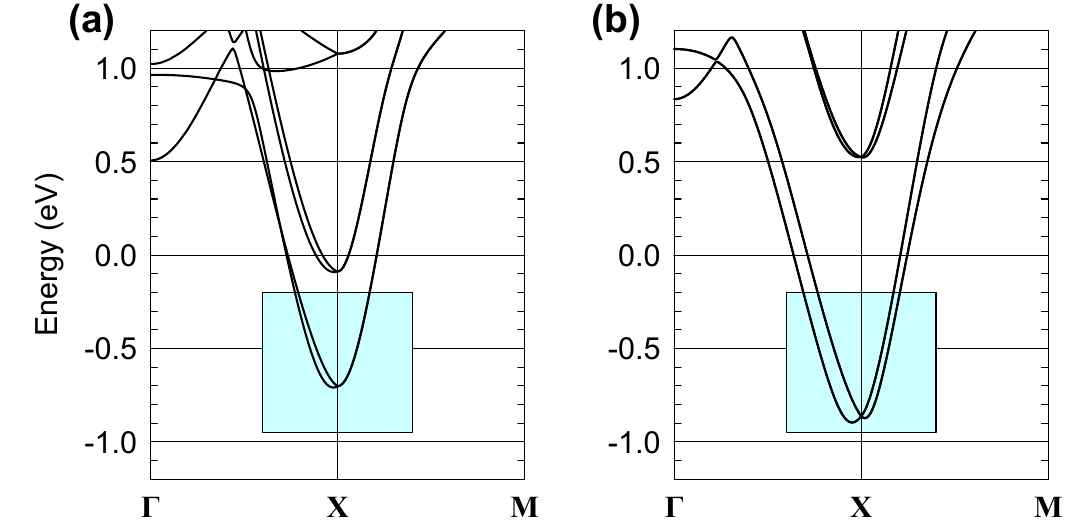}
	\caption{\label{Fig:BiS2} DFT bands of BiS$_2$ near the X point (a) without and (b) with the SOC\@.  The bands highlighted in the box are our focus.}
\end{figure*}

Here we turn to recent  experimental results on BiS$_2$-based superconductors  \cite{Mizuguchi:2015,Hoshi:2022}. This material has the tetragonal space group 129 (P4/nmm) and it exhibits two electron pockets about the two equivalent $X$ points \cite{Usui:2012,Cobo-Lopez:2018}. When S is replaced with Se, it has been observed that the in-plane upper critical field surpasses the usual Pauli limiting  field by a factor of 7 \cite{Hoshi:2022}. While it has been suggested that the local non-centrosymmetric structure is the source of this large critical field \cite{Hoshi:2022}, there has been no quantitative calculation for this. Here we apply Eq.~\ref{Eq:Pauli} to the {\it kp} theory at the $X$-point  to see if it is possible to account for this large critical field.  The $X$ point in space group 129 belongs to class $D_{2h,3}^{\text{type}1}$.For BiS$_2$, the dispersion is known to be strongly two-dimensional  (2D) \cite{Mizuguchi:2015,Usui:2012} so we consider the {\it kp} theory in the 2D limit.  This {\it kp} theory is 
\begin{equation}
H_{\text{BiS}_2}=\frac{\hbar^2}{2m} \left(k_x^2 +\gamma^2 k_y^2\right)-\mu+t_2k_y\tau_2+\lambda_xk_y\tau_3\sigma_x+\lambda_yk_x\tau_3\sigma_y.
\end{equation}
Assuming $s$-wave superconductivity and accounting for the two equivalent pockets yields 
\begin{equation}
h_{P,\bm{\hat{x}}}=h_{0}\frac{\sqrt{t_2^2+\lambda_x^2}+|\gamma \lambda_y|}{\sqrt{|t_2|+|\gamma \lambda_y|}(t_2^2+\lambda_x^2)^{1/4}}
\label{Eq:BiS2}
\end{equation}
where $h_{0}$ is the usual Pauli limiting field. For simplicity we consider $\gamma=1$ in the following. Eq.~\ref{Eq:BiS2} reveals that a large enhancement of the limiting field is possible and requires two conditions. The first is that  $t_2<<\lambda_x,\lambda_y$ and second is that these is substantial anisotropy in $\lambda_x$ and $\lambda_y$. To understand if these conditions are reasonable, we have carried out density-functional theory (DFT) calculations on LaO$_{1/2}$F$_{1/2}$BiS$_2$ with and without SOC.  DFT calculations for LaO$_{1/2}$F$_{1/2}$BiS$_2$ were carried out by 
the full-potential linearized augmented plane wave method
\cite{FLAPW2009}.
The Perdew-Burke-Ernzerhof form of the exchange correlation functional \cite{PBE}, 
wave function and potential energy cutoffs of 14 and 200 Ry, respectively, muffin-tin sphere radii of 
1.15, 1.2, 1.3, 1.0 \AA{} for Bi, S, La, O atoms, respectively, 
the experimental lattice parameters \cite{Mizuguchi2012}, and an $15\times15 
\times 5$ $k$-point mesh were employed for the self-consistent field calculation.  The virtual crystal approximation was used by setting the nuclear charge $Z=8.5$ at O(F) sites. The resultant bands are shown in Fig.~\ref{Fig:BiS2}. Without SOC, the band splitting along $\Gamma$ to $X$ yields an estimate for $t_2$. When SOC is present, the band splitting along the $X$ to $M$  yields $\lambda_y$ and the band splitting along  $\Gamma$ to $X$ yields $\sqrt{\lambda_x^2+t_2^2}$. The DFT calculated splittings suggest that $\lambda_x$ is the largest parameter by a factor of 3-4, while $t_2$ and $\lambda_y$ are comparable. This suggests that the conditions to achieve a large critical field are realistic in BiS$_2$-based superconductors. Note that the largest observed Pauli fields are found when the S is substituted by Se \cite{Hoshi:2022}. Se has a larger SOC than S, suggesting that the $\lambda_i$ parameters will be increased from what we estimate here. This is currently under exploration.

It is worthwhile contrasting the above theory with that for Fe-based materials in which electron pockets exist near the $M$ point of space group 129. The M-point  is described by class $D_{4h,1}^{\text{type}1}$. In this case, an analysis similar to to BiS$_2$ gives an enhancement of only $\sqrt{2}$ of the Pauli field for in-plane fields. For $c$-axis fields, this class implies a significantly enhanced Pauli limiting field. These results are consistent with experimental fits to upper critical fields in Fe-based superconductors that reveal that the upper critical field for in-plane fields are Pauli suppressed while those for field along the $c$-axis are not \cite{Zhang:2011}. The contrast bewteen Fe-based materials and BiS$_2$-based materials highlights the importance of the different classes. In particular, the lower orthorhombic symmetry of the $X$ point allows protection to in-plane fields not afforded to the $M$ point, where the theory is strongly constrained by tetragonal symmetry.

\subsubsection{Pair density wave states}

\begin{figure}[h]
		\centering
		\includegraphics[height=6cm]{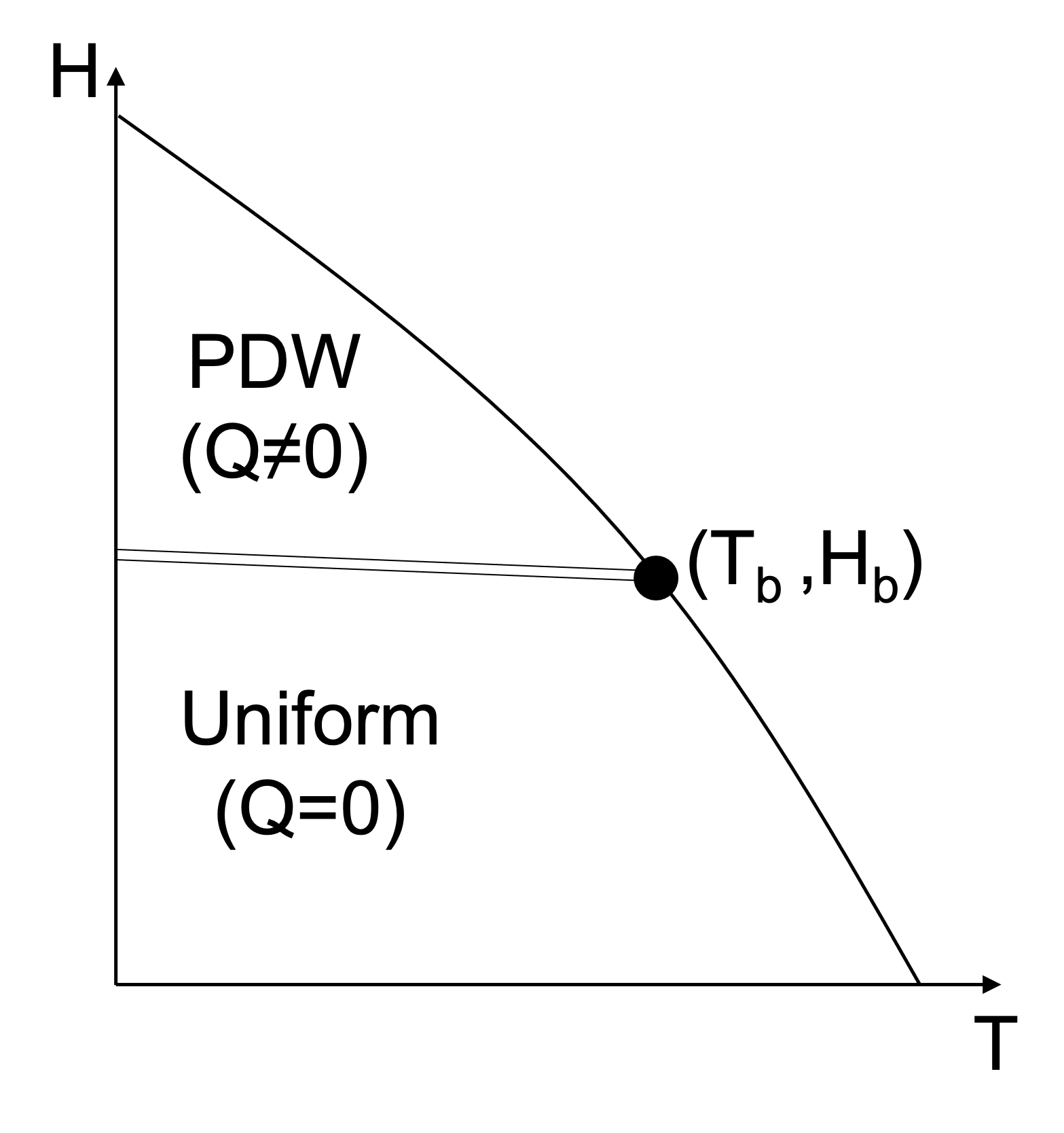}
		\caption{Schematic phase diagram for a spin-singlet superconductor under Zeeman effect. Single solid lines denote continuous phase transitions while double solid lines denote first-order phase transitions.}
		\label{F_PDW}
	\end{figure}

In BCS theory, a spin-singlet superconductor is suppressed by the Zeeman effect. Under a sufficiently strong magnetic field, the pairing susceptibility can be peaked at non-zero Cooper pair momenta, leading to a pair density wave or FFLO state \cite{Fulde:1964,Larkin:1965,PDW}. A schematic phase diagram for a centrosymmetric system is shown in the left panel of Fig.\ref{F_PDW}. The typically first order phase transition (double solid line) between the uniform and FFLO state ends at a bicritical point $(T_b,H_b)$, i.e. FFLO state only exists for $T<T_b$. A weak-coupling calculation reveals that for the usual FFLO phase, $T_b/T_c=0.56$ 

 It is known that for locally non-centrosymmetric superconductors, FFLO-like phases can appear at lower fields $H_b$ and higher temperatures $T_b$ than the usual FFLO-like instability \cite{Fischer:2023}.    This is closely linked to the symmetry required instability to a pair density wave  state for non-centrosymmetric superconductors when a field is applied \cite{Smidman:2017}. For a non-centrosymmetric system under magnetic field, both inversion and time-reversal symmetry are broken. As a result, the pairing susceptibility is generically peaked at non-zero momentum and $T_b=T_c$. For locally non-centrosymmtric superconductors, inversion symmetry is locally broken on each sublattice. In an extreme case, if the two sublattices are decoupled, then the system effectively becomes non-centrosymmetric, and  under a small magnetic field, an FFLO state can exists right below the zero-field superconducting $T_c$. However, these sublattices are generically coupled so that $T_b=T_c$ is not realized in practice. Here we show that for type 1 Hamiltonians,  FFLO-like states can in principle exist up to $T_b=T_c$. 
	
To show this, we consider the 2D version of class $D_{4h,1}^{\text{type}1}$ and use the pairing susceptibility to calculate $T_b$ and $H_b$.  In 2D, class $D_{4h,1}^{\text{type}1}$ has the following normal state Hamiltonian:
	\begin{equation}
		H_{D_{4h,1}}=\frac{\hbar^2}{2m} (k_x^2+k_y^2)-\mu+t_1k_xk_y\tau_1+\lambda_x\tau_3(k_y\sigma_x+k_x\sigma_y)+H_x\sigma_x
	\end{equation}
	$\lambda_x$ denotes the strength of the local inversion symmetry breaking (local Rashba SOC), while $t_1$ is the inter-sublattice coupling. The pairing susceptibility for an $s$-wave state with gap function $\tau_0\psi_{\bm{k}}$ is
	\begin{equation}
		\begin{split}
			\chi_\text{pairing}({\bf Q})=-\frac{1}{\beta}\sum_{\omega_n}\sum_{(\bf p,\bf p+Q)\in\text{FS}}\text{Tr}\left[G_0({\bf Q+p},\omega_n)G_0({\bf p},\omega_n)\right],
		\end{split}
	\end{equation}
 where $G_0$ is the normal state Green's function written in Nambu space. The FFLO state is favored, if the pairing susceptibility is peaked at non-zero $\bf Q$.  We examine the position of the bicritical point $(T_b,H_b)$, as a function of $\lambda_x/(t_1 k_F)$. We use the following two equations to locate the bicritical point: (1) The bicritical point lies on the BCS transition for the uniform superconductivity. (2) The bicritical point is a continuous phase transition between uniform and FFLO superconductivity, where $\nabla^2_{\bf Q}\chi_\text{pairing}({\bf Q})=0$. The result is in Fig.~4. $1000\times1000$ points are sampled in the 2D Brillouin zone. Other parameters are $t_1=0.2$, $t=\mu=1$. An energy cutoff of $E_c=0.1$ is applied to determine the position of the Fermi surface.

\begin{figure}[htb]
	\centering
	\includegraphics[height=6cm]{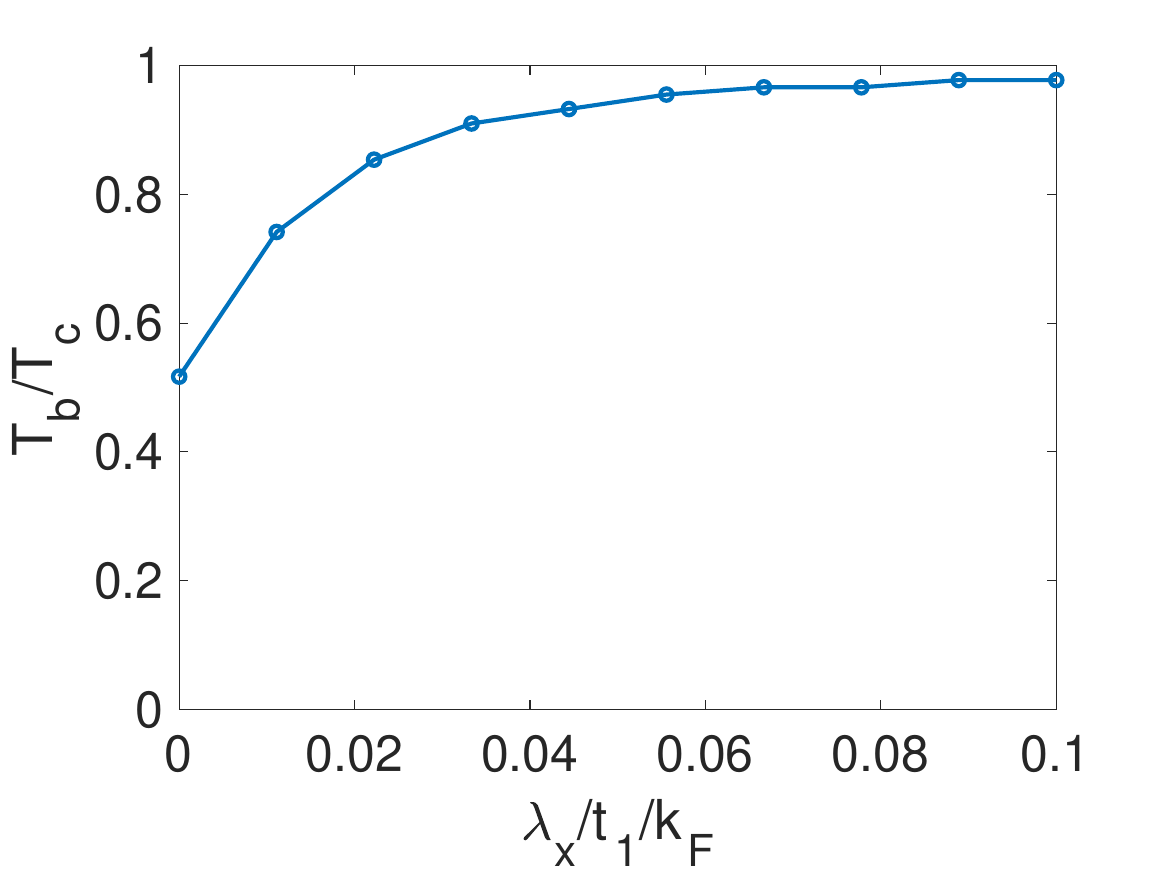}
	\includegraphics[height=6cm]{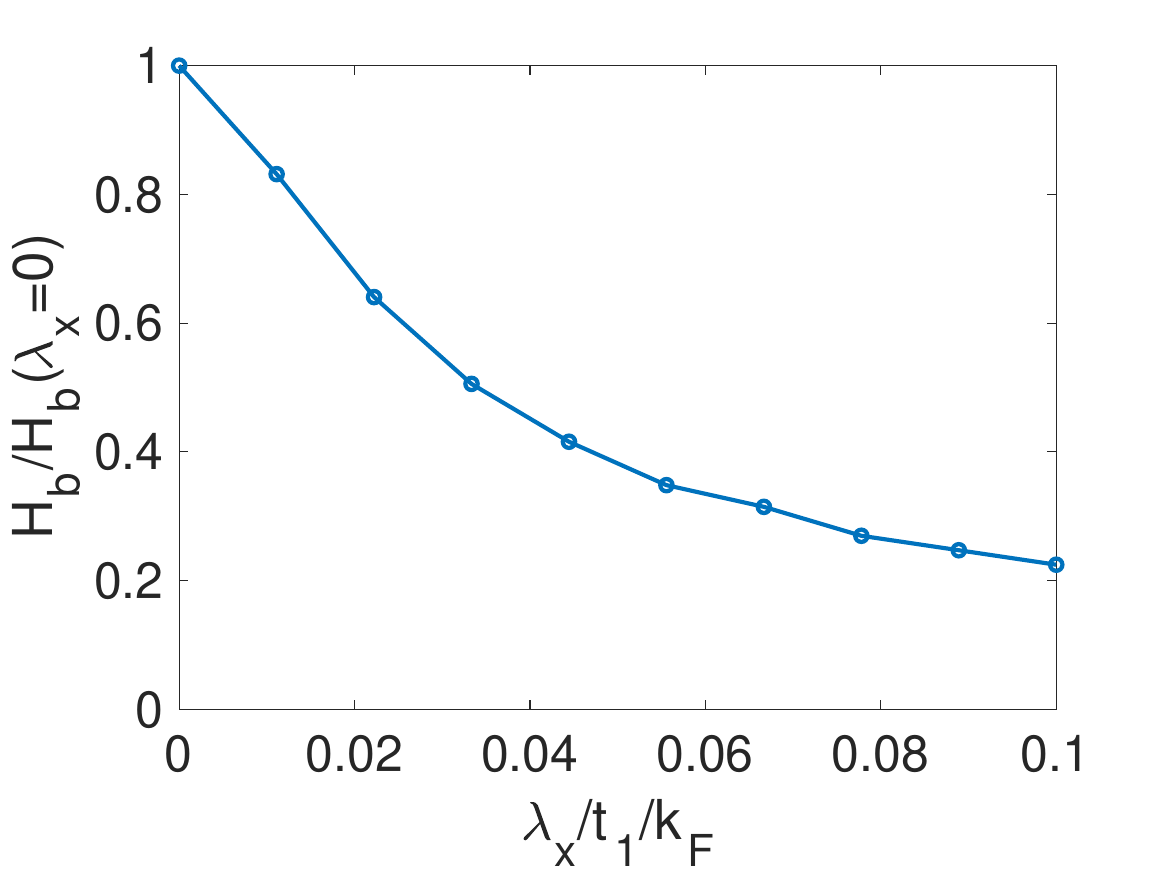}
	\caption{The position of the bicritical point $(T_b,H_b)$, as a function of $\lambda_x/k_Ft_1$.}
\end{figure}

These results show that for zero $\lambda_x/k_Ft_1$, a usual FFLO phase is found (that is $T_b/T_c\approx 0.56$). As the SOC $\lambda_x$ increases or equivalently, as $k_F$ decreases, $T_b$ increases and approaches the zero-field critical temperature. In the meantime, $H_b$ monotonically decreases.

We have shown that the FFLO phase can exist up to $T_b=T_c$ for a 2D version of class $D_{4h,1}^{\text{type}1}$. Key is that SOC is the leading order term in the $kp$ theory and this is also the case for other type 1 Hamiltonians. Hence the optimal conditions for an enhanced FFLO phase to occur are when fields are applied in-plane (perpendicular to the $c$-axis) for classes $D_{2h,1}^{\text{type}1}$, $D_{4h,1}^{\text{type}1}$, $D_{4h,3}^{\text{type}1}$, and $D_{4h,5}^{\text{type}1}$. 

 \subsection{Odd-parity superconductors}

For odd parity superconductors, the field fitness parameter $\tilde{F}_{{\bm k},\bm{\hat{h}}}$ can become less than 1 \cite{Cavanagh2:2022}. Of particular interest is when $\tilde{F}_{\bm {k},\bm{\hat{h}}}=0$ since this implies that $T_c$ is unchanged by the time-reversal symmetry breaking field (this is independent of the effective $g$-factor) \cite{Cavanagh2:2022}. For anomalous pseudospin this possibility leads to two consequences not expected for spin-triplet states made from usual spin-1/2 fermions. The first is a field induced transition from an even to an odd parity state. The second is that, in spite of the presence of strong SOC, the superconducting state is immune to magnetic fields for all field orientations. We discuss these each in turn.

\subsubsection{Field induced even to odd parity transitions}

In CeRh$_2$As$_2$, a field induced even to odd parity transition has been observed for the field oriented along the $c$-axis in this tetragonal material \cite{Khim:2021,Landaeta:2022}. Earlier, we argued that this was due the anomalous pseudospin that arises on the Brillouin zone faces in the  non-symmorphic space group P4/nmm \cite{Cavanagh:2022}. Here we show how this can be generalized to other space groups that admit type 1 {\it kp} theories and determine which classes are optimal for observing such a transition.  As discussed in Section IV C, an attractive electron-phonon like interaction gives rise to both both a usual $s$-wave $\tau_0\psi_{\bm{k}}$ state and an odd-parity  $\tau_3\psi_{\bm{k}}$ state. These two states have the same pairing interaction, but the gap projected onto the band basis is generally smaller for the $\tau_3\psi_{\bm{k}}$ state than for the $\tau_0\psi_{\bm{k}}$ state, implying that $\tau_0\psi_{\bm{k}}$ state has the higher $T_c$. For the type 1 classes $D_{2h,1}^{\text{type}1}$, $D_{4h,1}^{\text{type}1}$, $D_{4h,3}^{\text{type}1}$, and $D_{4h,5}^{\text{type}1}$,  anomalous pseudospin leads to $T_c$'s that are nearly the same for the even $\tau_0\psi$ and odd-parity $\tau_3\psi$ states. These classes are therefore promising  for observing a field induced transition from an even-parity to an odd-parity state.

To determine if a such a field induced transition occurs we compute $\tilde{F}_{\bm{k},\bm{\hat{h}}}$ for a pairing state $\tilde{\Delta}=\tau_3$. We find for type 1 {\it kp} theories
\begin{equation}
\tilde{F}_{\bm{k},\bm{\hat{h}}}=\frac{(\bm{\hat{h}}\cdot\bm{\lambda}_{\bm{k}})^2(t_{1,\bm{k}}^2+t_{2,\bm{k}}^2+|\bm{\lambda}_{\bm{k}}|^2)}{|\bm{\lambda}_{\bm{k}}|^2[\bm{\hat{h}}^2(t_{1,\bm{k}}^2+t_{2,\bm{k}}^2)+(\bm{\hat{h}}\cdot\bm{\lambda}_{\bm{k}})^2]}.
\end{equation}
Notice if $\bm{\hat{h}}\cdot\bm{\lambda}_{\bm{k}}=0$, then $\tilde{F}_{\bm{k},\bm{\hat{h}}}=0$ which maximizes $T_c$.
To determine the field orientations for which  $\tilde{F}_{\bm{k},\bm{\hat{h}}}=0$, we examine the form of $\bm{\lambda}_{\bm{k}}$ in the type 1 classes discussed above. In all these classes, the $\lambda_{z,\bm{k}}$ component appears with a higher power of momenta than the other components. Consequently, the field should be applied along the $\hat{z}$  direction. As an example, consider the class $D_{4h,3}^{\text{type}1}$. Here $\lambda_{z,\bm{k}}\propto k_xk_yk_z(k_x^2-k_y^2)$ while $\lambda_{x,\bm{k}}\propto k_y$ and $\lambda_{y,\bm{k}}\propto k_y$. In this case $\bm{\lambda}_{\bm{k}}$ will be in-plane to an excellent approximation, and an even to odd-parity transition can be expected for the field along the $c$-axis. Consequently, classes $D_{2h,1}^{\text{type}1}$, $D_{4h,1}^{\text{type}1}$, $D_{4h,3}^{\text{type}1}$, and $D_{4h,5}^{\text{type}1}$ and, hence, space groups 56, 58, 59, 62, 128, 129, 130, 136, 137, and 138 are promising for realizing a field-induced even to odd parity transition. 

 \subsubsection{Field immune odd-parity superconductivity}

For a conventional spin-triplet superconductor (with $\Delta=\bm{d}_{\bm{k}}\cdot\bm{\sigma}$) formed from usual spin-1/2 pseudospin, SOC typically pins the direction of the vector $\bm{d}_{\bm{k}}$. If the applied field is perpendicular to $\bm{d}_{\bm{k}}$, that is if  $\bm{d}_{\bm{k}}\cdot \bm{\hat{h}}=0$, then the $T_c$ for this field orientation is unchanged \cite{Sigrist:2005,Mineev:1999,Machida:1985}. Since there exists at least one field  direction for which $\bm{d}_{\bm{k}}\cdot \bm{\hat{h}}\ne 0$, it is not expected that usual spin-triplet superconductors are immune to fields applied in all  directions. For anomalous pseudopsin,  this is not the case, it is possible for an odd-parity state to be robust against suppression for arbitrarily oriented magnetic fields. To show how this is possible, we calculate $\tilde{F}_{\bm{k},\bm{\hat{h}}}$ for $\Delta=\tau_0(\bm{d}_{\bm{k}}\cdot{\bm{\sigma}})$ for type 1 {\it kp} theories, this yields
\begin{equation}
    \tilde{F}_{\bm{k},\bm{\hat{h}}}=\frac{[(t_{1,\bm{k}}^2+t_{2,\bm{k}}^2)\bm{d}_{\bm{k}}\cdot\bm{\hat{h}}+(\bm{d}_{\bm{k}}\cdot\bm{\lambda}_{\bm{k}})(\bm{\lambda}_{\bm{k}}\cdot\bm{\hat{h}})]^2}{[(t_{1,\bm{k}}^2+t_{2,\bm{k}}^2)\bm{\hat{h}}^2+(\bm{\lambda}_{\bm{k}}\cdot\bm{\hat{h}})^2][(t_{1,\bm{k}}^2+t_{2,\bm{k}}^2)|\bm{d}_{\bm{k}}|^2+(\bm{d}_{\bm{k}}\cdot\bm{\lambda}_{\bm{k}})^2]}.
 \label{Eq_immune}
\end{equation}
We first note that near the nodal plane, the effective $g$-factor is small for in-plane fields $\bm{\hat{n}}\cdot {\bm{ h}}=0$, so that for these field orientations superconductivity is not strongly suppressed (this is true for both even and odd-parity superconducting states). Hence, to show that an odd-parity state survives for all  field orientations, we need to show that $\tilde{F}_{\bm{k},\bm{\hat{h}}}\approx 0$ for a field applied along the nodal plane normal where $\bm{\lambda}_{\bm{k}}\cdot\bm{\hat{h}}$ becomes maximal.  Near the plane we expect that $\bm{\lambda}_{\bm{k}}\cdot\bm{\hat{h}} \gg \sqrt{t_{1,\bm{k}}^2+t_{2,\bm{k}}^2}$. 
Also, $(t_{1,\bm{k}}^2+t_{2,\bm{k}}^2)$ is small compared to $\bm{\lambda}_{\bm{k}}^2$, so $\tilde{F}_{\bm{k},\bm{\hat{h}}}$ is dominated by the $\bm{d}_{\bm{k}}\cdot\bm{\lambda}_{\bm{k}}$ term in the numerator. Hence if the denominator $|t_{1,2}\bm{d}_{\bm{k}}|$ is much bigger than $\bm{d}_{\bm{k}}\cdot\bm{\lambda}_{\bm{k}}$, then $\tilde{F}_{\bm{k},\bm{\hat{h}}}\approx 0$. Given that $\lambda_{\hat{n}}$ is the largest SOC component, this requirement is equivalent to $\lambda_{\perp}\ll t_{1,2}$ and $\bm{d}_{\bm{k}}\perp\hat{n}$ (where $\lambda_{\perp}$ is the magnitude of the SOC perpendicular to $\hat{n}$).



As a relevant example of the above mechanism we consider UPt$_3$ \cite{Joynt:2002}.  The superconducting state  in UPt$_3$ is believed to be an $E_{2u}$ state, with order parameter $\Delta=\eta_p(\sigma_xk_y+\sigma_yk_x)+\eta_f\sigma_z k_zk_xk_y$ (we only include one component of this two-component order parameter since similar arguments hold for the second component). In general, since the p-wave and f-wave components have the same symmetry, both $\eta_p$ and $\eta_f$ are non-zero. However, theories based on the usual pseudospin typically require $\eta_p=0$  due to the experimental observations discussed below \cite{Sauls:1994,Yanase:2016,Choi:1991}. Below we further show that $\eta_p=0$ is not required for these experimental observations when anomalous pseudospin is considered. Indeed, these experiments are consistent with $\eta_f=0$ and $\eta_p\ne 0$ if pairing occurs predominantly near the nodal plane $k_z=\pi/c$.

Thermal conductivity experiments suggest the existence of line nodes \cite{Joynt:2002}. For  usual pseudospin, the state $\sigma_xk_y+\sigma_yk_x$  is either fully gapped or has only point nodes.  This is one reason to expect that $\eta_p=0$. However, as illustrated in Table II, line nodes are expected for this state on the $k_z=\pi/c$ plane (note this conclusion also follows from Refs~\cite{Sumita:2018,Kobayashi:2016,Micklitz:2009}). This is relevant for UPt$_3$ since it is known to have the `starfish' Fermi surface near this nodal plane \cite{Joynt:2002} which belongs to class $D_{6h}^{\text{type}1}$

In terms of paramagnetic suppression, the superconducting state is known to be more robust under $\bm{B}\perp \hat{\bm{z}}$ compared to $\bm{B}\parallel \hat{\bm{z}}$ \cite{Choi:1991}. For the usual pseudospin, this requires $\bm{d}_{\bm{k}} \parallel \hat{\bm{z}}$, and thus $\eta_p=0$. However, on the `starfish' Fermi surface, the small $g$-factor for $\bm{B}\perp \hat{\bm{z}}$ can serve to protect the p-wave state against paramagnetic suppression. As discussed above, the suppression from $\bm{B} 
 \parallel \hat{\bm{z}}$ depends on the ratio $\lambda_{x,y}/t_{1,2}$, while the $g$-factor for $\bm{B}\perp \hat{\bm{z}}$ depends on the ratio $(t_{1,2},\lambda_{x,y})/\lambda_z$. The requirement $\lambda_{x,y}/t_{1,2}>(t_{1,2},\lambda_{x,y})/\lambda_z$ is thus sufficient to match the observations on the upper critical fields.
If both ratios are much smaller than one, the p-wave state is immune to paramagnetic suppression for field along arbitrary directions. This could be relevant to the approximately unchanged Knight shift in the superconducting state \cite{Tou:1996}.
We note that the use of $\tilde{F}_{\bm{k},\bm{\hat{h}}}$ to determine the magnetic response relies on the validity of projection to a single band. However, for class $D_{6h}^{\text{type}1}$ band degeneracies exist along three Dirac lines for which this projection is not valid. In Appendix B  we include a detailed numerical calculation that includes interband effects.

\section{8-fold degenerate points: application to UCoGe}

The arguments presented above relied on the 4-fold degeneracy at TRIM points when SOC is not present. However, some of these TRIM points have an 8-fold degeneracy without SOC. It is reasonable to ask if the conclusions found for {\it kp} theories of 4-fold degenerate points discussed above survive to 8-fold degenerate points.  To address this, we have determined the symmetries of all orbital operators in Appendix C. We find that in most cases, the 8-fold degeneracy at these TRIM is split by a single SOC term of the form $O\sigma_i$ where $O$ is a momentum independent 4 by 4 orbital matrix. In Table.\ref{tab:8fold}, we give the direction of the spin component $\sigma_i$ that appears in this SOC term at the TRIM point. The existence of this single SOC term ensures small effective $g$-factors for fields perpendicular to the spin-component direction. Consequently, the conclusions associated with the effective $g$-factor anisotropy discussed in Section V still hold for these 8-fold degenerate points.  We note that the 8-fold degeneracy at the $A$ point of space groups 130 and 135 are not split by SOC and these points provide examples of double Dirac points examined in \cite{Wieder:2016,Bradlyn:2016}.

\begin{table}[h]
	\begin{tabular}{|c|c|}
		\hline
		Spin Alignment    & Space Group Momenta          \\ \hline
		$\sigma_x$       & 54($U_1U_2$),54($R_1R_2$),56($U_1U_2$),60($R_1R_2$),61($S_1S_2$),62($S_1S_2$),205($M_1M_2$) \\ \hline
		$\sigma_y$       & 52($S_1S_2$),56($T_1T_2$),57($T_1T_2$),57($R_1R_2$),61($T_1T_2$),130($R_1R_2$),138($R_1R_2$)     \\ \hline
		$\sigma_z$        & 60($T_1T_2$),60($U_1U_2$),61($U_1U_2$),62($R_1R_2$),128($A_3A_4$),137($A_3A_4$),176($A_2A_3$),193($A_3$),194($A_3$)   \\ \hline
	\end{tabular}
	\caption{Spin alignment of 8-fold degenerate TRIM. }
 \label{tab:8fold}
\end{table}

One material for which these 8-fold degenerate points are likely to be  relevant is  the ferromagetic superconductor UCoGe, which crystalizes in space group 62 (Pnma) \cite{Aoki:2019}. UCoGe is believed to be a possibly topological  odd-parity superconductor \cite{Aoki:2019,Daido:2019}.  Our Fermi surface (given in Figure 3) reveals that all Fermi surface sheets lie near nodal planes with anomalous pseudospin and further reveal tube-shaped pockets that enclose the zone-boundary S point and stretch along the S-R axis. Here we focus on these Fermi surfaces.  This feature reasonably agrees with previous works 
\cite{Samsel2010,PhysRevB.91.174503,PhysRevLett.122.227001} using local density approximation and the existence of these tube shaped Fermi surfaces is consistent with quantum oscillation measurements \cite{Bastien:2016}. Here density-functional theory calculations for UCoGe were carried out by 
the full-potential linearized augmented plane wave method
\cite{FLAPW2009}.
Perdew-Burke-Ernzerhof form of exchange correlation functional \cite{PBE}, 
wave function and potential energy cutoffs of 16 and 200 Ry, respectively, muffin-tin sphere radii of 1.4 \AA{} for U and 1.2 \AA{} for Co and Ge, respectively, 
the experimental lattice parameters \cite{CANEPA1996225}, and an $8\times12 
\times 8$ $k$-point mesh were employed for the self-consistent field calculation.  Spin-orbit was fully taken into account in the assumed nonmagnetic state. Fermi surface was determined on a dense $30\times 50 \times 30$ $k$-point mesh and visualized by using {\tt FermiSurfer} \cite{KAWAMURA2019197}.   

Both the $R$ and $S$ points are 8-fold degenerate TRIM when SOC is not included for space group 62. Interestingly, from Table IV, the effective $g$-factors for fields along $\bm{\hat{y}}$ and $\bm{\hat{z}}$ directions are zero at the S-point and are zero for fields along $\bm{\hat{x}}$ and $\bm{\hat{y}}$ directions at the R-point. This indicates that superconductivity (both even and odd-parity) on the tube-shaped Fermi surfaces will be robust against magnetic fields applied along the $\bm{\hat{y}}$ direction. This is the field direction for which the upper critical field is observed to be the highest and for which an unusual S-shaped critical field curve appears \cite{Aoki:2019}. We leave a detailed examination of the consequences of anomalous pseudospin in space group 62 on superconductivity to a later work.

\begin{figure*}[tt]
	\centering
	\includegraphics[width=0.5\linewidth]{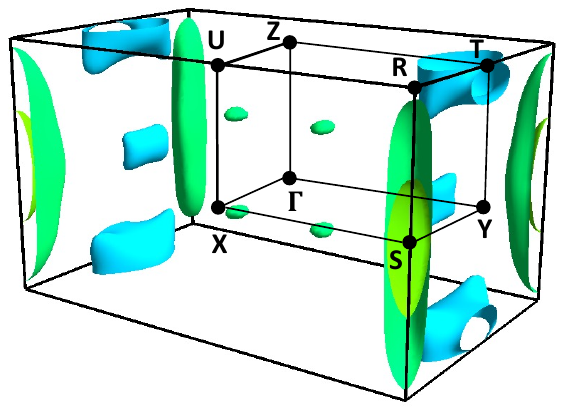}
	\caption{\label{Fig:UCoGe_FS} DFT Fermi surface of UCoGe.}
\end{figure*}

\section{Conclusions}

Non-symmorphic symmetries allow the existence of nodal planes at Brillouin zone edges when no SOC is present. When SOC is added, the pseudospin on these nodal planes has different symmetry properties than usual pseudospin-1/2. Here we have classified all space groups and effective single-particle theories near TRIM points on these nodal planes and examined the consequences of this anomalous pseudospin on the superconducting state. We have shown how this enhances the $T_c$ for  odd-parity superconducting states due to attractive interactions, leads to  unexpected superconducting nodal properties, allows large Pauli limiting fields and pair density wave  states for spin-singlet superconductors, gives rise to field immune odd-parity superconductivity, and to  field driven even to odd-parity superconducting transitions. Some of these properties have also been predicted for locally non-centrosymmetric superconductors, however anomalous pseudospin applies even when the crystal site symmetry contains inversion symmetry. This greatly extends the   number of materials that can exhibit this superconducting response. While we have emphasized nodal planes on which anomalous pseudospin exists, there are also materials for which anomalous pseudospin develops on nodal lines and not on nodal planes. Some such materials also exhibit unusual response to magnetic fields \cite{Shimizu:2019,Ma:2021,Ruan:2022}, suggesting a broader range of applicability for anomalous pseudospin superconductivity.

\section{Acknowledgements} 
DFA, HGS, and YY 
were supported by the US Department of Energy, Office
of Basic Energy Sciences, Division of Materials Sciences
and Engineering under Award DE-SC0021971 and by a UWM Discovery and Innovation Grant.
MW and TS were supported by
the US Department of Energy, Office of Basic Energy Sciences, Division of Materials Sciences and Engineering under Award DE-SC0017632. PMRB was supported by the Marsden Fund Council from Government
funding, managed by Royal Society Te Aparangi.  We acknowledge useful discussions with Mark
Fischer, Elena Hassinger, Seunghyun
Khim, Igor Mazin, Johnpierre Paglione, and Manfred Sigrist.



\bibliography{refs}

\begin{thebibliography}{89}%
\makeatletter
\providecommand \@ifxundefined [1]{%
 \@ifx{#1\undefined}
}%
\providecommand \@ifnum [1]{%
 \ifnum #1\expandafter \@firstoftwo
 \else \expandafter \@secondoftwo
 \fi
}%
\providecommand \@ifx [1]{%
 \ifx #1\expandafter \@firstoftwo
 \else \expandafter \@secondoftwo
 \fi
}%
\providecommand \natexlab [1]{#1}%
\providecommand \enquote  [1]{``#1''}%
\providecommand \bibnamefont  [1]{#1}%
\providecommand \bibfnamefont [1]{#1}%
\providecommand \citenamefont [1]{#1}%
\providecommand \href@noop [0]{\@secondoftwo}%
\providecommand \href [0]{\begingroup \@sanitize@url \@href}%
\providecommand \@href[1]{\@@startlink{#1}\@@href}%
\providecommand \@@href[1]{\endgroup#1\@@endlink}%
\providecommand \@sanitize@url [0]{\catcode `\\12\catcode `\$12\catcode
  `\&12\catcode `\#12\catcode `\^12\catcode `\_12\catcode `\%12\relax}%
\providecommand \@@startlink[1]{}%
\providecommand \@@endlink[0]{}%
\providecommand \url  [0]{\begingroup\@sanitize@url \@url }%
\providecommand \@url [1]{\endgroup\@href {#1}{\urlprefix }}%
\providecommand \urlprefix  [0]{URL }%
\providecommand \Eprint [0]{\href }%
\providecommand \doibase [0]{https://doi.org/}%
\providecommand \selectlanguage [0]{\@gobble}%
\providecommand \bibinfo  [0]{\@secondoftwo}%
\providecommand \bibfield  [0]{\@secondoftwo}%
\providecommand \translation [1]{[#1]}%
\providecommand \BibitemOpen [0]{}%
\providecommand \bibitemStop [0]{}%
\providecommand \bibitemNoStop [0]{.\EOS\space}%
\providecommand \EOS [0]{\spacefactor3000\relax}%
\providecommand \BibitemShut  [1]{\csname bibitem#1\endcsname}%
\let\auto@bib@innerbib\@empty
\bibitem [{\citenamefont {Manchon}\ \emph {et~al.}(2015)\citenamefont
  {Manchon}, \citenamefont {Koo}, \citenamefont {Nitta}, \citenamefont
  {Frolov},\ and\ \citenamefont {Duine}}]{Manchon:2015}%
  \BibitemOpen
  \bibfield  {author} {\bibinfo {author} {\bibfnamefont {A.}~\bibnamefont
  {Manchon}}, \bibinfo {author} {\bibfnamefont {H.~C.}\ \bibnamefont {Koo}},
  \bibinfo {author} {\bibfnamefont {J.}~\bibnamefont {Nitta}}, \bibinfo
  {author} {\bibfnamefont {S.~M.}\ \bibnamefont {Frolov}},\ and\ \bibinfo
  {author} {\bibfnamefont {R.~A.}\ \bibnamefont {Duine}},\ }\bibfield  {title}
  {\bibinfo {title} {{New perspectives for Rashba spin{\textendash}orbit
  coupling}},\ }\href {https://doi.org/10.1038/nmat4360} {\bibfield  {journal}
  {\bibinfo  {journal} {Nature Materials}\ }\textbf {\bibinfo {volume} {14}},\
  \bibinfo {pages} {871} (\bibinfo {year} {2015})}\BibitemShut {NoStop}%
\bibitem [{\citenamefont {Smidman}\ \emph {et~al.}(2017)\citenamefont
  {Smidman}, \citenamefont {Salamon}, \citenamefont {Yuan},\ and\ \citenamefont
  {Agterberg}}]{Smidman:2017}%
  \BibitemOpen
  \bibfield  {author} {\bibinfo {author} {\bibfnamefont {M.}~\bibnamefont
  {Smidman}}, \bibinfo {author} {\bibfnamefont {M.~B.}\ \bibnamefont
  {Salamon}}, \bibinfo {author} {\bibfnamefont {H.~Q.}\ \bibnamefont {Yuan}},\
  and\ \bibinfo {author} {\bibfnamefont {D.~F.}\ \bibnamefont {Agterberg}},\
  }\bibfield  {title} {\bibinfo {title} {Superconductivity and spin--orbit
  coupling in non-centrosymmetric materials: a review},\ }\href
  {http://stacks.iop.org/0034-4885/80/i=3/a=036501} {\bibfield  {journal}
  {\bibinfo  {journal} {Reports on Progress in Physics}\ }\textbf {\bibinfo
  {volume} {80}},\ \bibinfo {pages} {036501} (\bibinfo {year}
  {2017})}\BibitemShut {NoStop}%
\bibitem [{\citenamefont {Baltz}\ \emph {et~al.}(2018)\citenamefont {Baltz},
  \citenamefont {Manchon}, \citenamefont {Tsoi}, \citenamefont {Moriyama},
  \citenamefont {Ono},\ and\ \citenamefont {Tserkovnyak}}]{Baltz:2018}%
  \BibitemOpen
  \bibfield  {author} {\bibinfo {author} {\bibfnamefont {V.}~\bibnamefont
  {Baltz}}, \bibinfo {author} {\bibfnamefont {A.}~\bibnamefont {Manchon}},
  \bibinfo {author} {\bibfnamefont {M.}~\bibnamefont {Tsoi}}, \bibinfo {author}
  {\bibfnamefont {T.}~\bibnamefont {Moriyama}}, \bibinfo {author}
  {\bibfnamefont {T.}~\bibnamefont {Ono}},\ and\ \bibinfo {author}
  {\bibfnamefont {Y.}~\bibnamefont {Tserkovnyak}},\ }\bibfield  {title}
  {\bibinfo {title} {Antiferromagnetic spintronics},\ }\href
  {https://doi.org/10.1103/RevModPhys.90.015005} {\bibfield  {journal}
  {\bibinfo  {journal} {Rev. Mod. Phys.}\ }\textbf {\bibinfo {volume} {90}},\
  \bibinfo {pages} {015005} (\bibinfo {year} {2018})}\BibitemShut {NoStop}%
\bibitem [{\citenamefont {Zhang}\ \emph {et~al.}(2014)\citenamefont {Zhang},
  \citenamefont {Liu}, \citenamefont {Luo}, \citenamefont {Freeman},\ and\
  \citenamefont {Zunger}}]{Zhang:2014}%
  \BibitemOpen
  \bibfield  {author} {\bibinfo {author} {\bibfnamefont {X.}~\bibnamefont
  {Zhang}}, \bibinfo {author} {\bibfnamefont {Q.}~\bibnamefont {Liu}}, \bibinfo
  {author} {\bibfnamefont {J.-W.}\ \bibnamefont {Luo}}, \bibinfo {author}
  {\bibfnamefont {A.~J.}\ \bibnamefont {Freeman}},\ and\ \bibinfo {author}
  {\bibfnamefont {A.}~\bibnamefont {Zunger}},\ }\bibfield  {title} {\bibinfo
  {title} {Hidden spin polarization in inversion-symmetric bulk crystals},\
  }\href@noop {} {\bibfield  {journal} {\bibinfo  {journal} {Nature Physics}\
  }\textbf {\bibinfo {volume} {10}},\ \bibinfo {pages} {387} (\bibinfo {year}
  {2014})}\BibitemShut {NoStop}%
\bibitem [{\citenamefont {Fischer}\ \emph {et~al.}(2023)\citenamefont
  {Fischer}, \citenamefont {Sigrist}, \citenamefont {Agterberg},\ and\
  \citenamefont {Yanase}}]{Fischer:2023}%
  \BibitemOpen
  \bibfield  {author} {\bibinfo {author} {\bibfnamefont {M.~H.}\ \bibnamefont
  {Fischer}}, \bibinfo {author} {\bibfnamefont {M.}~\bibnamefont {Sigrist}},
  \bibinfo {author} {\bibfnamefont {D.~F.}\ \bibnamefont {Agterberg}},\ and\
  \bibinfo {author} {\bibfnamefont {Y.}~\bibnamefont {Yanase}},\ }\bibfield
  {title} {\bibinfo {title} {Superconductivity and local inversion-symmetry
  breaking},\ }\bibfield  {journal} {\bibinfo  {journal} {Annual Review of
  Condensed Matter Physics}\ }\textbf {\bibinfo {volume} {14}},\ \href
  {https://doi.org/10.1146/annurev-conmatphys-040521-042511}
  {10.1146/annurev-conmatphys-040521-042511} (\bibinfo {year} {2023}),\ \Eprint
  {https://arxiv.org/abs/https://doi.org/10.1146/annurev-conmatphys-040521-042511}
  {https://doi.org/10.1146/annurev-conmatphys-040521-042511} \BibitemShut
  {NoStop}%
\bibitem [{\citenamefont {Yoshida}\ \emph {et~al.}(2012)\citenamefont
  {Yoshida}, \citenamefont {Sigrist},\ and\ \citenamefont
  {Yanase}}]{Yoshida:2012}%
  \BibitemOpen
  \bibfield  {author} {\bibinfo {author} {\bibfnamefont {T.}~\bibnamefont
  {Yoshida}}, \bibinfo {author} {\bibfnamefont {M.}~\bibnamefont {Sigrist}},\
  and\ \bibinfo {author} {\bibfnamefont {Y.}~\bibnamefont {Yanase}},\
  }\bibfield  {title} {\bibinfo {title} {Pair-density wave states through
  spin-orbit coupling in multilayer superconductors},\ }\href@noop {}
  {\bibfield  {journal} {\bibinfo  {journal} {Phys. Rev. B}\ }\textbf {\bibinfo
  {volume} {86}},\ \bibinfo {pages} {134514} (\bibinfo {year}
  {2012})}\BibitemShut {NoStop}%
\bibitem [{\citenamefont {Khim}\ \emph {et~al.}(2021)\citenamefont {Khim},
  \citenamefont {Landaeta}, \citenamefont {Banda}, \citenamefont {Bannor},
  \citenamefont {Brando}, \citenamefont {Brydon}, \citenamefont {Hafner},
  \citenamefont {K{\"u}chler}, \citenamefont {Cardoso-Gil}, \citenamefont
  {Stockert}, \citenamefont {Mackenzie}, \citenamefont {Agterberg},
  \citenamefont {Geibel},\ and\ \citenamefont {Hassinger}}]{Khim:2021}%
  \BibitemOpen
  \bibfield  {author} {\bibinfo {author} {\bibfnamefont {S.}~\bibnamefont
  {Khim}}, \bibinfo {author} {\bibfnamefont {J.~F.}\ \bibnamefont {Landaeta}},
  \bibinfo {author} {\bibfnamefont {J.}~\bibnamefont {Banda}}, \bibinfo
  {author} {\bibfnamefont {N.}~\bibnamefont {Bannor}}, \bibinfo {author}
  {\bibfnamefont {M.}~\bibnamefont {Brando}}, \bibinfo {author} {\bibfnamefont
  {P.~M.~R.}\ \bibnamefont {Brydon}}, \bibinfo {author} {\bibfnamefont
  {D.}~\bibnamefont {Hafner}}, \bibinfo {author} {\bibfnamefont
  {R.}~\bibnamefont {K{\"u}chler}}, \bibinfo {author} {\bibfnamefont
  {R.}~\bibnamefont {Cardoso-Gil}}, \bibinfo {author} {\bibfnamefont
  {U.}~\bibnamefont {Stockert}}, \bibinfo {author} {\bibfnamefont {A.~P.}\
  \bibnamefont {Mackenzie}}, \bibinfo {author} {\bibfnamefont {D.~F.}\
  \bibnamefont {Agterberg}}, \bibinfo {author} {\bibfnamefont {C.}~\bibnamefont
  {Geibel}},\ and\ \bibinfo {author} {\bibfnamefont {E.}~\bibnamefont
  {Hassinger}},\ }\bibfield  {title} {\bibinfo {title} {{Field-induced
  transition within the superconducting state of CeRh$_2$As$_2$}},\ }\href
  {https://doi.org/10.1126/science.abe7518} {\bibfield  {journal} {\bibinfo
  {journal} {Science}\ }\textbf {\bibinfo {volume} {373}},\ \bibinfo {pages}
  {1012} (\bibinfo {year} {2021})},\ \Eprint
  {https://arxiv.org/abs/https://www.science.org/doi/pdf/10.1126/science.abe7518}
  {https://www.science.org/doi/pdf/10.1126/science.abe7518} \BibitemShut
  {NoStop}%
\bibitem [{\citenamefont {Landaeta}\ \emph {et~al.}(2022)\citenamefont
  {Landaeta}, \citenamefont {Khanenko}, \citenamefont {Cavanagh}, \citenamefont
  {Geibel}, \citenamefont {Khim}, \citenamefont {Mishra}, \citenamefont
  {Sheikin}, \citenamefont {Brydon}, \citenamefont {Agterberg}, \citenamefont
  {Brando},\ and\ \citenamefont {Hassinger}}]{Landaeta:2022}%
  \BibitemOpen
  \bibfield  {author} {\bibinfo {author} {\bibfnamefont {J.~F.}\ \bibnamefont
  {Landaeta}}, \bibinfo {author} {\bibfnamefont {P.}~\bibnamefont {Khanenko}},
  \bibinfo {author} {\bibfnamefont {D.~C.}\ \bibnamefont {Cavanagh}}, \bibinfo
  {author} {\bibfnamefont {C.}~\bibnamefont {Geibel}}, \bibinfo {author}
  {\bibfnamefont {S.}~\bibnamefont {Khim}}, \bibinfo {author} {\bibfnamefont
  {S.}~\bibnamefont {Mishra}}, \bibinfo {author} {\bibfnamefont
  {I.}~\bibnamefont {Sheikin}}, \bibinfo {author} {\bibfnamefont {P.~M.~R.}\
  \bibnamefont {Brydon}}, \bibinfo {author} {\bibfnamefont {D.~F.}\
  \bibnamefont {Agterberg}}, \bibinfo {author} {\bibfnamefont {M.}~\bibnamefont
  {Brando}},\ and\ \bibinfo {author} {\bibfnamefont {E.}~\bibnamefont
  {Hassinger}},\ }\bibfield  {title} {\bibinfo {title} {{Field-Angle Dependence
  Reveals Odd-Parity Superconductivity in
  ${\mathrm{CeRh}}_{2}{\mathrm{As}}_{2}$}},\ }\href
  {https://doi.org/10.1103/PhysRevX.12.031001} {\bibfield  {journal} {\bibinfo
  {journal} {Phys. Rev. X}\ }\textbf {\bibinfo {volume} {12}},\ \bibinfo
  {pages} {031001} (\bibinfo {year} {2022})}\BibitemShut {NoStop}%
\bibitem [{\citenamefont {Cavanagh}\ \emph
  {et~al.}(2022{\natexlab{a}})\citenamefont {Cavanagh}, \citenamefont
  {Shishidou}, \citenamefont {Weinert}, \citenamefont {Brydon},\ and\
  \citenamefont {Agterberg}}]{Cavanagh:2022}%
  \BibitemOpen
  \bibfield  {author} {\bibinfo {author} {\bibfnamefont {D.~C.}\ \bibnamefont
  {Cavanagh}}, \bibinfo {author} {\bibfnamefont {T.}~\bibnamefont {Shishidou}},
  \bibinfo {author} {\bibfnamefont {M.}~\bibnamefont {Weinert}}, \bibinfo
  {author} {\bibfnamefont {P.~M.~R.}\ \bibnamefont {Brydon}},\ and\ \bibinfo
  {author} {\bibfnamefont {D.~F.}\ \bibnamefont {Agterberg}},\ }\bibfield
  {title} {\bibinfo {title} {{Nonsymmorphic Symmetry and Field-Driven
  Odd-Parity Pairing in CeRh$_2$As$_2$}},\ }\bibfield  {journal} {\bibinfo
  {journal} {Physical Review B}\ }\textbf {\bibinfo {volume} {105}},\ \href
  {https://doi.org/10.1103/PhysRevB.105.L020505} {10.1103/PhysRevB.105.L020505}
  (\bibinfo {year} {2022}{\natexlab{a}})\BibitemShut {NoStop}%
\bibitem [{\citenamefont {Yuan}\ \emph {et~al.}(2019)\citenamefont {Yuan},
  \citenamefont {Liu}, \citenamefont {Zhang}, \citenamefont {Luo},
  \citenamefont {Li},\ and\ \citenamefont {Zunger}}]{Yuan:2019}%
  \BibitemOpen
  \bibfield  {author} {\bibinfo {author} {\bibfnamefont {L.}~\bibnamefont
  {Yuan}}, \bibinfo {author} {\bibfnamefont {Q.}~\bibnamefont {Liu}}, \bibinfo
  {author} {\bibfnamefont {X.}~\bibnamefont {Zhang}}, \bibinfo {author}
  {\bibfnamefont {J.-W.}\ \bibnamefont {Luo}}, \bibinfo {author} {\bibfnamefont
  {S.-S.}\ \bibnamefont {Li}},\ and\ \bibinfo {author} {\bibfnamefont
  {A.}~\bibnamefont {Zunger}},\ }\bibfield  {title} {\bibinfo {title}
  {{Uncovering and tailoring hidden Rashba spin{\textendash}orbit splitting in
  centrosymmetric crystals}},\ }\bibfield  {journal} {\bibinfo  {journal}
  {Nature Communications}\ }\textbf {\bibinfo {volume} {10}},\ \href
  {https://doi.org/10.1038/s41467-019-08836-4} {10.1038/s41467-019-08836-4}
  (\bibinfo {year} {2019})\BibitemShut {NoStop}%
\bibitem [{\citenamefont {Li}\ and\ \citenamefont
  {Appelbaum}(2018)}]{Pengke:2018}%
  \BibitemOpen
  \bibfield  {author} {\bibinfo {author} {\bibfnamefont {P.}~\bibnamefont
  {Li}}\ and\ \bibinfo {author} {\bibfnamefont {I.}~\bibnamefont {Appelbaum}},\
  }\bibfield  {title} {\bibinfo {title} {Illuminating "spin-polarized" bloch
  wave-function projection from degenerate bands in decomposable
  centrosymmetric lattices},\ }\href
  {https://doi.org/10.1103/PhysRevB.97.125434} {\bibfield  {journal} {\bibinfo
  {journal} {Phys. Rev. B}\ }\textbf {\bibinfo {volume} {97}},\ \bibinfo
  {pages} {125434} (\bibinfo {year} {2018})}\BibitemShut {NoStop}%
\bibitem [{\citenamefont {Nakamura}\ and\ \citenamefont
  {Yanase}(2017)}]{Nakamura:2017}%
  \BibitemOpen
  \bibfield  {author} {\bibinfo {author} {\bibfnamefont {Y.}~\bibnamefont
  {Nakamura}}\ and\ \bibinfo {author} {\bibfnamefont {Y.}~\bibnamefont
  {Yanase}},\ }\bibfield  {title} {\bibinfo {title} {Odd-parity
  superconductivity in bilayer transition metal dichalcogenides},\ }\href
  {https://doi.org/10.1103/PhysRevB.96.054501} {\bibfield  {journal} {\bibinfo
  {journal} {Phys. Rev. B}\ }\textbf {\bibinfo {volume} {96}},\ \bibinfo
  {pages} {054501} (\bibinfo {year} {2017})}\BibitemShut {NoStop}%
\bibitem [{\citenamefont {Anderson}(1984)}]{Anderson:1984}%
  \BibitemOpen
  \bibfield  {author} {\bibinfo {author} {\bibfnamefont {P.~W.}\ \bibnamefont
  {Anderson}},\ }\bibfield  {title} {\bibinfo {title} {Structure of "triplet"
  superconducting energy gaps},\ }\href@noop {} {\bibfield  {journal} {\bibinfo
   {journal} {Phys. Rev. B}\ }\textbf {\bibinfo {volume} {30}},\ \bibinfo
  {pages} {4000} (\bibinfo {year} {1984})}\BibitemShut {NoStop}%
\bibitem [{\citenamefont {Bradley}\ and\ \citenamefont
  {Cracknell}(1972)}]{Bradley:1972}%
  \BibitemOpen
  \bibfield  {author} {\bibinfo {author} {\bibfnamefont {C.~J.}\ \bibnamefont
  {Bradley}}\ and\ \bibinfo {author} {\bibfnamefont {A.~P.}\ \bibnamefont
  {Cracknell}},\ }\href@noop {} {\emph {\bibinfo {title} {The Mathematical
  Theory of Symmetry in Solids}}}\ (\bibinfo  {publisher} {(Oxford University
  Press)},\ \bibinfo {year} {1972})\BibitemShut {NoStop}%
\bibitem [{\citenamefont {Zhang}\ \emph {et~al.}(2018)\citenamefont {Zhang},
  \citenamefont {Chan}, \citenamefont {Chiu}, \citenamefont {Vergniory},
  \citenamefont {Schoop},\ and\ \citenamefont {Schnyder}}]{Zhang:2018}%
  \BibitemOpen
  \bibfield  {author} {\bibinfo {author} {\bibfnamefont {J.}~\bibnamefont
  {Zhang}}, \bibinfo {author} {\bibfnamefont {Y.-H.}\ \bibnamefont {Chan}},
  \bibinfo {author} {\bibfnamefont {C.-K.}\ \bibnamefont {Chiu}}, \bibinfo
  {author} {\bibfnamefont {M.~G.}\ \bibnamefont {Vergniory}}, \bibinfo {author}
  {\bibfnamefont {L.~M.}\ \bibnamefont {Schoop}},\ and\ \bibinfo {author}
  {\bibfnamefont {A.~P.}\ \bibnamefont {Schnyder}},\ }\bibfield  {title}
  {\bibinfo {title} {Topological band crossings in hexagonal materials},\
  }\href {https://doi.org/10.1103/PhysRevMaterials.2.074201} {\bibfield
  {journal} {\bibinfo  {journal} {Phys. Rev. Mater.}\ }\textbf {\bibinfo
  {volume} {2}},\ \bibinfo {pages} {074201} (\bibinfo {year}
  {2018})}\BibitemShut {NoStop}%
\bibitem [{\citenamefont {Hirschmann}\ \emph {et~al.}(2021)\citenamefont
  {Hirschmann}, \citenamefont {Leonhardt}, \citenamefont {Kilic}, \citenamefont
  {Fabini},\ and\ \citenamefont {Schnyder}}]{Hirschmann:2021}%
  \BibitemOpen
  \bibfield  {author} {\bibinfo {author} {\bibfnamefont {M.~M.}\ \bibnamefont
  {Hirschmann}}, \bibinfo {author} {\bibfnamefont {A.}~\bibnamefont
  {Leonhardt}}, \bibinfo {author} {\bibfnamefont {B.}~\bibnamefont {Kilic}},
  \bibinfo {author} {\bibfnamefont {D.~H.}\ \bibnamefont {Fabini}},\ and\
  \bibinfo {author} {\bibfnamefont {A.~P.}\ \bibnamefont {Schnyder}},\
  }\bibfield  {title} {\bibinfo {title} {{Symmetry-enforced band crossings in
  tetragonal materials: Dirac and Weyl degeneracies on points, lines, and
  planes}},\ }\href {https://doi.org/10.1103/PhysRevMaterials.5.054202}
  {\bibfield  {journal} {\bibinfo  {journal} {Phys. Rev. Mater.}\ }\textbf
  {\bibinfo {volume} {5}},\ \bibinfo {pages} {054202} (\bibinfo {year}
  {2021})}\BibitemShut {NoStop}%
\bibitem [{\citenamefont {Leonhardt}\ \emph {et~al.}(2021)\citenamefont
  {Leonhardt}, \citenamefont {Hirschmann}, \citenamefont {Heinsdorf},
  \citenamefont {Wu}, \citenamefont {Fabini},\ and\ \citenamefont
  {Schnyder}}]{Leonhardt:2021}%
  \BibitemOpen
  \bibfield  {author} {\bibinfo {author} {\bibfnamefont {A.}~\bibnamefont
  {Leonhardt}}, \bibinfo {author} {\bibfnamefont {M.~M.}\ \bibnamefont
  {Hirschmann}}, \bibinfo {author} {\bibfnamefont {N.}~\bibnamefont
  {Heinsdorf}}, \bibinfo {author} {\bibfnamefont {X.}~\bibnamefont {Wu}},
  \bibinfo {author} {\bibfnamefont {D.~H.}\ \bibnamefont {Fabini}},\ and\
  \bibinfo {author} {\bibfnamefont {A.~P.}\ \bibnamefont {Schnyder}},\
  }\bibfield  {title} {\bibinfo {title} {Symmetry-enforced topological band
  crossings in orthorhombic crystals: Classification and materials discovery},\
  }\href {https://doi.org/10.1103/PhysRevMaterials.5.124202} {\bibfield
  {journal} {\bibinfo  {journal} {Phys. Rev. Mater.}\ }\textbf {\bibinfo
  {volume} {5}},\ \bibinfo {pages} {124202} (\bibinfo {year}
  {2021})}\BibitemShut {NoStop}%
\bibitem [{\citenamefont {Norman}(1995)}]{Norman:1995}%
  \BibitemOpen
  \bibfield  {author} {\bibinfo {author} {\bibfnamefont {M.~R.}\ \bibnamefont
  {Norman}},\ }\bibfield  {title} {\bibinfo {title} {Odd parity and line nodes
  in heavy-fermion superconductors},\ }\href
  {https://doi.org/10.1103/physrevb.52.15093} {\bibfield  {journal} {\bibinfo
  {journal} {Physical Review B}\ }\textbf {\bibinfo {volume} {52}},\ \bibinfo
  {pages} {15093} (\bibinfo {year} {1995})}\BibitemShut {NoStop}%
\bibitem [{\citenamefont {Micklitz}\ and\ \citenamefont
  {Norman}(2017{\natexlab{a}})}]{Micklitz:2017}%
  \BibitemOpen
  \bibfield  {author} {\bibinfo {author} {\bibfnamefont {T.}~\bibnamefont
  {Micklitz}}\ and\ \bibinfo {author} {\bibfnamefont {M.}~\bibnamefont
  {Norman}},\ }\bibfield  {title} {\bibinfo {title} {Symmetry-enforced line
  nodes in unconventional superconductors},\ }\bibfield  {journal} {\bibinfo
  {journal} {Physical Review Letters}\ }\textbf {\bibinfo {volume} {118}},\
  \href {https://doi.org/10.1103/physrevlett.118.207001}
  {10.1103/physrevlett.118.207001} (\bibinfo {year}
  {2017}{\natexlab{a}})\BibitemShut {NoStop}%
\bibitem [{\citenamefont {Micklitz}\ and\ \citenamefont
  {Norman}(2017{\natexlab{b}})}]{Micklitz:2017-2}%
  \BibitemOpen
  \bibfield  {author} {\bibinfo {author} {\bibfnamefont {T.}~\bibnamefont
  {Micklitz}}\ and\ \bibinfo {author} {\bibfnamefont {M.~R.}\ \bibnamefont
  {Norman}},\ }\bibfield  {title} {\bibinfo {title} {Nodal lines and nodal
  loops in nonsymmorphic odd-parity superconductors},\ }\bibfield  {journal}
  {\bibinfo  {journal} {Physical Review B}\ }\textbf {\bibinfo {volume} {95}},\
  \href {https://doi.org/10.1103/physrevb.95.024508}
  {10.1103/physrevb.95.024508} (\bibinfo {year}
  {2017}{\natexlab{b}})\BibitemShut {NoStop}%
\bibitem [{\citenamefont {Daido}\ \emph
  {et~al.}(2019{\natexlab{a}})\citenamefont {Daido}, \citenamefont {Yoshida},\
  and\ \citenamefont {Yanase}}]{Daido:2019}%
  \BibitemOpen
  \bibfield  {author} {\bibinfo {author} {\bibfnamefont {A.}~\bibnamefont
  {Daido}}, \bibinfo {author} {\bibfnamefont {T.}~\bibnamefont {Yoshida}},\
  and\ \bibinfo {author} {\bibfnamefont {Y.}~\bibnamefont {Yanase}},\
  }\bibfield  {title} {\bibinfo {title} {{${\bm{Z}}_{4}$ Topological
  Superconductivity in UCoGe}},\ }\href
  {https://doi.org/10.1103/PhysRevLett.122.227001} {\bibfield  {journal}
  {\bibinfo  {journal} {Phys. Rev. Lett.}\ }\textbf {\bibinfo {volume} {122}},\
  \bibinfo {pages} {227001} (\bibinfo {year} {2019}{\natexlab{a}})}\BibitemShut
  {NoStop}%
\bibitem [{\citenamefont {Yanase}(2016)}]{Yanase:2016}%
  \BibitemOpen
  \bibfield  {author} {\bibinfo {author} {\bibfnamefont {Y.}~\bibnamefont
  {Yanase}},\ }\bibfield  {title} {\bibinfo {title} {{Nonsymmorphic Weyl
  superconductivity in ${\mathrm{UPt}}_{3}$ based on ${E}_{2u}$
  representation}},\ }\href {https://doi.org/10.1103/PhysRevB.94.174502}
  {\bibfield  {journal} {\bibinfo  {journal} {Phys. Rev. B}\ }\textbf {\bibinfo
  {volume} {94}},\ \bibinfo {pages} {174502} (\bibinfo {year}
  {2016})}\BibitemShut {NoStop}%
\bibitem [{\citenamefont {Sumita}\ and\ \citenamefont
  {Yanase}(2018)}]{Sumita:2018}%
  \BibitemOpen
  \bibfield  {author} {\bibinfo {author} {\bibfnamefont {S.}~\bibnamefont
  {Sumita}}\ and\ \bibinfo {author} {\bibfnamefont {Y.}~\bibnamefont
  {Yanase}},\ }\bibfield  {title} {\bibinfo {title} {Unconventional
  superconducting gap structure protected by space group symmetry},\ }\href
  {https://doi.org/10.1103/PhysRevB.97.134512} {\bibfield  {journal} {\bibinfo
  {journal} {Phys. Rev. B}\ }\textbf {\bibinfo {volume} {97}},\ \bibinfo
  {pages} {134512} (\bibinfo {year} {2018})}\BibitemShut {NoStop}%
\bibitem [{\citenamefont {Kobayashi}\ \emph {et~al.}(2016)\citenamefont
  {Kobayashi}, \citenamefont {Yanase},\ and\ \citenamefont
  {Sato}}]{Kobayashi:2016}%
  \BibitemOpen
  \bibfield  {author} {\bibinfo {author} {\bibfnamefont {S.}~\bibnamefont
  {Kobayashi}}, \bibinfo {author} {\bibfnamefont {Y.}~\bibnamefont {Yanase}},\
  and\ \bibinfo {author} {\bibfnamefont {M.}~\bibnamefont {Sato}},\ }\bibfield
  {title} {\bibinfo {title} {Topologically stable gapless phases in
  nonsymmorphic superconductors},\ }\href
  {https://doi.org/10.1103/PhysRevB.94.134512} {\bibfield  {journal} {\bibinfo
  {journal} {Phys. Rev. B}\ }\textbf {\bibinfo {volume} {94}},\ \bibinfo
  {pages} {134512} (\bibinfo {year} {2016})}\BibitemShut {NoStop}%
\bibitem [{\citenamefont {Kobayashi}\ \emph {et~al.}(2014)\citenamefont
  {Kobayashi}, \citenamefont {Shiozaki}, \citenamefont {Tanaka},\ and\
  \citenamefont {Sato}}]{Kobayashi:2014}%
  \BibitemOpen
  \bibfield  {author} {\bibinfo {author} {\bibfnamefont {S.}~\bibnamefont
  {Kobayashi}}, \bibinfo {author} {\bibfnamefont {K.}~\bibnamefont {Shiozaki}},
  \bibinfo {author} {\bibfnamefont {Y.}~\bibnamefont {Tanaka}},\ and\ \bibinfo
  {author} {\bibfnamefont {M.}~\bibnamefont {Sato}},\ }\bibfield  {title}
  {\bibinfo {title} {Topological blount's theorem of odd-parity
  superconductors},\ }\href {https://doi.org/10.1103/PhysRevB.90.024516}
  {\bibfield  {journal} {\bibinfo  {journal} {Phys. Rev. B}\ }\textbf {\bibinfo
  {volume} {90}},\ \bibinfo {pages} {024516} (\bibinfo {year}
  {2014})}\BibitemShut {NoStop}%
\bibitem [{\citenamefont {Micklitz}\ and\ \citenamefont
  {Norman}(2009)}]{Micklitz:2009}%
  \BibitemOpen
  \bibfield  {author} {\bibinfo {author} {\bibfnamefont {T.}~\bibnamefont
  {Micklitz}}\ and\ \bibinfo {author} {\bibfnamefont {M.~R.}\ \bibnamefont
  {Norman}},\ }\bibfield  {title} {\bibinfo {title} {Odd parity and line nodes
  in nonsymmorphic superconductors},\ }\href
  {https://doi.org/10.1103/PhysRevB.80.100506} {\bibfield  {journal} {\bibinfo
  {journal} {Phys. Rev. B}\ }\textbf {\bibinfo {volume} {80}},\ \bibinfo
  {pages} {100506(R)} (\bibinfo {year} {2009})}\BibitemShut {NoStop}%
\bibitem [{\citenamefont {Badger}\ \emph {et~al.}(2022)\citenamefont {Badger},
  \citenamefont {Quan}, \citenamefont {Staab}, \citenamefont {Sumita},
  \citenamefont {Rossi}, \citenamefont {Devlin}, \citenamefont {Neubauer},
  \citenamefont {Shulman}, \citenamefont {Fettinger}, \citenamefont {Klavins},
  \citenamefont {Kauzlarich}, \citenamefont {Aoki}, \citenamefont {Vishik},
  \citenamefont {Pickett},\ and\ \citenamefont {Taufour}}]{Taufour:2022}%
  \BibitemOpen
  \bibfield  {author} {\bibinfo {author} {\bibfnamefont {J.~R.}\ \bibnamefont
  {Badger}}, \bibinfo {author} {\bibfnamefont {Y.}~\bibnamefont {Quan}},
  \bibinfo {author} {\bibfnamefont {M.~C.}\ \bibnamefont {Staab}}, \bibinfo
  {author} {\bibfnamefont {S.}~\bibnamefont {Sumita}}, \bibinfo {author}
  {\bibfnamefont {A.}~\bibnamefont {Rossi}}, \bibinfo {author} {\bibfnamefont
  {K.~P.}\ \bibnamefont {Devlin}}, \bibinfo {author} {\bibfnamefont
  {K.}~\bibnamefont {Neubauer}}, \bibinfo {author} {\bibfnamefont {D.~S.}\
  \bibnamefont {Shulman}}, \bibinfo {author} {\bibfnamefont {J.~C.}\
  \bibnamefont {Fettinger}}, \bibinfo {author} {\bibfnamefont {P.}~\bibnamefont
  {Klavins}}, \bibinfo {author} {\bibfnamefont {S.~M.}\ \bibnamefont
  {Kauzlarich}}, \bibinfo {author} {\bibfnamefont {D.}~\bibnamefont {Aoki}},
  \bibinfo {author} {\bibfnamefont {I.~M.}\ \bibnamefont {Vishik}}, \bibinfo
  {author} {\bibfnamefont {W.~E.}\ \bibnamefont {Pickett}},\ and\ \bibinfo
  {author} {\bibfnamefont {V.}~\bibnamefont {Taufour}},\ }\bibfield  {title}
  {\bibinfo {title} {Dirac lines and loop at the fermi level in the
  time-reversal symmetry breaking superconductor {LaNiGa$_2$}},\ }\bibfield
  {journal} {\bibinfo  {journal} {Communications Physics}\ }\textbf {\bibinfo
  {volume} {5}},\ \href {https://doi.org/10.1038/s42005-021-00771-5}
  {10.1038/s42005-021-00771-5} (\bibinfo {year} {2022})\BibitemShut {NoStop}%
\bibitem [{\citenamefont {Mizuguchi}(2015)}]{Mizuguchi:2015}%
  \BibitemOpen
  \bibfield  {author} {\bibinfo {author} {\bibfnamefont {Y.}~\bibnamefont
  {Mizuguchi}},\ }\bibfield  {title} {\bibinfo {title} {{Review of
  superconductivity in BiS$_2$-based layered materials}},\ }\href
  {https://doi.org/https://doi.org/10.1016/j.jpcs.2014.09.003} {\bibfield
  {journal} {\bibinfo  {journal} {Journal of Physics and Chemistry of Solids}\
  }\textbf {\bibinfo {volume} {84}},\ \bibinfo {pages} {34} (\bibinfo {year}
  {2015})},\ \bibinfo {note} {focus issue on the Study of matter at extreme
  conditions and related phenomena}\BibitemShut {NoStop}%
\bibitem [{\citenamefont {Stewart}(2011)}]{Stewart:2011}%
  \BibitemOpen
  \bibfield  {author} {\bibinfo {author} {\bibfnamefont {G.~R.}\ \bibnamefont
  {Stewart}},\ }\bibfield  {title} {\bibinfo {title} {Superconductivity in iron
  compounds},\ }\href {https://doi.org/10.1103/RevModPhys.83.1589} {\bibfield
  {journal} {\bibinfo  {journal} {Rev. Mod. Phys.}\ }\textbf {\bibinfo {volume}
  {83}},\ \bibinfo {pages} {1589} (\bibinfo {year} {2011})}\BibitemShut
  {NoStop}%
\bibitem [{\citenamefont {Joynt}\ and\ \citenamefont
  {Taillefer}(2002)}]{Joynt:2002}%
  \BibitemOpen
  \bibfield  {author} {\bibinfo {author} {\bibfnamefont {R.}~\bibnamefont
  {Joynt}}\ and\ \bibinfo {author} {\bibfnamefont {L.}~\bibnamefont
  {Taillefer}},\ }\bibfield  {title} {\bibinfo {title} {{The superconducting
  phases of ${\mathrm{UPt}}_{3}$}},\ }\href
  {https://doi.org/10.1103/RevModPhys.74.235} {\bibfield  {journal} {\bibinfo
  {journal} {Rev. Mod. Phys.}\ }\textbf {\bibinfo {volume} {74}},\ \bibinfo
  {pages} {235} (\bibinfo {year} {2002})}\BibitemShut {NoStop}%
\bibitem [{\citenamefont {Aoki}\ \emph {et~al.}(2019)\citenamefont {Aoki},
  \citenamefont {Ishida},\ and\ \citenamefont {Flouquet}}]{Aoki:2019}%
  \BibitemOpen
  \bibfield  {author} {\bibinfo {author} {\bibfnamefont {D.}~\bibnamefont
  {Aoki}}, \bibinfo {author} {\bibfnamefont {K.}~\bibnamefont {Ishida}},\ and\
  \bibinfo {author} {\bibfnamefont {J.}~\bibnamefont {Flouquet}},\ }\bibfield
  {title} {\bibinfo {title} {{Review of U-based Ferromagnetic Superconductors:
  Comparison between UGe$_2$, URhGe, and UCoGe}},\ }\href
  {https://doi.org/10.7566/jpsj.88.022001} {\bibfield  {journal} {\bibinfo
  {journal} {Journal of the Physical Society of Japan}\ }\textbf {\bibinfo
  {volume} {88}},\ \bibinfo {pages} {022001} (\bibinfo {year}
  {2019})}\BibitemShut {NoStop}%
\bibitem [{\citenamefont {Fu}(2015)}]{Fu:2015}%
  \BibitemOpen
  \bibfield  {author} {\bibinfo {author} {\bibfnamefont {L.}~\bibnamefont
  {Fu}},\ }\bibfield  {title} {\bibinfo {title} {Parity-breaking phases of
  spin-orbit-coupled metals with gyrotropic, ferroelectric, and multipolar
  orders},\ }\href {https://doi.org/10.1103/PhysRevLett.115.026401} {\bibfield
  {journal} {\bibinfo  {journal} {Phys. Rev. Lett.}\ }\textbf {\bibinfo
  {volume} {115}},\ \bibinfo {pages} {026401} (\bibinfo {year}
  {2015})}\BibitemShut {NoStop}%
\bibitem [{\citenamefont {Brydon}\ \emph {et~al.}(2016)\citenamefont {Brydon},
  \citenamefont {Wang}, \citenamefont {Weinert},\ and\ \citenamefont
  {Agterberg}}]{Brydon:2016}%
  \BibitemOpen
  \bibfield  {author} {\bibinfo {author} {\bibfnamefont {P.~M.~R.}\
  \bibnamefont {Brydon}}, \bibinfo {author} {\bibfnamefont {L.}~\bibnamefont
  {Wang}}, \bibinfo {author} {\bibfnamefont {M.}~\bibnamefont {Weinert}},\ and\
  \bibinfo {author} {\bibfnamefont {D.~F.}\ \bibnamefont {Agterberg}},\
  }\bibfield  {title} {\bibinfo {title} {Pairing of $j=3/2$ fermions in
  half-heusler superconductors},\ }\href
  {https://doi.org/10.1103/PhysRevLett.116.177001} {\bibfield  {journal}
  {\bibinfo  {journal} {Phys. Rev. Lett.}\ }\textbf {\bibinfo {volume} {116}},\
  \bibinfo {pages} {177001} (\bibinfo {year} {2016})}\BibitemShut {NoStop}%
\bibitem [{\citenamefont {Wang}\ \emph {et~al.}(2019)\citenamefont {Wang},
  \citenamefont {Lian}, \citenamefont {Guo}, \citenamefont {Mao}, \citenamefont
  {Zhang}, \citenamefont {Zhang}, \citenamefont {Gu}, \citenamefont {Xu},\ and\
  \citenamefont {Duan}}]{Wang:2019}%
  \BibitemOpen
  \bibfield  {author} {\bibinfo {author} {\bibfnamefont {C.}~\bibnamefont
  {Wang}}, \bibinfo {author} {\bibfnamefont {B.}~\bibnamefont {Lian}}, \bibinfo
  {author} {\bibfnamefont {X.}~\bibnamefont {Guo}}, \bibinfo {author}
  {\bibfnamefont {J.}~\bibnamefont {Mao}}, \bibinfo {author} {\bibfnamefont
  {Z.}~\bibnamefont {Zhang}}, \bibinfo {author} {\bibfnamefont
  {D.}~\bibnamefont {Zhang}}, \bibinfo {author} {\bibfnamefont {B.-L.}\
  \bibnamefont {Gu}}, \bibinfo {author} {\bibfnamefont {Y.}~\bibnamefont
  {Xu}},\ and\ \bibinfo {author} {\bibfnamefont {W.}~\bibnamefont {Duan}},\
  }\bibfield  {title} {\bibinfo {title} {{Type-II Ising} superconductivity in
  two-dimensional materials with spin-orbit coupling},\ }\href
  {https://doi.org/10.1103/PhysRevLett.123.126402} {\bibfield  {journal}
  {\bibinfo  {journal} {Phys. Rev. Lett.}\ }\textbf {\bibinfo {volume} {123}},\
  \bibinfo {pages} {126402} (\bibinfo {year} {2019})}\BibitemShut {NoStop}%
\bibitem [{\citenamefont {Falson}\ \emph {et~al.}(2020)\citenamefont {Falson},
  \citenamefont {Xu}, \citenamefont {Liao}, \citenamefont {Zang}, \citenamefont
  {Zhu}, \citenamefont {Wang}, \citenamefont {Zhang}, \citenamefont {Liu},
  \citenamefont {Duan}, \citenamefont {He}, \citenamefont {Liu}, \citenamefont
  {Smet}, \citenamefont {Zhang},\ and\ \citenamefont {Xue}}]{Falson:2020}%
  \BibitemOpen
  \bibfield  {author} {\bibinfo {author} {\bibfnamefont {J.}~\bibnamefont
  {Falson}}, \bibinfo {author} {\bibfnamefont {Y.}~\bibnamefont {Xu}}, \bibinfo
  {author} {\bibfnamefont {M.}~\bibnamefont {Liao}}, \bibinfo {author}
  {\bibfnamefont {Y.}~\bibnamefont {Zang}}, \bibinfo {author} {\bibfnamefont
  {K.}~\bibnamefont {Zhu}}, \bibinfo {author} {\bibfnamefont {C.}~\bibnamefont
  {Wang}}, \bibinfo {author} {\bibfnamefont {Z.}~\bibnamefont {Zhang}},
  \bibinfo {author} {\bibfnamefont {H.}~\bibnamefont {Liu}}, \bibinfo {author}
  {\bibfnamefont {W.}~\bibnamefont {Duan}}, \bibinfo {author} {\bibfnamefont
  {K.}~\bibnamefont {He}}, \bibinfo {author} {\bibfnamefont {H.}~\bibnamefont
  {Liu}}, \bibinfo {author} {\bibfnamefont {J.~H.}\ \bibnamefont {Smet}},
  \bibinfo {author} {\bibfnamefont {D.}~\bibnamefont {Zhang}},\ and\ \bibinfo
  {author} {\bibfnamefont {Q.-K.}\ \bibnamefont {Xue}},\ }\bibfield  {title}
  {\bibinfo {title} {{Type-II Ising pairing in few-layer stanene}},\ }\href
  {https://doi.org/10.1126/science.aax3873} {\bibfield  {journal} {\bibinfo
  {journal} {Science}\ }\textbf {\bibinfo {volume} {367}},\ \bibinfo {pages}
  {1454} (\bibinfo {year} {2020})}\BibitemShut {NoStop}%
\bibitem [{\citenamefont {Samokhin}(2019{\natexlab{a}})}]{Samokhin:2019-2}%
  \BibitemOpen
  \bibfield  {author} {\bibinfo {author} {\bibfnamefont {K.~V.}\ \bibnamefont
  {Samokhin}},\ }\bibfield  {title} {\bibinfo {title} {Symmetry of
  superconducting pairing in non-pseudospin electron bands},\ }\href
  {https://doi.org/10.1103/PhysRevB.100.054501} {\bibfield  {journal} {\bibinfo
   {journal} {Phys. Rev. B}\ }\textbf {\bibinfo {volume} {100}},\ \bibinfo
  {pages} {054501} (\bibinfo {year} {2019}{\natexlab{a}})}\BibitemShut
  {NoStop}%
\bibitem [{\citenamefont {Samokhin}(2020)}]{Samokhin:2020}%
  \BibitemOpen
  \bibfield  {author} {\bibinfo {author} {\bibfnamefont {K.~V.}\ \bibnamefont
  {Samokhin}},\ }\bibfield  {title} {\bibinfo {title} {Exotic interband pairing
  in multiband superconductors},\ }\href
  {https://doi.org/10.1103/PhysRevB.101.214524} {\bibfield  {journal} {\bibinfo
   {journal} {Phys. Rev. B}\ }\textbf {\bibinfo {volume} {101}},\ \bibinfo
  {pages} {214524} (\bibinfo {year} {2020})}\BibitemShut {NoStop}%
\bibitem [{\citenamefont {Samokhin}(2021)}]{Samokhin:2021}%
  \BibitemOpen
  \bibfield  {author} {\bibinfo {author} {\bibfnamefont {K.~V.}\ \bibnamefont
  {Samokhin}},\ }\bibfield  {title} {\bibinfo {title} {{Spin Susceptibility of
  Superconductors with Strong Spin-Orbit Coupling}},\ }\href
  {https://doi.org/10.1103/physrevb.103.174505} {\bibfield  {journal} {\bibinfo
   {journal} {Physical Review B}\ }\textbf {\bibinfo {volume} {103}},\ \bibinfo
  {pages} {174505} (\bibinfo {year} {2021})}\BibitemShut {NoStop}%
\bibitem [{\citenamefont {Sigrist}\ and\ \citenamefont
  {Ueda}(1991)}]{Sigrist:1991}%
  \BibitemOpen
  \bibfield  {author} {\bibinfo {author} {\bibfnamefont {M.}~\bibnamefont
  {Sigrist}}\ and\ \bibinfo {author} {\bibfnamefont {K.}~\bibnamefont {Ueda}},\
  }\bibfield  {title} {\bibinfo {title} {Phenomenological theory of
  unconventional superconductivity},\ }\href
  {https://doi.org/10.1103/RevModPhys.63.239} {\bibfield  {journal} {\bibinfo
  {journal} {Rev. Mod. Phys.}\ }\textbf {\bibinfo {volume} {63}},\ \bibinfo
  {pages} {239} (\bibinfo {year} {1991})}\BibitemShut {NoStop}%
\bibitem [{\citenamefont {Gor'kov}\ and\ \citenamefont
  {Rashba}(2001)}]{Gorkov:2001}%
  \BibitemOpen
  \bibfield  {author} {\bibinfo {author} {\bibfnamefont {L.~P.}\ \bibnamefont
  {Gor'kov}}\ and\ \bibinfo {author} {\bibfnamefont {E.~I.}\ \bibnamefont
  {Rashba}},\ }\bibfield  {title} {\bibinfo {title} {Superconducting 2d system
  with lifted spin degeneracy: Mixed singlet-triplet state},\ }\href
  {https://doi.org/10.1103/PhysRevLett.87.037004} {\bibfield  {journal}
  {\bibinfo  {journal} {Phys. Rev. Lett.}\ }\textbf {\bibinfo {volume} {87}},\
  \bibinfo {pages} {037004} (\bibinfo {year} {2001})}\BibitemShut {NoStop}%
\bibitem [{\citenamefont {Aroyo}\ \emph
  {et~al.}(2006{\natexlab{a}})\citenamefont {Aroyo}, \citenamefont
  {Perez-Mato}, \citenamefont {Capillas}, \citenamefont {Kroumova},
  \citenamefont {Ivantchev}, \citenamefont {Madariaga}, \citenamefont {Kirov},\
  and\ \citenamefont {Wondratschek}}]{Aroyo:2006}%
  \BibitemOpen
  \bibfield  {author} {\bibinfo {author} {\bibfnamefont {M.~I.}\ \bibnamefont
  {Aroyo}}, \bibinfo {author} {\bibfnamefont {J.~M.}\ \bibnamefont
  {Perez-Mato}}, \bibinfo {author} {\bibfnamefont {C.}~\bibnamefont
  {Capillas}}, \bibinfo {author} {\bibfnamefont {E.}~\bibnamefont {Kroumova}},
  \bibinfo {author} {\bibfnamefont {S.}~\bibnamefont {Ivantchev}}, \bibinfo
  {author} {\bibfnamefont {G.}~\bibnamefont {Madariaga}}, \bibinfo {author}
  {\bibfnamefont {A.}~\bibnamefont {Kirov}},\ and\ \bibinfo {author}
  {\bibfnamefont {H.}~\bibnamefont {Wondratschek}},\ }\bibfield  {title}
  {\bibinfo {title} {Bilbao crystallographic server: I. databases and
  crystallographic computing programs},\ }\href
  {https://doi.org/doi:10.1524/zkri.2006.221.1.15} {\bibfield  {journal}
  {\bibinfo  {journal} {Zeitschrift für Kristallographie - Crystalline
  Materials}\ }\textbf {\bibinfo {volume} {221}},\ \bibinfo {pages} {15}
  (\bibinfo {year} {2006}{\natexlab{a}})}\BibitemShut {NoStop}%
\bibitem [{\citenamefont {Aroyo}\ \emph
  {et~al.}(2006{\natexlab{b}})\citenamefont {Aroyo}, \citenamefont {Kirov},
  \citenamefont {Capillas}, \citenamefont {Perez-Mato},\ and\ \citenamefont
  {Wondratschek}}]{Aroyo2:2006}%
  \BibitemOpen
  \bibfield  {author} {\bibinfo {author} {\bibfnamefont {M.~I.}\ \bibnamefont
  {Aroyo}}, \bibinfo {author} {\bibfnamefont {A.}~\bibnamefont {Kirov}},
  \bibinfo {author} {\bibfnamefont {C.}~\bibnamefont {Capillas}}, \bibinfo
  {author} {\bibfnamefont {J.~M.}\ \bibnamefont {Perez-Mato}},\ and\ \bibinfo
  {author} {\bibfnamefont {H.}~\bibnamefont {Wondratschek}},\ }\bibfield
  {title} {\bibinfo {title} {{Bilbao Crystallographic Server. II.
  Representations of crystallographic point groups and space groups}},\ }\href
  {https://doi.org/10.1107/S0108767305040286} {\bibfield  {journal} {\bibinfo
  {journal} {Acta Crystallographica Section A}\ }\textbf {\bibinfo {volume}
  {62}},\ \bibinfo {pages} {115} (\bibinfo {year}
  {2006}{\natexlab{b}})}\BibitemShut {NoStop}%
\bibitem [{\citenamefont {Stokes}\ \emph {et~al.}(2013)\citenamefont {Stokes},
  \citenamefont {Campbell},\ and\ \citenamefont {Cordes}}]{Stokes:2013}%
  \BibitemOpen
  \bibfield  {author} {\bibinfo {author} {\bibfnamefont {H.~T.}\ \bibnamefont
  {Stokes}}, \bibinfo {author} {\bibfnamefont {B.~J.}\ \bibnamefont
  {Campbell}},\ and\ \bibinfo {author} {\bibfnamefont {R.}~\bibnamefont
  {Cordes}},\ }\bibfield  {title} {\bibinfo {title} {{Tabulation of irreducible
  representations of the crystallographic space groups and their superspace
  extensions}},\ }\href {https://doi.org/10.1107/S0108767313007538} {\bibfield
  {journal} {\bibinfo  {journal} {Acta Crystallographica Section A}\ }\textbf
  {\bibinfo {volume} {69}},\ \bibinfo {pages} {388} (\bibinfo {year}
  {2013})}\BibitemShut {NoStop}%
\bibitem [{\citenamefont {Zhao}\ and\ \citenamefont
  {Schnyder}(2016)}]{Zhao:2016}%
  \BibitemOpen
  \bibfield  {author} {\bibinfo {author} {\bibfnamefont {Y.~X.}\ \bibnamefont
  {Zhao}}\ and\ \bibinfo {author} {\bibfnamefont {A.~P.}\ \bibnamefont
  {Schnyder}},\ }\bibfield  {title} {\bibinfo {title} {Nonsymmorphic
  symmetry-required band crossings in topological semimetals},\ }\href
  {https://doi.org/10.1103/PhysRevB.94.195109} {\bibfield  {journal} {\bibinfo
  {journal} {Phys. Rev. B}\ }\textbf {\bibinfo {volume} {94}},\ \bibinfo
  {pages} {195109} (\bibinfo {year} {2016})}\BibitemShut {NoStop}%
\bibitem [{\citenamefont {Huang}\ and\ \citenamefont
  {Hoffman}(2017)}]{Huang:2017}%
  \BibitemOpen
  \bibfield  {author} {\bibinfo {author} {\bibfnamefont {D.}~\bibnamefont
  {Huang}}\ and\ \bibinfo {author} {\bibfnamefont {J.~E.}\ \bibnamefont
  {Hoffman}},\ }\bibfield  {title} {\bibinfo {title} {{Monolayer FeSe on
  SrTiO$_3$}},\ }\href
  {https://doi.org/10.1146/annurev-conmatphys-031016-025242} {\bibfield
  {journal} {\bibinfo  {journal} {Annual Review of Condensed Matter Physics}\
  }\textbf {\bibinfo {volume} {8}},\ \bibinfo {pages} {311} (\bibinfo {year}
  {2017})},\ \Eprint
  {https://arxiv.org/abs/https://doi.org/10.1146/annurev-conmatphys-031016-025242}
  {https://doi.org/10.1146/annurev-conmatphys-031016-025242} \BibitemShut
  {NoStop}%
\bibitem [{\citenamefont {Ruf}\ \emph {et~al.}(2021)\citenamefont {Ruf},
  \citenamefont {Paik}, \citenamefont {Schreiber}, \citenamefont {Nair},
  \citenamefont {Miao}, \citenamefont {Kawasaki}, \citenamefont {Nelson},
  \citenamefont {Faeth}, \citenamefont {Lee}, \citenamefont {Goodge},
  \citenamefont {Pamuk}, \citenamefont {Fennie}, \citenamefont {Kourkoutis},
  \citenamefont {Schlom},\ and\ \citenamefont {Shen}}]{Ruf:2021}%
  \BibitemOpen
  \bibfield  {author} {\bibinfo {author} {\bibfnamefont {J.~P.}\ \bibnamefont
  {Ruf}}, \bibinfo {author} {\bibfnamefont {H.}~\bibnamefont {Paik}}, \bibinfo
  {author} {\bibfnamefont {N.~J.}\ \bibnamefont {Schreiber}}, \bibinfo {author}
  {\bibfnamefont {H.~P.}\ \bibnamefont {Nair}}, \bibinfo {author}
  {\bibfnamefont {L.}~\bibnamefont {Miao}}, \bibinfo {author} {\bibfnamefont
  {J.~K.}\ \bibnamefont {Kawasaki}}, \bibinfo {author} {\bibfnamefont {J.~N.}\
  \bibnamefont {Nelson}}, \bibinfo {author} {\bibfnamefont {B.~D.}\
  \bibnamefont {Faeth}}, \bibinfo {author} {\bibfnamefont {Y.}~\bibnamefont
  {Lee}}, \bibinfo {author} {\bibfnamefont {B.~H.}\ \bibnamefont {Goodge}},
  \bibinfo {author} {\bibfnamefont {B.}~\bibnamefont {Pamuk}}, \bibinfo
  {author} {\bibfnamefont {C.~J.}\ \bibnamefont {Fennie}}, \bibinfo {author}
  {\bibfnamefont {L.~F.}\ \bibnamefont {Kourkoutis}}, \bibinfo {author}
  {\bibfnamefont {D.~G.}\ \bibnamefont {Schlom}},\ and\ \bibinfo {author}
  {\bibfnamefont {K.~M.}\ \bibnamefont {Shen}},\ }\bibfield  {title} {\bibinfo
  {title} {Strain-stabilized superconductivity},\ }\bibfield  {journal}
  {\bibinfo  {journal} {Nature Communications}\ }\textbf {\bibinfo {volume}
  {12}},\ \href {https://doi.org/10.1038/s41467-020-20252-7}
  {10.1038/s41467-020-20252-7} (\bibinfo {year} {2021})\BibitemShut {NoStop}%
\bibitem [{\citenamefont {Uchida}\ \emph {et~al.}(2020)\citenamefont {Uchida},
  \citenamefont {Nomoto}, \citenamefont {Musashi}, \citenamefont {Arita},\ and\
  \citenamefont {Kawasaki}}]{Uchida:2020}%
  \BibitemOpen
  \bibfield  {author} {\bibinfo {author} {\bibfnamefont {M.}~\bibnamefont
  {Uchida}}, \bibinfo {author} {\bibfnamefont {T.}~\bibnamefont {Nomoto}},
  \bibinfo {author} {\bibfnamefont {M.}~\bibnamefont {Musashi}}, \bibinfo
  {author} {\bibfnamefont {R.}~\bibnamefont {Arita}},\ and\ \bibinfo {author}
  {\bibfnamefont {M.}~\bibnamefont {Kawasaki}},\ }\bibfield  {title} {\bibinfo
  {title} {{Superconductivity in Uniquely Strained ${\mathrm{RuO}}_{2}$
  Films}},\ }\href {https://doi.org/10.1103/PhysRevLett.125.147001} {\bibfield
  {journal} {\bibinfo  {journal} {Phys. Rev. Lett.}\ }\textbf {\bibinfo
  {volume} {125}},\ \bibinfo {pages} {147001} (\bibinfo {year}
  {2020})}\BibitemShut {NoStop}%
\bibitem [{\citenamefont {\ifmmode~\check{S}\else \v{S}\fi{}mejkal}\ \emph
  {et~al.}(2022)\citenamefont {\ifmmode~\check{S}\else \v{S}\fi{}mejkal},
  \citenamefont {Sinova},\ and\ \citenamefont {Jungwirth}}]{Smejkal:2022}%
  \BibitemOpen
  \bibfield  {author} {\bibinfo {author} {\bibfnamefont {L.}~\bibnamefont
  {\ifmmode~\check{S}\else \v{S}\fi{}mejkal}}, \bibinfo {author} {\bibfnamefont
  {J.}~\bibnamefont {Sinova}},\ and\ \bibinfo {author} {\bibfnamefont
  {T.}~\bibnamefont {Jungwirth}},\ }\bibfield  {title} {\bibinfo {title}
  {Emerging research landscape of altermagnetism},\ }\href
  {https://doi.org/10.1103/PhysRevX.12.040501} {\bibfield  {journal} {\bibinfo
  {journal} {Phys. Rev. X}\ }\textbf {\bibinfo {volume} {12}},\ \bibinfo
  {pages} {040501} (\bibinfo {year} {2022})}\BibitemShut {NoStop}%
\bibitem [{\citenamefont {Blount}(1985)}]{Blount:1985}%
  \BibitemOpen
  \bibfield  {author} {\bibinfo {author} {\bibfnamefont {E.~I.}\ \bibnamefont
  {Blount}},\ }\bibfield  {title} {\bibinfo {title} {Symmetry properties of
  triplet superconductors},\ }\href {https://doi.org/10.1103/PhysRevB.32.2935}
  {\bibfield  {journal} {\bibinfo  {journal} {Phys. Rev. B}\ }\textbf {\bibinfo
  {volume} {32}},\ \bibinfo {pages} {2935} (\bibinfo {year}
  {1985})}\BibitemShut {NoStop}%
\bibitem [{\citenamefont {Samokhin}(2019{\natexlab{b}})}]{Samokhin:2019}%
  \BibitemOpen
  \bibfield  {author} {\bibinfo {author} {\bibfnamefont {K.}~\bibnamefont
  {Samokhin}},\ }\bibfield  {title} {\bibinfo {title} {{On the pseudospin
  description of the electron Bloch bands}},\ }\href
  {https://doi.org/10.1016/j.aop.2019.04.016} {\bibfield  {journal} {\bibinfo
  {journal} {Annals of Physics}\ }\textbf {\bibinfo {volume} {407}},\ \bibinfo
  {pages} {179} (\bibinfo {year} {2019}{\natexlab{b}})}\BibitemShut {NoStop}%
\bibitem [{\citenamefont {Cavanagh}\ \emph
  {et~al.}(2022{\natexlab{b}})\citenamefont {Cavanagh}, \citenamefont
  {Agterberg},\ and\ \citenamefont {Brydon}}]{Cavanagh2:2022}%
  \BibitemOpen
  \bibfield  {author} {\bibinfo {author} {\bibfnamefont {D.~C.}\ \bibnamefont
  {Cavanagh}}, \bibinfo {author} {\bibfnamefont {D.~F.}\ \bibnamefont
  {Agterberg}},\ and\ \bibinfo {author} {\bibfnamefont {P.~M.~R.}\ \bibnamefont
  {Brydon}},\ }\href {https://doi.org/10.48550/ARXIV.2207.01191} {\bibinfo
  {title} {Pair-breaking in superconductors with strong spin-orbit coupling}}
  (\bibinfo {year} {2022}{\natexlab{b}})\BibitemShut {NoStop}%
\bibitem [{\citenamefont {Ramires}\ and\ \citenamefont
  {Sigrist}(2016)}]{Ramires:2016}%
  \BibitemOpen
  \bibfield  {author} {\bibinfo {author} {\bibfnamefont {A.}~\bibnamefont
  {Ramires}}\ and\ \bibinfo {author} {\bibfnamefont {M.}~\bibnamefont
  {Sigrist}},\ }\bibfield  {title} {\bibinfo {title} {{Identifying detrimental
  effects for multiorbital superconductivity: Application to
  ${\mathrm{Sr}}_{2}{\mathrm{RuO}}_{4}$}},\ }\href
  {https://doi.org/10.1103/PhysRevB.94.104501} {\bibfield  {journal} {\bibinfo
  {journal} {Phys. Rev. B}\ }\textbf {\bibinfo {volume} {94}},\ \bibinfo
  {pages} {104501} (\bibinfo {year} {2016})}\BibitemShut {NoStop}%
\bibitem [{\citenamefont {Ramires}\ \emph {et~al.}(2018)\citenamefont
  {Ramires}, \citenamefont {Agterberg},\ and\ \citenamefont
  {Sigrist}}]{Ramires:2018}%
  \BibitemOpen
  \bibfield  {author} {\bibinfo {author} {\bibfnamefont {A.}~\bibnamefont
  {Ramires}}, \bibinfo {author} {\bibfnamefont {D.~F.}\ \bibnamefont
  {Agterberg}},\ and\ \bibinfo {author} {\bibfnamefont {M.}~\bibnamefont
  {Sigrist}},\ }\bibfield  {title} {\bibinfo {title} {Tailoring ${T}_{c}$ by
  symmetry principles: The concept of superconducting fitness},\ }\href
  {https://doi.org/10.1103/PhysRevB.98.024501} {\bibfield  {journal} {\bibinfo
  {journal} {Phys. Rev. B}\ }\textbf {\bibinfo {volume} {98}},\ \bibinfo
  {pages} {024501} (\bibinfo {year} {2018})}\BibitemShut {NoStop}%
\bibitem [{\citenamefont {Yanase}\ and\ \citenamefont
  {Shiozaki}(2017)}]{Yanase:2017}%
  \BibitemOpen
  \bibfield  {author} {\bibinfo {author} {\bibfnamefont {Y.}~\bibnamefont
  {Yanase}}\ and\ \bibinfo {author} {\bibfnamefont {K.}~\bibnamefont
  {Shiozaki}},\ }\bibfield  {title} {\bibinfo {title} {{M\"obius topological
  superconductivity in ${\mathrm{UPt}}_{3}$}},\ }\href
  {https://doi.org/10.1103/PhysRevB.95.224514} {\bibfield  {journal} {\bibinfo
  {journal} {Phys. Rev. B}\ }\textbf {\bibinfo {volume} {95}},\ \bibinfo
  {pages} {224514} (\bibinfo {year} {2017})}\BibitemShut {NoStop}%
\bibitem [{\citenamefont {Chubukov}\ \emph {et~al.}(2016)\citenamefont
  {Chubukov}, \citenamefont {Vafek},\ and\ \citenamefont
  {Fernandes}}]{PhysRevB.94.174518}%
  \BibitemOpen
  \bibfield  {author} {\bibinfo {author} {\bibfnamefont {A.~V.}\ \bibnamefont
  {Chubukov}}, \bibinfo {author} {\bibfnamefont {O.}~\bibnamefont {Vafek}},\
  and\ \bibinfo {author} {\bibfnamefont {R.~M.}\ \bibnamefont {Fernandes}},\
  }\bibfield  {title} {\bibinfo {title} {Displacement and annihilation of dirac
  gap nodes in $d$-wave iron-based superconductors},\ }\href
  {https://doi.org/10.1103/PhysRevB.94.174518} {\bibfield  {journal} {\bibinfo
  {journal} {Phys. Rev. B}\ }\textbf {\bibinfo {volume} {94}},\ \bibinfo
  {pages} {174518} (\bibinfo {year} {2016})}\BibitemShut {NoStop}%
\bibitem [{\citenamefont {Agterberg}\ \emph {et~al.}(2017)\citenamefont
  {Agterberg}, \citenamefont {Shishidou}, \citenamefont {O'Halloran},
  \citenamefont {Brydon},\ and\ \citenamefont {Weinert}}]{Agterberg:2017}%
  \BibitemOpen
  \bibfield  {author} {\bibinfo {author} {\bibfnamefont {D.~F.}\ \bibnamefont
  {Agterberg}}, \bibinfo {author} {\bibfnamefont {T.}~\bibnamefont
  {Shishidou}}, \bibinfo {author} {\bibfnamefont {J.}~\bibnamefont
  {O'Halloran}}, \bibinfo {author} {\bibfnamefont {P.~M.~R.}\ \bibnamefont
  {Brydon}},\ and\ \bibinfo {author} {\bibfnamefont {M.}~\bibnamefont
  {Weinert}},\ }\bibfield  {title} {\bibinfo {title} {{Resilient Nodeless
  $d$-Wave Superconductivity in Monolayer FeSe}},\ }\href
  {https://doi.org/10.1103/PhysRevLett.119.267001} {\bibfield  {journal}
  {\bibinfo  {journal} {Phys. Rev. Lett.}\ }\textbf {\bibinfo {volume} {119}},\
  \bibinfo {pages} {267001} (\bibinfo {year} {2017})}\BibitemShut {NoStop}%
\bibitem [{\citenamefont {Fu}\ and\ \citenamefont {Berg}(2010)}]{Fu:2010}%
  \BibitemOpen
  \bibfield  {author} {\bibinfo {author} {\bibfnamefont {L.}~\bibnamefont
  {Fu}}\ and\ \bibinfo {author} {\bibfnamefont {E.}~\bibnamefont {Berg}},\
  }\bibfield  {title} {\bibinfo {title} {{Odd-Parity Topological
  Superconductors: Theory and Application to
  ${\mathrm{Cu}}_{x}{\mathrm{Bi}}_{2}{\mathrm{Se}}_{3}$}},\ }\href@noop {}
  {\bibfield  {journal} {\bibinfo  {journal} {Phys. Rev. Lett.}\ }\textbf
  {\bibinfo {volume} {105}},\ \bibinfo {pages} {097001} (\bibinfo {year}
  {2010})}\BibitemShut {NoStop}%
\bibitem [{\citenamefont {Qin}\ \emph {et~al.}(2022)\citenamefont {Qin},
  \citenamefont {Fang}, \citenamefont {Zhang},\ and\ \citenamefont
  {Hu}}]{Qin:2022}%
  \BibitemOpen
  \bibfield  {author} {\bibinfo {author} {\bibfnamefont {S.}~\bibnamefont
  {Qin}}, \bibinfo {author} {\bibfnamefont {C.}~\bibnamefont {Fang}}, \bibinfo
  {author} {\bibfnamefont {F.-C.}\ \bibnamefont {Zhang}},\ and\ \bibinfo
  {author} {\bibfnamefont {J.}~\bibnamefont {Hu}},\ }\href
  {https://doi.org/10.48550/ARXIV.2208.09409} {\bibinfo {title} {Spin-triplet
  superconductivity in nonsymmorphic crystals}} (\bibinfo {year}
  {2022})\BibitemShut {NoStop}%
\bibitem [{\citenamefont {Vafek}\ and\ \citenamefont
  {Chubukov}(2017)}]{Vafek:2017}%
  \BibitemOpen
  \bibfield  {author} {\bibinfo {author} {\bibfnamefont {O.}~\bibnamefont
  {Vafek}}\ and\ \bibinfo {author} {\bibfnamefont {A.~V.}\ \bibnamefont
  {Chubukov}},\ }\bibfield  {title} {\bibinfo {title} {Hund interaction,
  spin-orbit coupling, and the mechanism of superconductivity in strongly
  hole-doped iron pnictides},\ }\href
  {https://doi.org/10.1103/PhysRevLett.118.087003} {\bibfield  {journal}
  {\bibinfo  {journal} {Phys. Rev. Lett.}\ }\textbf {\bibinfo {volume} {118}},\
  \bibinfo {pages} {087003} (\bibinfo {year} {2017})}\BibitemShut {NoStop}%
\bibitem [{\citenamefont {Cheung}\ and\ \citenamefont
  {Agterberg}(2019)}]{Cheung:2019}%
  \BibitemOpen
  \bibfield  {author} {\bibinfo {author} {\bibfnamefont {A.~K.~C.}\
  \bibnamefont {Cheung}}\ and\ \bibinfo {author} {\bibfnamefont {D.~F.}\
  \bibnamefont {Agterberg}},\ }\bibfield  {title} {\bibinfo {title}
  {{Superconductivity in the presence of spin-orbit interactions stabilized by
  Hund coupling}},\ }\href {https://doi.org/10.1103/PhysRevB.99.024516}
  {\bibfield  {journal} {\bibinfo  {journal} {Phys. Rev. B}\ }\textbf {\bibinfo
  {volume} {99}},\ \bibinfo {pages} {024516} (\bibinfo {year}
  {2019})}\BibitemShut {NoStop}%
\bibitem [{\citenamefont {Wang}\ \emph {et~al.}(2021)\citenamefont {Wang},
  \citenamefont {Xu},\ and\ \citenamefont {Duan}}]{Wang:2021}%
  \BibitemOpen
  \bibfield  {author} {\bibinfo {author} {\bibfnamefont {C.}~\bibnamefont
  {Wang}}, \bibinfo {author} {\bibfnamefont {Y.}~\bibnamefont {Xu}},\ and\
  \bibinfo {author} {\bibfnamefont {W.}~\bibnamefont {Duan}},\ }\bibfield
  {title} {\bibinfo {title} {{Ising Superconductivity and Its Hidden
  Variants}},\ }\href {https://doi.org/10.1021/accountsmr.1c00068} {\bibfield
  {journal} {\bibinfo  {journal} {Accounts of Materials Research}\ }\textbf
  {\bibinfo {volume} {2}},\ \bibinfo {pages} {526} (\bibinfo {year} {2021})},\
  \Eprint {https://arxiv.org/abs/https://doi.org/10.1021/accountsmr.1c00068}
  {https://doi.org/10.1021/accountsmr.1c00068} \BibitemShut {NoStop}%
\bibitem [{\citenamefont {Skurativska}\ \emph {et~al.}(2021)\citenamefont
  {Skurativska}, \citenamefont {Sigrist},\ and\ \citenamefont
  {Fischer}}]{Skurativska:2021}%
  \BibitemOpen
  \bibfield  {author} {\bibinfo {author} {\bibfnamefont {A.}~\bibnamefont
  {Skurativska}}, \bibinfo {author} {\bibfnamefont {M.}~\bibnamefont
  {Sigrist}},\ and\ \bibinfo {author} {\bibfnamefont {M.~H.}\ \bibnamefont
  {Fischer}},\ }\bibfield  {title} {\bibinfo {title} {{Spin response and
  topology of a staggered-Rashba superconductor}},\ }\href
  {https://doi.org/10.1103/PhysRevResearch.3.033133} {\bibfield  {journal}
  {\bibinfo  {journal} {Phys. Rev. Res.}\ }\textbf {\bibinfo {volume} {3}},\
  \bibinfo {pages} {033133} (\bibinfo {year} {2021})}\BibitemShut {NoStop}%
\bibitem [{\citenamefont {Hoshi}\ \emph {et~al.}(2022)\citenamefont {Hoshi},
  \citenamefont {Kurihara}, \citenamefont {Goto}, \citenamefont {Tokunaga},\
  and\ \citenamefont {Mizuguchi}}]{Hoshi:2022}%
  \BibitemOpen
  \bibfield  {author} {\bibinfo {author} {\bibfnamefont {K.}~\bibnamefont
  {Hoshi}}, \bibinfo {author} {\bibfnamefont {R.}~\bibnamefont {Kurihara}},
  \bibinfo {author} {\bibfnamefont {Y.}~\bibnamefont {Goto}}, \bibinfo {author}
  {\bibfnamefont {M.}~\bibnamefont {Tokunaga}},\ and\ \bibinfo {author}
  {\bibfnamefont {Y.}~\bibnamefont {Mizuguchi}},\ }\bibfield  {title} {\bibinfo
  {title} {{Extremely high upper critical field in {BiCh}$_2$-based (Ch: S and
  Se) layered superconductor {LaO}$_{0.5}$F$_{0.5}$BiS$_{2-x}$Se$_x$
  (x{\hspace{0.167em}}={\hspace{0.167em}}0.22 and 0.69)}},\ }\bibfield
  {journal} {\bibinfo  {journal} {Scientific Reports}\ }\textbf {\bibinfo
  {volume} {12}},\ \href {https://doi.org/10.1038/s41598-021-04393-3}
  {10.1038/s41598-021-04393-3} (\bibinfo {year} {2022})\BibitemShut {NoStop}%
\bibitem [{\citenamefont {Usui}\ \emph {et~al.}(2012)\citenamefont {Usui},
  \citenamefont {Suzuki},\ and\ \citenamefont {Kuroki}}]{Usui:2012}%
  \BibitemOpen
  \bibfield  {author} {\bibinfo {author} {\bibfnamefont {H.}~\bibnamefont
  {Usui}}, \bibinfo {author} {\bibfnamefont {K.}~\bibnamefont {Suzuki}},\ and\
  \bibinfo {author} {\bibfnamefont {K.}~\bibnamefont {Kuroki}},\ }\bibfield
  {title} {\bibinfo {title} {{Minimal electronic models for superconducting
  BiS${}_{2}$ layers}},\ }\href {https://doi.org/10.1103/PhysRevB.86.220501}
  {\bibfield  {journal} {\bibinfo  {journal} {Phys. Rev. B}\ }\textbf {\bibinfo
  {volume} {86}},\ \bibinfo {pages} {220501(R)} (\bibinfo {year}
  {2012})}\BibitemShut {NoStop}%
\bibitem [{\citenamefont {Cobo-Lopez}\ \emph {et~al.}(2018)\citenamefont
  {Cobo-Lopez}, \citenamefont {Bahramy}, \citenamefont {Arita}, \citenamefont
  {Akbari},\ and\ \citenamefont {Eremin}}]{Cobo-Lopez:2018}%
  \BibitemOpen
  \bibfield  {author} {\bibinfo {author} {\bibfnamefont {S.}~\bibnamefont
  {Cobo-Lopez}}, \bibinfo {author} {\bibfnamefont {M.~S.}\ \bibnamefont
  {Bahramy}}, \bibinfo {author} {\bibfnamefont {R.}~\bibnamefont {Arita}},
  \bibinfo {author} {\bibfnamefont {A.}~\bibnamefont {Akbari}},\ and\ \bibinfo
  {author} {\bibfnamefont {I.}~\bibnamefont {Eremin}},\ }\bibfield  {title}
  {\bibinfo {title} {{Spin-orbit coupling, minimal model and potential
  Cooper-pairing from repulsion in BiS$_2$-superconductors}},\ }\href
  {https://doi.org/10.1088/1367-2630/aaaf87} {\bibfield  {journal} {\bibinfo
  {journal} {New Journal of Physics}\ }\textbf {\bibinfo {volume} {20}},\
  \bibinfo {pages} {043029} (\bibinfo {year} {2018})}\BibitemShut {NoStop}%
\bibitem [{\citenamefont {Weinert}\ \emph {et~al.}(2009)\citenamefont
  {Weinert}, \citenamefont {Schneider}, \citenamefont {Podloucky},\ and\
  \citenamefont {Redinger}}]{FLAPW2009}%
  \BibitemOpen
  \bibfield  {author} {\bibinfo {author} {\bibfnamefont {M.}~\bibnamefont
  {Weinert}}, \bibinfo {author} {\bibfnamefont {G.}~\bibnamefont {Schneider}},
  \bibinfo {author} {\bibfnamefont {R.}~\bibnamefont {Podloucky}},\ and\
  \bibinfo {author} {\bibfnamefont {J.}~\bibnamefont {Redinger}},\ }\bibfield
  {title} {\bibinfo {title} {{FLAPW}: Applications and implementations},\
  }\href {https://doi.org/10.1088/0953-8984/21/8/084201} {\bibfield  {journal}
  {\bibinfo  {journal} {J. Phys. Condens. Matter}\ }\textbf {\bibinfo {volume}
  {21}},\ \bibinfo {pages} {084201} (\bibinfo {year} {2009})}\BibitemShut
  {NoStop}%
\bibitem [{\citenamefont {Perdew}\ \emph {et~al.}(1996)\citenamefont {Perdew},
  \citenamefont {Burke},\ and\ \citenamefont {Ernzerhof}}]{PBE}%
  \BibitemOpen
  \bibfield  {author} {\bibinfo {author} {\bibfnamefont {J.~P.}\ \bibnamefont
  {Perdew}}, \bibinfo {author} {\bibfnamefont {K.}~\bibnamefont {Burke}},\ and\
  \bibinfo {author} {\bibfnamefont {M.}~\bibnamefont {Ernzerhof}},\ }\bibfield
  {title} {\bibinfo {title} {{Generalized Gradient Approximation Made
  Simple}},\ }\href {https://doi.org/10.1103/PhysRevLett.77.3865} {\bibfield
  {journal} {\bibinfo  {journal} {Phys. Rev. Lett.}\ }\textbf {\bibinfo
  {volume} {77}},\ \bibinfo {pages} {3865} (\bibinfo {year}
  {1996})}\BibitemShut {NoStop}%
\bibitem [{\citenamefont {Mizuguchi}\ \emph {et~al.}(2012)\citenamefont
  {Mizuguchi}, \citenamefont {Demura}, \citenamefont {Deguchi}, \citenamefont
  {Takano}, \citenamefont {Fujihisa}, \citenamefont {Gotoh}, \citenamefont
  {Izawa},\ and\ \citenamefont {Miura}}]{Mizuguchi2012}%
  \BibitemOpen
  \bibfield  {author} {\bibinfo {author} {\bibfnamefont {Y.}~\bibnamefont
  {Mizuguchi}}, \bibinfo {author} {\bibfnamefont {S.}~\bibnamefont {Demura}},
  \bibinfo {author} {\bibfnamefont {K.}~\bibnamefont {Deguchi}}, \bibinfo
  {author} {\bibfnamefont {Y.}~\bibnamefont {Takano}}, \bibinfo {author}
  {\bibfnamefont {H.}~\bibnamefont {Fujihisa}}, \bibinfo {author}
  {\bibfnamefont {Y.}~\bibnamefont {Gotoh}}, \bibinfo {author} {\bibfnamefont
  {H.}~\bibnamefont {Izawa}},\ and\ \bibinfo {author} {\bibfnamefont
  {O.}~\bibnamefont {Miura}},\ }\bibfield  {title} {\bibinfo {title}
  {{Superconductivity in Novel BiS$_2$-Based Layered Superconductor
  LaO$_{1-x}$F$_x$BiS$_2$}},\ }\href {https://doi.org/10.1143/JPSJ.81.114725}
  {\bibfield  {journal} {\bibinfo  {journal} {Journal of the Physical Society
  of Japan}\ }\textbf {\bibinfo {volume} {81}},\ \bibinfo {pages} {114725}
  (\bibinfo {year} {2012})},\ \Eprint
  {https://arxiv.org/abs/https://doi.org/10.1143/JPSJ.81.114725}
  {https://doi.org/10.1143/JPSJ.81.114725} \BibitemShut {NoStop}%
\bibitem [{\citenamefont {lei Zhang}\ \emph {et~al.}(2011)\citenamefont {lei
  Zhang}, \citenamefont {Jiao}, \citenamefont {Chen},\ and\ \citenamefont {qiu
  Yuan}}]{Zhang:2011}%
  \BibitemOpen
  \bibfield  {author} {\bibinfo {author} {\bibfnamefont {J.}~\bibnamefont {lei
  Zhang}}, \bibinfo {author} {\bibfnamefont {L.}~\bibnamefont {Jiao}}, \bibinfo
  {author} {\bibfnamefont {Y.}~\bibnamefont {Chen}},\ and\ \bibinfo {author}
  {\bibfnamefont {H.}~\bibnamefont {qiu Yuan}},\ }\bibfield  {title} {\bibinfo
  {title} {Universal behavior of the upper critical field in iron-based
  superconductors},\ }\href {https://doi.org/10.1007/s11467-011-0235-7}
  {\bibfield  {journal} {\bibinfo  {journal} {Frontiers of Physics}\ }\textbf
  {\bibinfo {volume} {6}},\ \bibinfo {pages} {463} (\bibinfo {year}
  {2011})}\BibitemShut {NoStop}%
\bibitem [{\citenamefont {Fulde}\ and\ \citenamefont
  {Ferrell}(1964)}]{Fulde:1964}%
  \BibitemOpen
  \bibfield  {author} {\bibinfo {author} {\bibfnamefont {P.}~\bibnamefont
  {Fulde}}\ and\ \bibinfo {author} {\bibfnamefont {R.~A.}\ \bibnamefont
  {Ferrell}},\ }\bibfield  {title} {\bibinfo {title} {Superconductivity in a
  strong spin-exchange field},\ }\href
  {https://doi.org/10.1103/PhysRev.135.A550} {\bibfield  {journal} {\bibinfo
  {journal} {Phys. Rev.}\ }\textbf {\bibinfo {volume} {135}},\ \bibinfo {pages}
  {A550} (\bibinfo {year} {1964})}\BibitemShut {NoStop}%
\bibitem [{\citenamefont {Larkin}\ and\ \citenamefont
  {Ovchinnikov}(1965)}]{Larkin:1965}%
  \BibitemOpen
  \bibfield  {author} {\bibinfo {author} {\bibfnamefont {A.}~\bibnamefont
  {Larkin}}\ and\ \bibinfo {author} {\bibfnamefont {Y.}~\bibnamefont
  {Ovchinnikov}},\ }\bibfield  {title} {\bibinfo {title} {Inhomogeneous state
  of superconductors},\ }\href@noop {} {\bibfield  {journal} {\bibinfo
  {journal} {Sov. Phys. JETP}\ }\textbf {\bibinfo {volume} {20}},\ \bibinfo
  {pages} {762} (\bibinfo {year} {1965})}\BibitemShut {NoStop}%
\bibitem [{\citenamefont {Agterberg}\ \emph {et~al.}(2020)\citenamefont
  {Agterberg}, \citenamefont {Davis}, \citenamefont {Edkins}, \citenamefont
  {Fradkin}, \citenamefont {Van~Harlingen}, \citenamefont {Kivelson},
  \citenamefont {Lee}, \citenamefont {Radzihovsky}, \citenamefont {Tranquada},\
  and\ \citenamefont {Wang}}]{PDW}%
  \BibitemOpen
  \bibfield  {author} {\bibinfo {author} {\bibfnamefont {D.~F.}\ \bibnamefont
  {Agterberg}}, \bibinfo {author} {\bibfnamefont {J.~S.}\ \bibnamefont
  {Davis}}, \bibinfo {author} {\bibfnamefont {S.~D.}\ \bibnamefont {Edkins}},
  \bibinfo {author} {\bibfnamefont {E.}~\bibnamefont {Fradkin}}, \bibinfo
  {author} {\bibfnamefont {D.~J.}\ \bibnamefont {Van~Harlingen}}, \bibinfo
  {author} {\bibfnamefont {S.~A.}\ \bibnamefont {Kivelson}}, \bibinfo {author}
  {\bibfnamefont {P.~A.}\ \bibnamefont {Lee}}, \bibinfo {author} {\bibfnamefont
  {L.}~\bibnamefont {Radzihovsky}}, \bibinfo {author} {\bibfnamefont {J.~M.}\
  \bibnamefont {Tranquada}},\ and\ \bibinfo {author} {\bibfnamefont
  {Y.}~\bibnamefont {Wang}},\ }\bibfield  {title} {\bibinfo {title} {The
  physics of pair-density waves: Cuprate superconductors and beyond},\ }\href
  {https://doi.org/10.1146/annurev-conmatphys-031119-050711} {\bibfield
  {journal} {\bibinfo  {journal} {Annual Review of Condensed Matter Physics}\
  }\textbf {\bibinfo {volume} {11}},\ \bibinfo {pages} {231} (\bibinfo {year}
  {2020})}\BibitemShut {NoStop}%
\bibitem [{\citenamefont {Sigrist}(2005)}]{Sigrist:2005}%
  \BibitemOpen
  \bibfield  {author} {\bibinfo {author} {\bibfnamefont {M.}~\bibnamefont
  {Sigrist}},\ }\bibfield  {title} {\bibinfo {title} {{Introduction to
  Unconventional Superconductivity}},\ }\href
  {https://doi.org/10.1063/1.2080350} {\bibfield  {journal} {\bibinfo
  {journal} {AIP Conference Proceedings}\ }\textbf {\bibinfo {volume} {789}},\
  \bibinfo {pages} {165} (\bibinfo {year} {2005})},\ \Eprint
  {https://arxiv.org/abs/https://aip.scitation.org/doi/pdf/10.1063/1.2080350}
  {https://aip.scitation.org/doi/pdf/10.1063/1.2080350} \BibitemShut {NoStop}%
\bibitem [{\citenamefont {Mineev}\ and\ \citenamefont
  {Samokhin}(1999)}]{Mineev:1999}%
  \BibitemOpen
  \bibfield  {author} {\bibinfo {author} {\bibfnamefont {V.}~\bibnamefont
  {Mineev}}\ and\ \bibinfo {author} {\bibfnamefont {K.}~\bibnamefont
  {Samokhin}},\ }\href@noop {} {\emph {\bibinfo {title} {{Introduction to
  Unconventional Superconductivity}}}}\ (\bibinfo  {publisher} {Gordon and
  Breach Science Publishers},\ \bibinfo {year} {1999})\BibitemShut {NoStop}%
\bibitem [{\citenamefont {Machida}\ \emph {et~al.}(1985)\citenamefont
  {Machida}, \citenamefont {Ohmi},\ and\ \citenamefont {Ozaki}}]{Machida:1985}%
  \BibitemOpen
  \bibfield  {author} {\bibinfo {author} {\bibfnamefont {K.}~\bibnamefont
  {Machida}}, \bibinfo {author} {\bibfnamefont {T.}~\bibnamefont {Ohmi}},\ and\
  \bibinfo {author} {\bibfnamefont {M.-A.}\ \bibnamefont {Ozaki}},\ }\bibfield
  {title} {\bibinfo {title} {{Anisotropy of Upper Critical Fields for d- and
  p-Wave Pairing Superconductivity}},\ }\href
  {https://doi.org/10.1143/JPSJ.54.1552} {\bibfield  {journal} {\bibinfo
  {journal} {Journal of the Physical Society of Japan}\ }\textbf {\bibinfo
  {volume} {54}},\ \bibinfo {pages} {1552} (\bibinfo {year} {1985})},\ \Eprint
  {https://arxiv.org/abs/https://doi.org/10.1143/JPSJ.54.1552}
  {https://doi.org/10.1143/JPSJ.54.1552} \BibitemShut {NoStop}%
\bibitem [{\citenamefont {Sauls}(1994)}]{Sauls:1994}%
  \BibitemOpen
  \bibfield  {author} {\bibinfo {author} {\bibfnamefont {J.~A.}\ \bibnamefont
  {Sauls}},\ }\bibfield  {title} {\bibinfo {title} {{The order parameter for
  the superconducting phases of UPt$_3$}},\ }\href
  {https://doi.org/10.1080/00018739400101475} {\bibfield  {journal} {\bibinfo
  {journal} {Advances in Physics}\ }\textbf {\bibinfo {volume} {43}},\ \bibinfo
  {pages} {113} (\bibinfo {year} {1994})}\BibitemShut {NoStop}%
\bibitem [{\citenamefont {Choi}\ and\ \citenamefont {Sauls}(1991)}]{Choi:1991}%
  \BibitemOpen
  \bibfield  {author} {\bibinfo {author} {\bibfnamefont {C.~H.}\ \bibnamefont
  {Choi}}\ and\ \bibinfo {author} {\bibfnamefont {J.~A.}\ \bibnamefont
  {Sauls}},\ }\bibfield  {title} {\bibinfo {title} {{Identification of
  odd-parity superconductivity in ${\mathrm{UPt}}_{3}$ from paramagnetic
  effects on the upper critical field}},\ }\href
  {https://doi.org/10.1103/PhysRevLett.66.484} {\bibfield  {journal} {\bibinfo
  {journal} {Phys. Rev. Lett.}\ }\textbf {\bibinfo {volume} {66}},\ \bibinfo
  {pages} {484} (\bibinfo {year} {1991})}\BibitemShut {NoStop}%
\bibitem [{\citenamefont {Tou}\ \emph {et~al.}(1996)\citenamefont {Tou},
  \citenamefont {Kitaoka}, \citenamefont {Asayama}, \citenamefont {Kimura},
  \citenamefont {Onuki}, \citenamefont {Yamamoto},\ and\ \citenamefont
  {Maezawa}}]{Tou:1996}%
  \BibitemOpen
  \bibfield  {author} {\bibinfo {author} {\bibfnamefont {H.}~\bibnamefont
  {Tou}}, \bibinfo {author} {\bibfnamefont {Y.}~\bibnamefont {Kitaoka}},
  \bibinfo {author} {\bibfnamefont {K.}~\bibnamefont {Asayama}}, \bibinfo
  {author} {\bibfnamefont {N.}~\bibnamefont {Kimura}}, \bibinfo {author}
  {\bibfnamefont {Y.}~\bibnamefont {Onuki}}, \bibinfo {author} {\bibfnamefont
  {E.}~\bibnamefont {Yamamoto}},\ and\ \bibinfo {author} {\bibfnamefont
  {K.}~\bibnamefont {Maezawa}},\ }\bibfield  {title} {\bibinfo {title}
  {{Odd-Parity Superconductivity with Parallel Spin Pairing in
  ${\mathrm{UPt}}_{3}$: Evidence from ${}^{195}\mathrm{Pt}$ Knight Shift
  Study}},\ }\href {https://doi.org/10.1103/PhysRevLett.77.1374} {\bibfield
  {journal} {\bibinfo  {journal} {Phys. Rev. Lett.}\ }\textbf {\bibinfo
  {volume} {77}},\ \bibinfo {pages} {1374} (\bibinfo {year}
  {1996})}\BibitemShut {NoStop}%
\bibitem [{\citenamefont {Wieder}\ \emph {et~al.}(2016)\citenamefont {Wieder},
  \citenamefont {Kim}, \citenamefont {Rappe},\ and\ \citenamefont
  {Kane}}]{Wieder:2016}%
  \BibitemOpen
  \bibfield  {author} {\bibinfo {author} {\bibfnamefont {B.~J.}\ \bibnamefont
  {Wieder}}, \bibinfo {author} {\bibfnamefont {Y.}~\bibnamefont {Kim}},
  \bibinfo {author} {\bibfnamefont {A.~M.}\ \bibnamefont {Rappe}},\ and\
  \bibinfo {author} {\bibfnamefont {C.~L.}\ \bibnamefont {Kane}},\ }\bibfield
  {title} {\bibinfo {title} {Double dirac semimetals in three dimensions},\
  }\href {https://doi.org/10.1103/PhysRevLett.116.186402} {\bibfield  {journal}
  {\bibinfo  {journal} {Phys. Rev. Lett.}\ }\textbf {\bibinfo {volume} {116}},\
  \bibinfo {pages} {186402} (\bibinfo {year} {2016})}\BibitemShut {NoStop}%
\bibitem [{\citenamefont {Bradlyn}\ \emph {et~al.}(2016)\citenamefont
  {Bradlyn}, \citenamefont {Cano}, \citenamefont {Wang}, \citenamefont
  {Vergniory}, \citenamefont {Felser}, \citenamefont {Cava},\ and\
  \citenamefont {Bernevig}}]{Bradlyn:2016}%
  \BibitemOpen
  \bibfield  {author} {\bibinfo {author} {\bibfnamefont {B.}~\bibnamefont
  {Bradlyn}}, \bibinfo {author} {\bibfnamefont {J.}~\bibnamefont {Cano}},
  \bibinfo {author} {\bibfnamefont {Z.}~\bibnamefont {Wang}}, \bibinfo {author}
  {\bibfnamefont {M.~G.}\ \bibnamefont {Vergniory}}, \bibinfo {author}
  {\bibfnamefont {C.}~\bibnamefont {Felser}}, \bibinfo {author} {\bibfnamefont
  {R.~J.}\ \bibnamefont {Cava}},\ and\ \bibinfo {author} {\bibfnamefont
  {B.~A.}\ \bibnamefont {Bernevig}},\ }\bibfield  {title} {\bibinfo {title}
  {Beyond dirac and weyl fermions: Unconventional quasiparticles in
  conventional crystals},\ }\href {https://doi.org/10.1126/science.aaf5037}
  {\bibfield  {journal} {\bibinfo  {journal} {Science}\ }\textbf {\bibinfo
  {volume} {353}},\ \bibinfo {pages} {aaf5037} (\bibinfo {year} {2016})},\
  \Eprint
  {https://arxiv.org/abs/https://www.science.org/doi/pdf/10.1126/science.aaf5037}
  {https://www.science.org/doi/pdf/10.1126/science.aaf5037} \BibitemShut
  {NoStop}%
\bibitem [{\citenamefont {Samsel-Czekała}\ \emph {et~al.}(2009)\citenamefont
  {Samsel-Czekała}, \citenamefont {Elgazzar}, \citenamefont {Oppeneer},
  \citenamefont {Talik}, \citenamefont {Walerczyk},\ and\ \citenamefont
  {Troć}}]{Samsel2010}%
  \BibitemOpen
  \bibfield  {author} {\bibinfo {author} {\bibfnamefont {M.}~\bibnamefont
  {Samsel-Czekała}}, \bibinfo {author} {\bibfnamefont {S.}~\bibnamefont
  {Elgazzar}}, \bibinfo {author} {\bibfnamefont {P.~M.}\ \bibnamefont
  {Oppeneer}}, \bibinfo {author} {\bibfnamefont {E.}~\bibnamefont {Talik}},
  \bibinfo {author} {\bibfnamefont {W.}~\bibnamefont {Walerczyk}},\ and\
  \bibinfo {author} {\bibfnamefont {R.}~\bibnamefont {Troć}},\ }\bibfield
  {title} {\bibinfo {title} {{The electronic structure of UCoGe by ab initio
  calculations and XPS experiment}},\ }\href
  {https://doi.org/10.1088/0953-8984/22/1/015503} {\bibfield  {journal}
  {\bibinfo  {journal} {Journal of Physics: Condensed Matter}\ }\textbf
  {\bibinfo {volume} {22}},\ \bibinfo {pages} {015503} (\bibinfo {year}
  {2009})}\BibitemShut {NoStop}%
\bibitem [{\citenamefont {Fujimori}\ \emph {et~al.}(2015)\citenamefont
  {Fujimori}, \citenamefont {Ohkochi}, \citenamefont {Kawasaki}, \citenamefont
  {Yasui}, \citenamefont {Takeda}, \citenamefont {Okane}, \citenamefont
  {Saitoh}, \citenamefont {Fujimori}, \citenamefont {Yamagami}, \citenamefont
  {Haga}, \citenamefont {Yamamoto},\ and\ \citenamefont
  {Onuki}}]{PhysRevB.91.174503}%
  \BibitemOpen
  \bibfield  {author} {\bibinfo {author} {\bibfnamefont {S.-I.}\ \bibnamefont
  {Fujimori}}, \bibinfo {author} {\bibfnamefont {T.}~\bibnamefont {Ohkochi}},
  \bibinfo {author} {\bibfnamefont {I.}~\bibnamefont {Kawasaki}}, \bibinfo
  {author} {\bibfnamefont {A.}~\bibnamefont {Yasui}}, \bibinfo {author}
  {\bibfnamefont {Y.}~\bibnamefont {Takeda}}, \bibinfo {author} {\bibfnamefont
  {T.}~\bibnamefont {Okane}}, \bibinfo {author} {\bibfnamefont
  {Y.}~\bibnamefont {Saitoh}}, \bibinfo {author} {\bibfnamefont
  {A.}~\bibnamefont {Fujimori}}, \bibinfo {author} {\bibfnamefont
  {H.}~\bibnamefont {Yamagami}}, \bibinfo {author} {\bibfnamefont
  {Y.}~\bibnamefont {Haga}}, \bibinfo {author} {\bibfnamefont {E.}~\bibnamefont
  {Yamamoto}},\ and\ \bibinfo {author} {\bibfnamefont {Y.}~\bibnamefont
  {Onuki}},\ }\bibfield  {title} {\bibinfo {title} {{Electronic structures of
  ferromagnetic superconductors ${\text{UGe}}_{2}$ and UCoGe studied by
  angle-resolved photoelectron spectroscopy}},\ }\href
  {https://doi.org/10.1103/PhysRevB.91.174503} {\bibfield  {journal} {\bibinfo
  {journal} {Phys. Rev. B}\ }\textbf {\bibinfo {volume} {91}},\ \bibinfo
  {pages} {174503} (\bibinfo {year} {2015})}\BibitemShut {NoStop}%
\bibitem [{\citenamefont {Daido}\ \emph
  {et~al.}(2019{\natexlab{b}})\citenamefont {Daido}, \citenamefont {Yoshida},\
  and\ \citenamefont {Yanase}}]{PhysRevLett.122.227001}%
  \BibitemOpen
  \bibfield  {author} {\bibinfo {author} {\bibfnamefont {A.}~\bibnamefont
  {Daido}}, \bibinfo {author} {\bibfnamefont {T.}~\bibnamefont {Yoshida}},\
  and\ \bibinfo {author} {\bibfnamefont {Y.}~\bibnamefont {Yanase}},\
  }\bibfield  {title} {\bibinfo {title} {{${\bm{Z}}_{4}$ Topological
  Superconductivity in UCoGe}},\ }\href
  {https://doi.org/10.1103/PhysRevLett.122.227001} {\bibfield  {journal}
  {\bibinfo  {journal} {Phys. Rev. Lett.}\ }\textbf {\bibinfo {volume} {122}},\
  \bibinfo {pages} {227001} (\bibinfo {year} {2019}{\natexlab{b}})}\BibitemShut
  {NoStop}%
\bibitem [{\citenamefont {Bastien}\ \emph {et~al.}(2016)\citenamefont
  {Bastien}, \citenamefont {Gourgout}, \citenamefont {Aoki}, \citenamefont
  {Pourret}, \citenamefont {Sheikin}, \citenamefont {Seyfarth}, \citenamefont
  {Flouquet},\ and\ \citenamefont {Knebel}}]{Bastien:2016}%
  \BibitemOpen
  \bibfield  {author} {\bibinfo {author} {\bibfnamefont {G.}~\bibnamefont
  {Bastien}}, \bibinfo {author} {\bibfnamefont {A.}~\bibnamefont {Gourgout}},
  \bibinfo {author} {\bibfnamefont {D.}~\bibnamefont {Aoki}}, \bibinfo {author}
  {\bibfnamefont {A.}~\bibnamefont {Pourret}}, \bibinfo {author} {\bibfnamefont
  {I.}~\bibnamefont {Sheikin}}, \bibinfo {author} {\bibfnamefont
  {G.}~\bibnamefont {Seyfarth}}, \bibinfo {author} {\bibfnamefont
  {J.}~\bibnamefont {Flouquet}},\ and\ \bibinfo {author} {\bibfnamefont
  {G.}~\bibnamefont {Knebel}},\ }\bibfield  {title} {\bibinfo {title} {Lifshitz
  transitions in the ferromagnetic superconductor ucoge},\ }\href
  {https://doi.org/10.1103/PhysRevLett.117.206401} {\bibfield  {journal}
  {\bibinfo  {journal} {Phys. Rev. Lett.}\ }\textbf {\bibinfo {volume} {117}},\
  \bibinfo {pages} {206401} (\bibinfo {year} {2016})}\BibitemShut {NoStop}%
\bibitem [{\citenamefont {Canepa}\ \emph {et~al.}(1996)\citenamefont {Canepa},
  \citenamefont {Manfrinetti}, \citenamefont {Pani},\ and\ \citenamefont
  {Palenzona}}]{CANEPA1996225}%
  \BibitemOpen
  \bibfield  {author} {\bibinfo {author} {\bibfnamefont {F.}~\bibnamefont
  {Canepa}}, \bibinfo {author} {\bibfnamefont {P.}~\bibnamefont {Manfrinetti}},
  \bibinfo {author} {\bibfnamefont {M.}~\bibnamefont {Pani}},\ and\ \bibinfo
  {author} {\bibfnamefont {A.}~\bibnamefont {Palenzona}},\ }\bibfield  {title}
  {\bibinfo {title} {{Structural and transport properties of some UTX compounds
  where T = Fe, Co, Ni and X = Si, Ge}},\ }\href
  {https://doi.org/https://doi.org/10.1016/0925-8388(95)02037-3} {\bibfield
  {journal} {\bibinfo  {journal} {Journal of Alloys and Compounds}\ }\textbf
  {\bibinfo {volume} {234}},\ \bibinfo {pages} {225} (\bibinfo {year}
  {1996})}\BibitemShut {NoStop}%
\bibitem [{\citenamefont {Kawamura}(2019)}]{KAWAMURA2019197}%
  \BibitemOpen
  \bibfield  {author} {\bibinfo {author} {\bibfnamefont {M.}~\bibnamefont
  {Kawamura}},\ }\bibfield  {title} {\bibinfo {title} {{FermiSurfer:
  Fermi-surface viewer providing multiple representation schemes}},\ }\href
  {https://doi.org/https://doi.org/10.1016/j.cpc.2019.01.017} {\bibfield
  {journal} {\bibinfo  {journal} {Computer Physics Communications}\ }\textbf
  {\bibinfo {volume} {239}},\ \bibinfo {pages} {197} (\bibinfo {year}
  {2019})}\BibitemShut {NoStop}%
\bibitem [{\citenamefont {Shimizu}\ \emph {et~al.}(2019)\citenamefont
  {Shimizu}, \citenamefont {Braithwaite}, \citenamefont {Aoki}, \citenamefont
  {Salce},\ and\ \citenamefont {Brison}}]{Shimizu:2019}%
  \BibitemOpen
  \bibfield  {author} {\bibinfo {author} {\bibfnamefont {Y.}~\bibnamefont
  {Shimizu}}, \bibinfo {author} {\bibfnamefont {D.}~\bibnamefont
  {Braithwaite}}, \bibinfo {author} {\bibfnamefont {D.}~\bibnamefont {Aoki}},
  \bibinfo {author} {\bibfnamefont {B.}~\bibnamefont {Salce}},\ and\ \bibinfo
  {author} {\bibfnamefont {J.-P.}\ \bibnamefont {Brison}},\ }\bibfield  {title}
  {\bibinfo {title} {{Spin-Triplet $p$-Wave Superconductivity Revealed under
  High Pressure in ${\mathrm{UBe}}_{13}$}},\ }\href
  {https://doi.org/10.1103/PhysRevLett.122.067001} {\bibfield  {journal}
  {\bibinfo  {journal} {Phys. Rev. Lett.}\ }\textbf {\bibinfo {volume} {122}},\
  \bibinfo {pages} {067001} (\bibinfo {year} {2019})}\BibitemShut {NoStop}%
\bibitem [{\citenamefont {Ma}\ \emph {et~al.}(2021)\citenamefont {Ma},
  \citenamefont {Gornicka}, \citenamefont {Lef{\`{e} }vre}, \citenamefont
  {Yang}, \citenamefont {R{\o}nnow}, \citenamefont {Jeschke}, \citenamefont
  {Klimczuk},\ and\ \citenamefont {von Rohr}}]{Ma:2021}%
  \BibitemOpen
  \bibfield  {author} {\bibinfo {author} {\bibfnamefont {K.}~\bibnamefont
  {Ma}}, \bibinfo {author} {\bibfnamefont {K.}~\bibnamefont {Gornicka}},
  \bibinfo {author} {\bibfnamefont {R.}~\bibnamefont {Lef{\`{e} }vre}},
  \bibinfo {author} {\bibfnamefont {Y.}~\bibnamefont {Yang}}, \bibinfo {author}
  {\bibfnamefont {H.~M.}\ \bibnamefont {R{\o}nnow}}, \bibinfo {author}
  {\bibfnamefont {H.~O.}\ \bibnamefont {Jeschke}}, \bibinfo {author}
  {\bibfnamefont {T.}~\bibnamefont {Klimczuk}},\ and\ \bibinfo {author}
  {\bibfnamefont {F.~O.}\ \bibnamefont {von Rohr}},\ }\bibfield  {title}
  {\bibinfo {title} {{Superconductivity with High Upper Critical Field in the
  Cubic Centrosymmetric $\eta$-Carbide Nb$_4$Rh$_2$C$_{1-\delta}$}},\ }\href
  {https://doi.org/10.1021/acsmaterialsau.1c00011} {\bibfield  {journal}
  {\bibinfo  {journal} {{ACS} Materials Au}\ }\textbf {\bibinfo {volume} {1}},\
  \bibinfo {pages} {55} (\bibinfo {year} {2021})}\BibitemShut {NoStop}%
\bibitem [{\citenamefont {Ruan}\ \emph {et~al.}(2022)\citenamefont {Ruan},
  \citenamefont {Zhou}, \citenamefont {Yang}, \citenamefont {Gu}, \citenamefont
  {Ma}, \citenamefont {Chen},\ and\ \citenamefont {Ren}}]{Ruan:2022}%
  \BibitemOpen
  \bibfield  {author} {\bibinfo {author} {\bibfnamefont {B.-B.}\ \bibnamefont
  {Ruan}}, \bibinfo {author} {\bibfnamefont {M.-H.}\ \bibnamefont {Zhou}},
  \bibinfo {author} {\bibfnamefont {Q.-S.}\ \bibnamefont {Yang}}, \bibinfo
  {author} {\bibfnamefont {Y.-D.}\ \bibnamefont {Gu}}, \bibinfo {author}
  {\bibfnamefont {M.-W.}\ \bibnamefont {Ma}}, \bibinfo {author} {\bibfnamefont
  {G.-F.}\ \bibnamefont {Chen}},\ and\ \bibinfo {author} {\bibfnamefont
  {Z.-A.}\ \bibnamefont {Ren}},\ }\bibfield  {title} {\bibinfo {title}
  {{Superconductivity with a Violation of Pauli Limit and Evidences for
  Multigap in $\eta$-Carbide Type Ti$_4$Ir$_2$O}},\ }\href
  {https://doi.org/10.1088/0256-307x/39/2/027401} {\bibfield  {journal}
  {\bibinfo  {journal} {Chinese Physics Letters}\ }\textbf {\bibinfo {volume}
  {39}},\ \bibinfo {pages} {027401} (\bibinfo {year} {2022})}\BibitemShut
  {NoStop}%
\end{thebibliography}%

\clearpage
	\appendix

\section{Full excitation spectrum on the nodal plane}

On the nodal plane, the Bogoliubov de-Gennes Hamiltonian takes the form
\begin{equation}
H=\sum_{\bm{k}}\bm{\Psi}_{\bm{k}}^{\dagger}\begin{pmatrix} {\varepsilon_{0,\bm{k}}+\tau_3 (\bm{\lambda}_{\bm{k}}\cdot \bm{\hat{n}})(\bm{\sigma}\cdot\bm{\hat{n}})  }  & \Delta_{\bm{k}}\\
\Delta_{\bm{k}}^\dagger & -\varepsilon_{0,\bm{k}}-\tau_3 (\bm{\lambda}_{\bm{k}}\cdot \bm{\hat{n}})(\bm{\sigma}\cdot \bm{\hat{n}}) 
\end{pmatrix}\bm{\Psi}_{\bm{k}},\label{eq:BdG_Ham}
\end{equation}
It is possible to classify the gap symmetry as even or odd under both inversion and mirror symmetries. 
For momenta on the nodal surface we have,
\begin{eqnarray}
U_P^{\dagger}\Delta_{\bm{k}}U_P=&\pm \Delta_{-{\bm{k}}} \nonumber\\
U_{M}^{\dagger}\Delta_{\bm{k}}U_{M}=&\pm \Delta_{\bm{k}}
\end{eqnarray}
where for type 1 TRIM $U_P=\tau_1$ and $U_{M}=-i\tau_3\sigma_z$ and for type 2 TRIM  $U_P=\tau_0$ and $U_{M}=-i\tau_2\sigma_z$.
We label the gaps as $\Delta_{1(2),i,j}$ where $i=\pm$ labels the parity symmetry and $j=\pm$ labels the mirror symmetry. Here, for clarity, we drop the ${\bm k}$ labels (note that $\bm{k}$ is unchanged by the mirror symmetry).  For the type 1 TRIM, we write the gap functions in terms of the complete set of gap functions with the correct symmetries given in Table \ref{tab:pairings} as
\begin{eqnarray}
\Delta_{1,++}=&\psi_0\tau_0+(\bm{d}_z\cdot\bm{\hat{n}})(\bm{\sigma}\cdot\bm{\hat{n}})\tau_3\nonumber \\
\Delta_{1,+-}=&\psi_x\tau_1+(\bm{d}_z\times\bm{\hat{n}})\cdot(\bm{\sigma}\times{\bm{\hat{n}}})\tau_3+d_y\tau_2 \nonumber\\
\Delta_{1,-+}=&(\bm{d}_0\cdot\bm{\hat{n}})(\bm{\sigma}\cdot\bm{\hat{n}})\tau_0+(\bm{d}_x\times\bm{\hat{n}})\cdot(\bm{\sigma}\times{\bm{\hat{n}}})\tau_1+\psi_z\tau_3+(\bm{\psi}\times\bm{\hat{n}})\cdot(\bm{\sigma}\times{\bm{\hat{n}}})\tau_2 \nonumber\\
\Delta_{1,--}=&(\bm{d}_0\times\bm{\hat{n}})\cdot(\bm{\sigma}\times{\bm{\hat{n}}})\tau_0+(\bm{d}_x\cdot\bm{\hat{n}})(\bm{\sigma}\cdot\bm{\hat{n}})\tau_1+(\bm{\psi}\cdot\bm{\hat{n}})(\bm{\sigma}\cdot\bm{\hat{n}})\tau_2
\end{eqnarray}
where $d_i$ are odd functions of $\bm{k}$ and $\psi_i$ are even functions of $\bm{k}$. Using Eq.~\ref{eq:BdG_Ham}, the corresponding quasiparticle excitation energies can be found to be
\begin{eqnarray}
E_{1,++}=&\pm^\prime\sqrt{(\epsilon_0\pm \bm{\lambda}\cdot{\bm{\hat{n}}})^2+(\psi_0\pm \bm{d}_z\cdot{\bm{\hat{n}}})^2}\nonumber \\
E_{1,+-}=&\pm^\prime\Big(\sqrt{\epsilon_0^2 +\psi_x^2+(\bm{d}_z\times\bm{\hat{n}})^2+d_y^2}\pm \bm{\lambda}\cdot{\bm{\hat{n}}}\Big)\nonumber\\
E_{1,-+}=&\pm^\prime\sqrt{(\epsilon_0\pm \bm{\lambda}\cdot \bm{\hat{n}})^2+(\psi_z \pm \bm{d}_0\cdot\bm{\hat{n}})^2+(\bm{d}_x\times\bm{\hat{n}})^2+(\bm{\psi}\times \bm{\hat{n}})^2\pm 2 (\bm{d}_x\times\bm{\psi})\cdot\bm{\hat{n}}}\nonumber\\
E_{1,--}=&\pm^\prime\Big(\sqrt{\epsilon_0^2 +(\bm{d}_0\times\bm{\hat{n}})^2+(\bm{d}_x\cdot\bm{\hat{n}})^2+(\bm{\psi}\cdot\bm{\hat{n}})^2}\pm \bm{\lambda}\cdot\bm{\hat{n}}\Big)
\end{eqnarray}
where the prime denotes independent choices of the sign.
For type 2 TRIM we similarly have
\begin{eqnarray}
\Delta_{2,++}=&\psi_0\tau_0+(\bm{\psi}\cdot\bm{\hat{n}})(\bm{\sigma}\cdot\bm{\hat{n}})\tau_2\nonumber \\
\Delta_{2,+-}=&\psi_x\tau_1+\psi_z\tau_3+(\bm{\psi}\times\bm{\hat{n}})\cdot(\bm{\sigma}\times{\bm{\hat{n}}})\tau_2  \nonumber\\
\Delta_{2,-+}=&(\bm{d}_0\cdot\bm{\hat{n}})(\bm{\sigma}\cdot\bm{\hat{n}})\tau_0+(\bm{d}_x\times\bm{\hat{n}})\cdot(\bm{\sigma}\times{\bm{\hat{n}}})\tau_1+(\bm{d}_z\times\bm{\hat{n}})\cdot(\bm{\sigma}\times{\bm{\hat{n}}})\tau_3+d_y\tau_2\nonumber\\
\Delta_{2,--}=&(\bm{d}_0\times\bm{\hat{n}})\cdot(\bm{\sigma}\times{\bm{\hat{n}}})\tau_0+(\bm{d}_x\cdot\bm{\hat{n}})(\bm{\sigma}\cdot\bm{\hat{n}})\tau_1+(\bm{d}_z\cdot\bm{\hat{n}})(\bm{\sigma}\cdot\bm{\hat{n}})\tau_3
\end{eqnarray}
The quasiparticle excitation spectra for these states are

\begin{eqnarray}
E_{2,++}=&\pm^\prime\sqrt{(\epsilon_0\pm \bm{\lambda}\cdot{\bm{\hat{n}}})^2+(\psi_0\pm \bm{\psi}\cdot\bm{\hat{n}})^2}\nonumber \\
E_{2,+-}=&\pm^\prime\Big(\sqrt{\epsilon_0^2 +\psi_x^2+\psi_z^2+(\bm{\psi}\times\bm{\hat{n}})^2}\pm \bm{\lambda}\cdot{\bm{\hat{n}}}\Big)\nonumber\\
E_{2,-+}=&\pm^\prime\sqrt{(\epsilon_0\pm \bm{\lambda}\cdot{\bm{\hat{n}}})^2+(d_y\pm\bm{d}_0\cdot\bm{\hat{n}})^2+(\bm{d}_x\times\bm{\hat{n}})^2+(\bm{d}_z\times\bm{\hat{n}})^2\pm 2(\bm{d}_x\times\bm{d}_z)\cdot\bm{\hat{n}}}\nonumber\\
E_{2,--}=&\pm^\prime\Big(\sqrt{\epsilon_0^2 +(\bm{d}_0\times\bm{\hat{n}})^2+(\bm{d}_x\cdot\bm{\hat{n}})^2+(\bm{d}_z\cdot\bm{\hat{n}})^2}\pm \bm{\lambda}\cdot{\bm{\hat{n}}}\Big)
\end{eqnarray}

\clearpage

	\section{Magnetic susceptibility UPt$_3$}
	In the main text, we illustrated how the p-wave state in UPt$_3$ is immune to the magnetic field along arbitrary directions. An important step is to consider the small $g$-factor for field $\bm{B}\perp \hat{\bm{z}}$. However, the discussion is not complete. In the normal state, there exist 4-fold degenerate Dirac lines on the plane $k_z=\pi/c$, where the $g$-factor is not small. In terms of the field fitness, Eq.\ref{Eq_immune} in the main text only considered doubly degenerate bands. In principle, extra terms in the field fitness are needed for to describe these Dirac lines. However, the Fermi surface is not right on the nodal plane. This can make the Dirac lines unimportant. In this section, we will explicitly check the field response in the superconducting state through a numerical calculation on a tight-binding model for UPt$_3$.

    In the following calculations, we will focus on the Knight shift (spin-susceptibility). Knight shift measures spin polarization at atom sites. By extracting spin susceptibility $\chi_s$, one can determine pairing functions of an unconventional superconductor.
	For a single-band spin-triplet superconductor, the change of Knight shift depends on the orientation of magnetic field with respect to the ${\bm d}$-vector of the superconducting state.
	If the magnetic field is perpendicular to the ${\bm d}$-vector, the Knight shift should be a constant across superconducting $T_c$. If the magnetic field is parallel to the ${\bm d}$-vector, the Knight shift will decrease to zero as temperature approaches zero. For the multi-band non-symmorphic superconductor UPt$_3$, Knight shift is almost unchanged for all field orientations, suggesting the importance of spin-orbit coupling in this heavy fermion material. 
	
	One of the Fermi surfaces (`starfish') of UPt$_3$ is flat and located near the high symmetry plane $k_z=\pi/c$. Zeeman terms $B_x\sigma_x$ and $B_y\sigma_y$ then becomes inter-band. From non-generate perturbation theory, spin susceptibilities are inversely proportional to the band gap. This is different from the intra-band Zeeman effect, where susceptibilities are proportional to the density of states on Fermi surface, according to degenerate perturbation theory. 
	
	Since the superconducting gap is much smaller than the band gap, inter-band susceptibilities will be unchanged across $T_c$. If the superconductivity is mainly developed on the above flat Fermi surface, then Knight shift is expected to be unchanged for in-plane magnetic fields, regardless of the superconducting pairing symmetry. If the ${\bm d}$-vector is in-plane, then Knight shift will also be unchanged for a perpendicular magnetic field. In this section, we will explicitly illustrate this idea to understand the experimental results on UPt$_3$.

	\begin{figure}[htb]
		\centering
		\includegraphics[height=5cm]{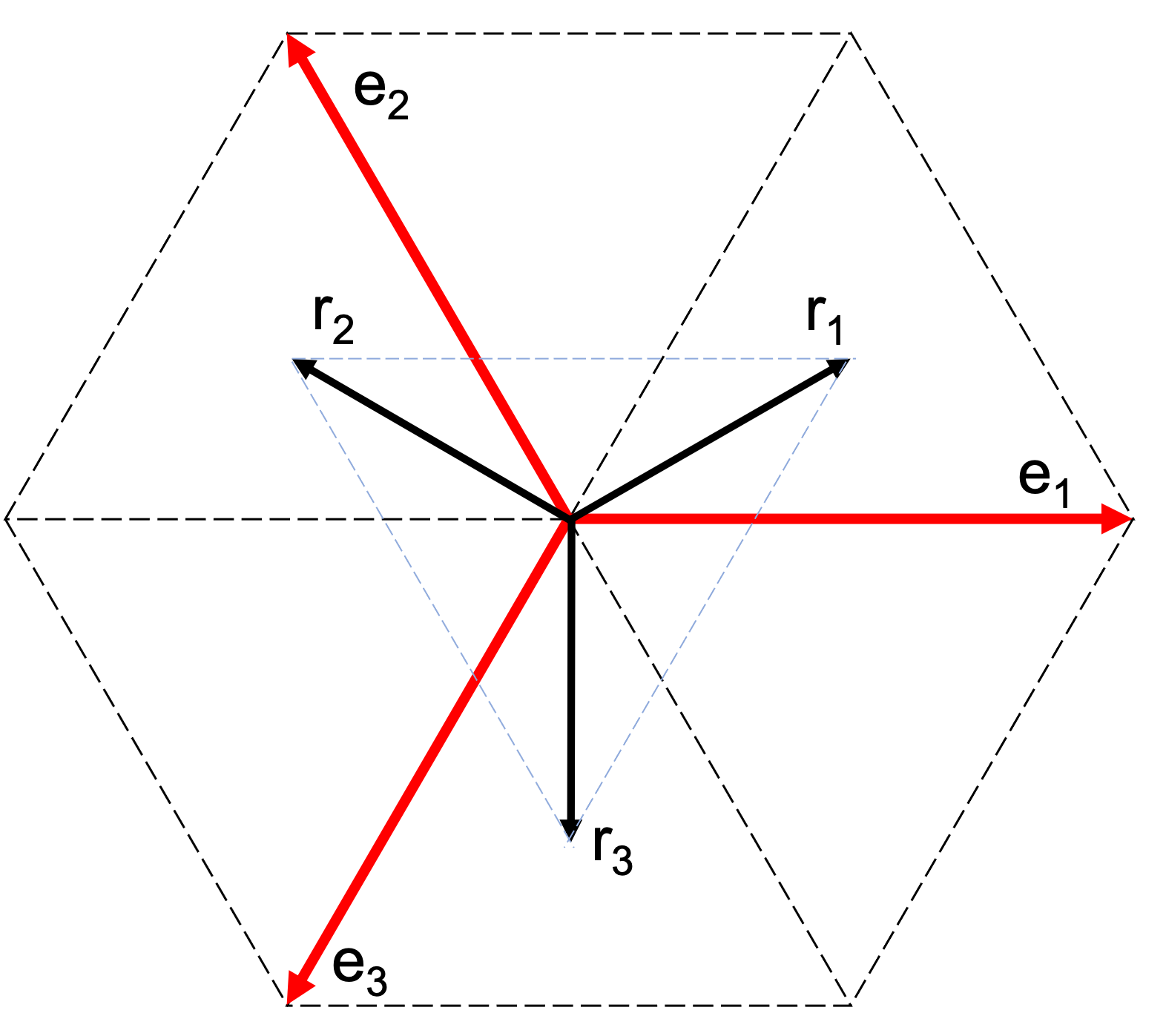}
		\caption{Crystal structure of UPt$_3$ with the unit vector ${\bf e_1}=(1,0,0)$.}
		\label{F_UPt3}
	\end{figure}
	
	The $4\times4$ normal state Hamiltonian reads \cite{Yanase:2016}:
	\begin{equation}
		\begin{split}
			&H=\varepsilon({\bf k})+g_z({\bf k})\sigma_z\tau_3+a_1({\bf k})\tau_1+a_2({\bf k})\tau_2+\left[g_x({\bf k})\sigma_x+g_y({\bf k})\sigma_y\right]\tau_3\\
			&\varepsilon_{\bf k}=2t\sum_{i=1,2,3}\cos{\bf k_\parallel\cdot{e_i}}+2t_3\cos{k_z}-\mu,\;\;\;
			g_z({\bf k})=g_{z0}\sum_i\sin{\bf k_\parallel\cdot{e_i}}\\
			&a_1({\bf k})=2t'\sin\frac{k_z}{2}\sum_{i=1,2,3}\sin{\bf k_\parallel\cdot r_i},\;\;\;
			a_2({\bf k})=2t'\sin\frac{k_z}{2}\sum_{i=1,2,3}\cos{\bf k_\parallel\cdot r_i}\\
			&g_x({\bf k})=g_{x0}(f_x^2-f_y^2)\sin{k_z},\;\;\;
			g_y({\bf k})=g_{y0}f_xf_y\sin{k_z} \\
			&f_x\equiv\sin{\bf k_\parallel\cdot e_1}-\frac{\sin{\bf k_\parallel\cdot e_2}+\sin{\bf k_\parallel\cdot e_3}}{2},\;\;\;
			f_y\equiv\sqrt{3}\sin{\bf k_\parallel\cdot e_2}-\sin{\bf k_\parallel\cdot e_3},
		\end{split}
		\label{E_model}
	\end{equation}
	 here $(k_x,k_y,k_z)$ are relative to the high symmetry point $(0,0,\pi)$. Relevant vectors $\bf e_i$ and $\bf r_i$ can be found in Fig.\ref{F_UPt3}. $\tau_i$ matrices live in the sublattice space.
	On the high-symmetry plane $k_z=0$, the inter-sublattice hopping $a_{1,2}$ and the spin-flip SOC $g_{x,y}$ vanish. $|{\bf k},m=1,\uparrow\rangle$ and $|{\bf k},m=2,\downarrow\rangle$ states form a pseudospin band, while $|{\bf k},m=2,\uparrow\rangle$ and $|{\bf k},m=1,\downarrow\rangle$ states form another band.

	We now study spin susceptibilities. We will focus on a p-wave state in the $E_{2u}$ channel. Its ${\bm d}$-vector is in-plane: ${\bm d}=\Delta(T)(f_x,-f_y,0)$. $f_x$ and $f_y$ are introduced in Eq.\ref{E_model}, and they transform as $k_x$ and $k_y$. The gap magnitude is taken to be $\Delta(T)=\Delta_0\sqrt{1-T/T_c}$.
	$t=1, t_3=-4, g_{z0}=2, \mu=12$ and $\Delta_0=T_c=0.001$ is taken in the calculation. 
	
	\begin{figure}[htb]
	\centering
	\includegraphics[width=5cm]{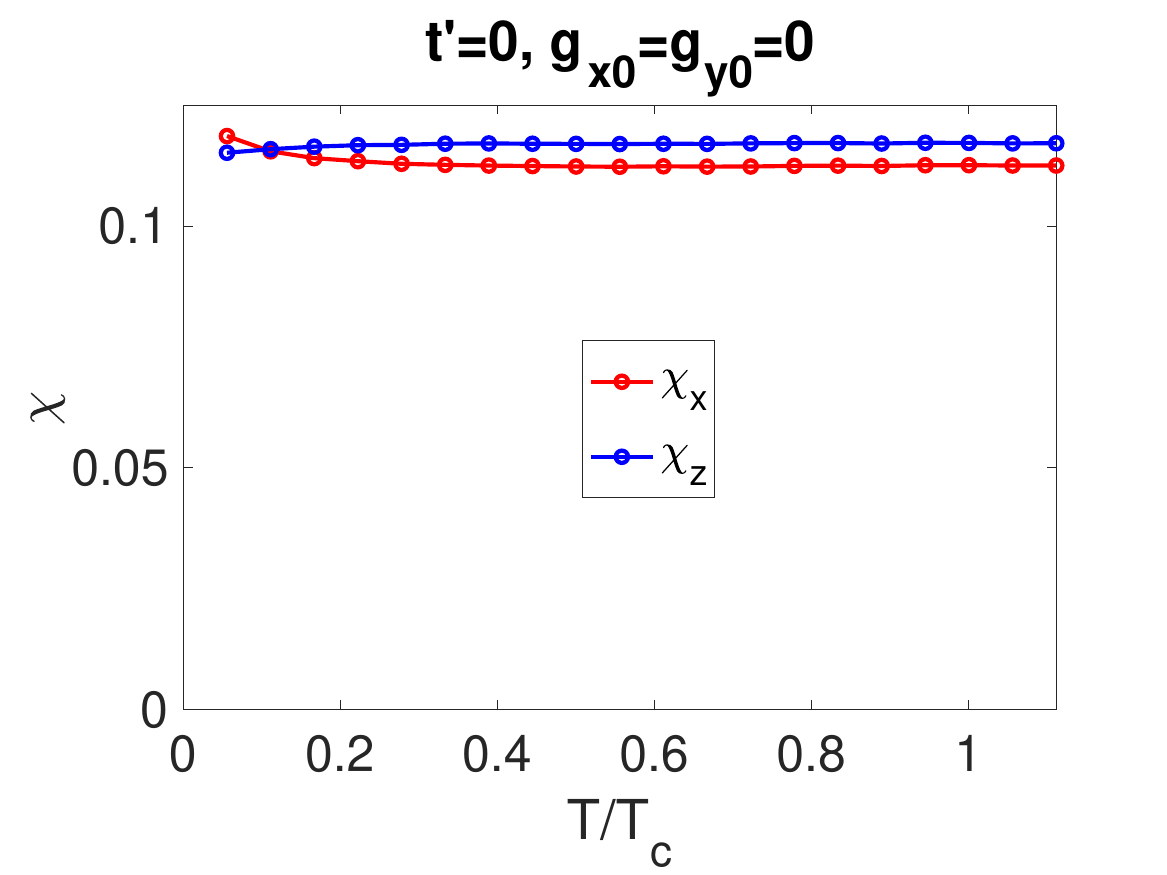}
	\includegraphics[width=5cm]{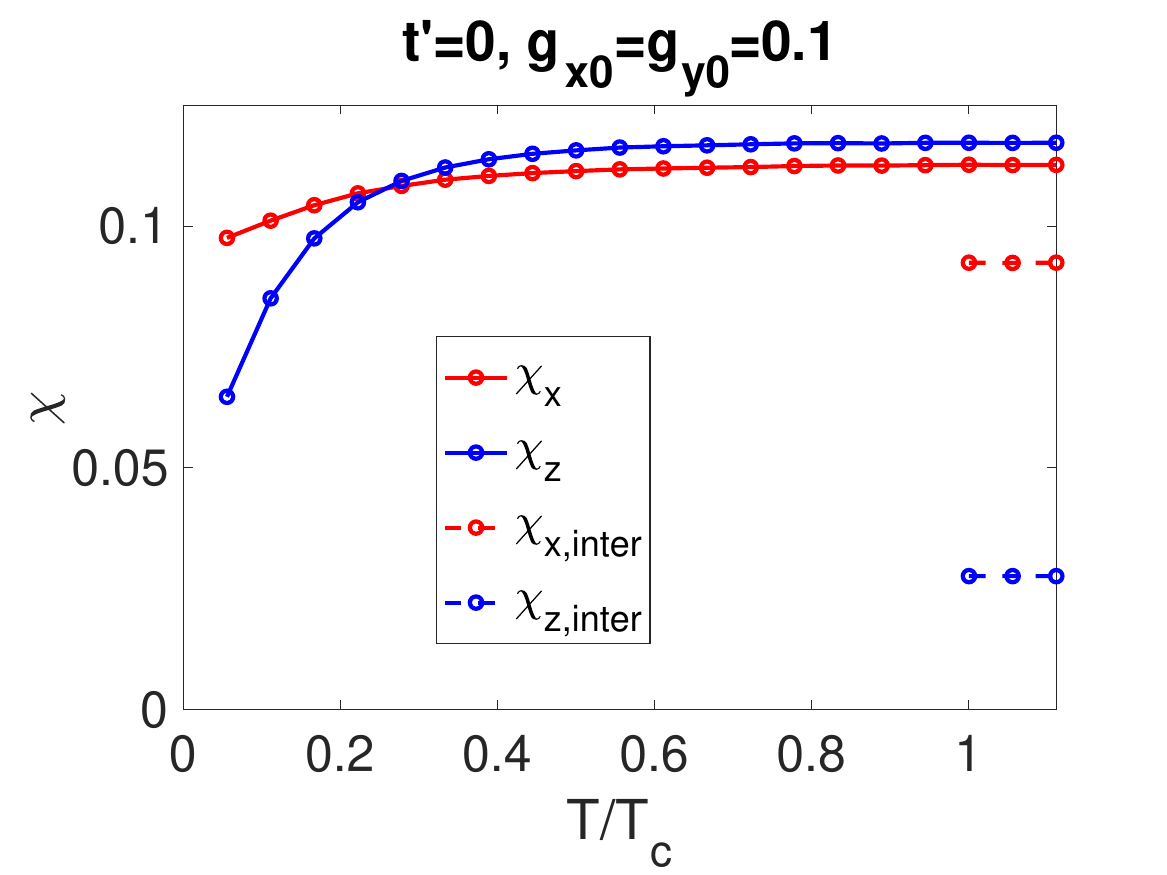}
	\includegraphics[width=5cm]{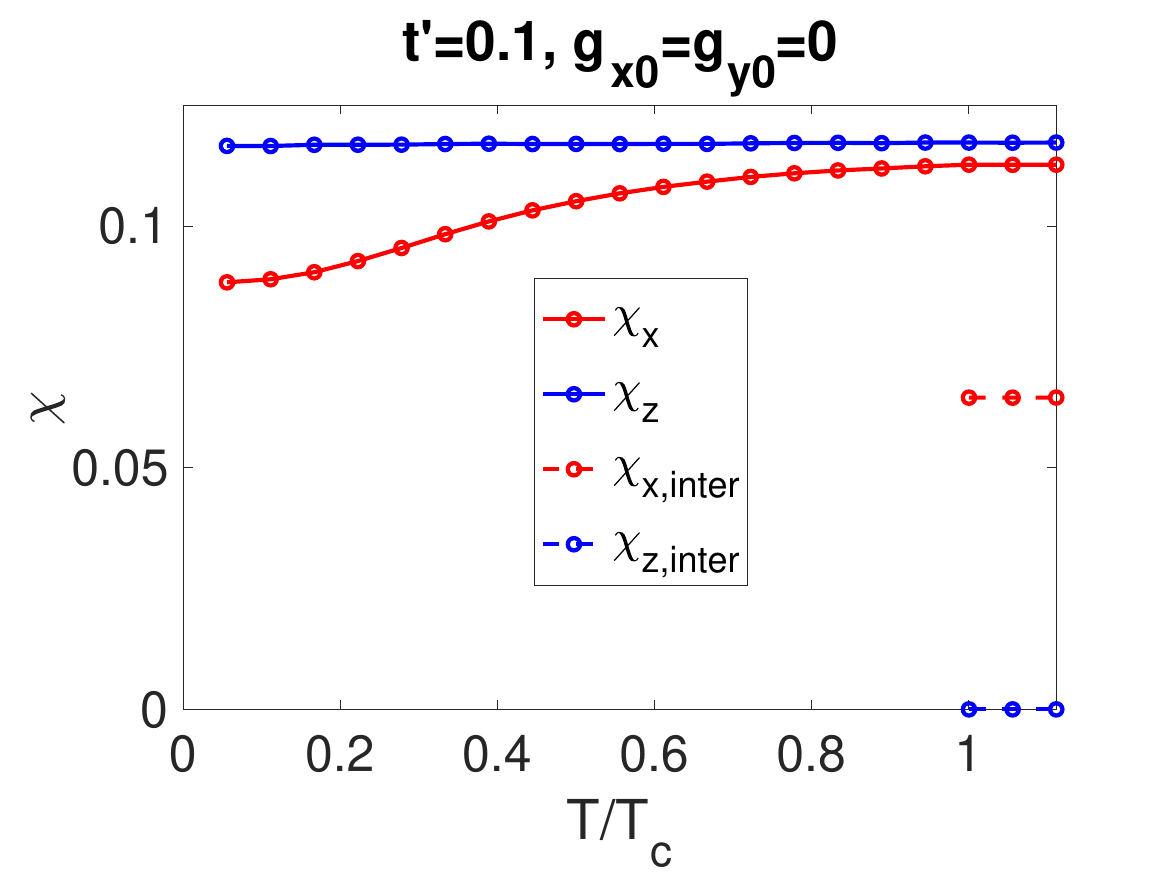}
	\caption{Spin susceptibilities as a function of temperature, for  (left) $a_1=a_2=g_{x}=g_{y}=0$, which would be the case if the Fermi surface exactly lied on the high-symmetry plane. (middle) non-zero spin-flip SOC but zero inter-sublattice hopping. (right) non-zero inter-sublattice hopping but zero spin-flip SOC.  }
	\label{F_Knight}
\end{figure}
	
	To illustrate the effect of the anomalous pseudospin, we start with a toy model with zero inter-sublattice hopping and spin-flip SOC: $t'=g_{x0}=g_{y0}$. The corresponding  four terms vanish in the normal state Hamiltonian: $a_1=a_2=g_{x}=g_{y}=0$. In this extreme case, the spin susceptibilities are unchanged across $T_c$, as shown in the left panel of Fig.\ref{F_Knight}. 
	
	We now turn on the spin-flip SOC ($g_{x0}$ and $g_{y0}$), while keeping the inter-sublattice hopping $t'$ to be zero. $h_x\sigma_x$ develops an intra-band component, which will be suppressed in the superconducting state. As a result, the total $\chi_x$ deep in the superconducting state starts to decrease as function of temperature. For $\chi_z$, spin-flip SOC induces higher-order terms in the $E_{2u}$ channel. The ${\bm d}$-vector develops non-zero z-component in the band basis. This causes a decrease in $\chi_z$. The result for $g_{x0}=g_{y0}$ can be found in the middle panel of Fig.\ref{F_Knight}. The inter-band susceptibilities in the normal state are included in dashed lines.
	
	We now turn on the inter-sublattice hopping $t'$, while keeping the spin-flip SOC ($g_{x0}$ and $g_{y0}$) to be zero. A similar effect is expected for $\chi_x$ due to the intra-band contribution. For $\chi_z$, since $\sigma_z$ is a good quantum number, $\chi_z$ will be unchanged. The result can be found in the right panel of Fig.\ref{F_Knight}. 

    Experimentally, the superconducting state is known to be more robust under ${\bf B}\parallel \hat{\bm{x}}$ compared to ${\bf B}\parallel \hat{\bm{z}}$. In other words, the decrease in $\chi_x$ needs to be smaller than $\chi_z$. This scenario is closer to the second limit.

\clearpage

\section{8-fold Representations}
Here, we list the symmetries of all orbital operators near the 8-fold degenerate points. The point group that keeps the TRIM point invariant can be found in the title. The bracket notation \([\cdot]\) is also used for antisymmetric operators which was \(\tau_2\) in the main context, but in 8-fold cases, the antisymmetric component is not unique due to the higher degrees of freedom.
\begin{table}[h]
	\begin{tabular}{|c|c|}
		\hline
		Space group momenta  & Point group $D_{2h}$                                          \\ \hline
			52($S_1S_2$)        & $A_g+2B_{1g}+B_{2g}+2B_{3g}+2A_u+B_{1u}+B_{3u}+[A_g]+[B_{2g}]+[B_{1u}]+2[B_{2u}]+[B_{3u}]$ \\ \hline
		54($U_1U_2$)        & $A_g+2B_{1g}+2B_{2g}+B_{3g}+2A_u+B_{1u}+B_{2u}+[A_g]+[B_{3g}]+[B_{1u}]+[B_{2u}]+2[B_{3u}]$ \\ \hline
		54($R_1R_2$)        & $A_g+2B_{1g}+2B_{2g}+B_{3g}+2A_u+B_{1u}+B_{2u}+[A_g]+[B_{3g}]+[B_{1u}]+[B_{2u}]+2[B_{3u}]$ \\ \hline
		56($U_1U_2$)        & $A_g+2B_{1g}+2B_{2g}+B_{3g}+2A_u+B_{1u}+B_{2u}+[A_g]+[B_{3g}]+[B_{1u}]+[B_{2u}]+2[B_{3u}]$ \\ \hline
		56($T_1T_2$)        & $A_g+2B_{1g}+B_{2g}+2B_{3g}+2A_u+B_{1u}+B_{3u}+[A_g]+[B_{2g}]+[B_{1u}]+2[B_{2u}]+[B_{3u}]$ \\ \hline
		57($T_1T_2$)        & $A_g+2B_{1g}+B_{2g}+2B_{3g}+A_u+B_{2u}+2B_{3u}+[A_g]+[B_{2g}]+[A_u]+2[B_{1u}]+[B_{2u}]$    \\ \hline
		57($R_1R_2$)        & $A_g+2B_{1g}+B_{2g}+2B_{3g}+A_u+B_{2u}+2B_{3u}+[A_g]+[B_{2g}]+[A_u]+2[B_{1u}]+[B_{2u}]$    \\ \hline
		60($R_1R_2$)        & $A_g+2B_{1g}+2B_{2g}+B_{3g}+A_u+2B_{2u}+B_{3u}+[A_g]+[B_{3g}]+[A_u]+2[B_{1u}]+[B_{3u}]$    \\ \hline
		60($T_1T_2$)        & $A_g+B_{1g}+2B_{2g}+2B_{3g}+2A_u+B_{2u}+B_{3u}+[A_g]+[B_{1g}]+2[B_{1u}]+[B_{2u}]+[B_{3u}]$ \\ \hline
		60($U_1U_2$)        & $A_g+B_{1g}+2B_{2g}+2B_{3g}+A_u+B_{1u}+2B_{2u}+[A_g]+[B_{1g}]+[A_u]+[B_{1u}]+2[B_{3u}]$    \\ \hline
		61($S_1S_2$)        & $A_g+2B_{1g}+2B_{2g}+B_{3g}+A_u+2B_{1u}+B_{3u}+[A_g]+[B_{3g}]+[A_u]+2[B_{2u}]+[B_{3u}]$    \\ \hline
		61($T_1T_2$)        & $A_g+2B_{1g}+B_{2g}+2B_{3g}+A_u+B_{2u}+2B_{3u}+[A_g]+[B_{2g}]+[A_u]+2[B_{1u}]+[B_{2u}]$    \\ \hline
		61($U_1U_2$)        & $A_g+B_{1g}+2B_{2g}+2B_{3g}+A_u+B_{1u}+2B_{2u}+[A_g]+[B_{1g}]+[A_u]+[B_{1u}]+2[B_{3u}]$    \\ \hline
		62($S_1S_2$)        & $A_g+2B_{1g}+2B_{2g}+B_{3g}+A_u+2B_{1u}+B_{3u}+[A_g]+[B_{3g}]+[A_u]+2[B_{2u}]+[B_{3u}]$    \\ \hline
		62($R_1R_2$)        & $A_g+B_{1g}+2B_{2g}+2B_{3g}+2B_{1u}+B_{2u}+B_{3u}+[A_g]+[B_{1g}]+2[A_u]+[B_{2u}]+[B_{3u}]$ \\ \hline		130($R_1R_2$)       & $A_g+2B_{1g}+B_{2g}+2B_{3g}+2A_u+B_{1u}+B_{3u}+[A_g]+[B_{2g}]+[B_{1u}]+2[B_{2u}]+[B_{3u}]$ \\ \hline
		138($R_1R_2$)       & $A_g+2B_{1g}+B_{2g}+2B_{3g}+2A_u+B_{1u}+B_{3u}+[A_g]+[B_{2g}]+[B_{1u}]+2[B_{2u}]+[B_{3u}]$ \\ \hline
		205($M_1M_2$)       & $A_g+2B_{1g}+2B_{2g}+B_{3g}+A_u+2B_{1u}+B_{3u}+[A_g]+[B_{3g}]+[A_u]+2[B_{2u}]+[B_{3u}]$    \\ \hline
	\end{tabular}
\vspace{0.5em}

	\begin{tabular}{|c|c|}
		\hline
		Space group momenta & Point group $D_{4h}$                                                                               \\ \hline
		128($A_3A_4$)       & $A_{1g}+A_{2g}+2B_{1g}+2B_{2g}+A_{1u}+A_{2u}+2B_{2u}+[A_{1g}]+[A_{2g}]+[A_{1u}]+[A_{2u}]+2[B_{1u}]$ \\ \hline
		137($A_3A_4$)       & $A_{1g}+A_{2g}+2B_{1g}+2B_{2g}+A_{1u}+A_{2u}+2B_{2u}+[A_{1g}]+[A_{2g}]+[A_{1u}]+[A_{2u}]+2[B_{1u}]$ \\ \hline
	\end{tabular}
\vspace{0.5em}

	\begin{tabular}{|c|c|}
		\hline
		Space group momenta & Point group $C_{6h}$                                                  \\ \hline
		176($A_2A_3$)       & $A_g+B_g+E_{1g}+E_{2g}+A_u+B_u+E_{1u}+[A_g]+[B_g]+[A_u]+[B_u]+[E_{2u}]$ \\ \hline
	\end{tabular}
\vspace{0.5em}

	\begin{tabular}{|c|c|}
		\hline
		Space group momenta& Point group $D_{6h}$                                                    \\ \hline
		193($A_3$)          & $A_{1g}+B_{2g}+E_{1g}+E_{2g}+A_{1u}+B_{1u}+E_{1u}+[A_{2g}]+[B_{1g}]+[A_{2u}]+[B_{2u}]+[E_{2u}]$ \\ \hline
              193($H_1H_3$, $H_2H_4$) &
            $3A_{1g} + 3B_{2g} + A_{2u} + 3B_{1u} + [A_{1g}] + [B_{2g}] + 3[A_{2u}] + [B_{1u}]$ \\ \hline
		194($A_3$)          & $A_{1g}+B_{1g}+E_{1g}+E_{2g}+A_{1u}+B_{2u}+E_{1u}+[A_{2g}]+[B_{2g}]+[A_{2u}]+[B_{1u}]+[E_{2u}]$ \\ \hline

	\end{tabular}
\vspace{0.5em}

        \begin{tabular}{|c|c|}
		\hline
		Space group momenta& Point group $T_{h}$                                                    \\ \hline
		205($R_1R_3$)          & $A_g + 3T_g + [A_g] + [E_g] + [T_g]$ \\ \hline
		205($R_2R_2$)          & $A_g + 3T_g + 3[A_g] + [T_g]$ \\ \hline
	\end{tabular}
 \caption{Symmetries of orbital operators at the 8-fold degenerate points.}
\end{table}

\clearpage

\end{document}